\documentclass{article}

\usepackage[T1]{fontenc}

\usepackage{fullpage}
\usepackage{setspace}
\usepackage{url}

\usepackage[dvipsnames]{xcolor}
\definecolor{linkblue}{HTML}{1B4F8A}
\usepackage[
  colorlinks = true,
  linkcolor  = linkblue,
  citecolor  = linkblue,
  urlcolor   = linkblue,
  bookmarksnumbered,
  pdfusetitle,
]{hyperref}

\usepackage[
  backend      = biber,
  style        = alphabetic,
  sorting      = nyt,
  maxcitenames = 2,
  mincitenames = 1,
  maxbibnames  = 99,   
  giveninits,
  doi          = false,
  url          = false,
  isbn         = false,
]{biblatex}
\usepackage{graphicx}
\usepackage{tabularx}
\usepackage[table]{xcolor}
\usepackage{amsmath, amssymb, amsthm}
\usepackage[capitalise,noabbrev]{cleveref}
\usepackage{longtable}
\usepackage{array}
\usepackage{listings}
\usepackage{multirow}
\usepackage{booktabs}
\usepackage{lmodern}
\usepackage{microtype}
\usepackage{makecell}
\usepackage{tikz}
\usetikzlibrary{arrows.meta,positioning,calc,fit}

\lstdefinestyle{promptfile}{
  basicstyle=\ttfamily\scriptsize,
  backgroundcolor=\color{promptbg},
  frame=single,
  rulecolor=\color{promptframe},
  framesep=4pt,
  xleftmargin=2.5em,
  xrightmargin=0.5em,
  numbers=left,
  numberstyle=\ttfamily\tiny\color{promptnum},
  numbersep=8pt,
  stepnumber=1,
  breaklines=true,
  breakatwhitespace=false,
  breakindent=0pt,
  postbreak=\mbox{\textcolor{promptnum}{$\hookrightarrow$}\space},
  columns=fullflexible,
  keepspaces=true,
  showstringspaces=false,
  tabsize=2,
  upquote=true,
  extendedchars=true,
  literate={—}{{\textemdash}}1,
}%

\definecolor{promptbg}{gray}{0.97}
\definecolor{promptframe}{gray}{0.75}
\definecolor{promptnum}{gray}{0.55}
\newsavebox{\panelbox}
\newcommand{\panel}[2]{%
  \sbox{\panelbox}{#2}%
  \begin{tikzpicture}[baseline=(img.base)]
    \node[inner sep=0pt, outer sep=0pt] (img) {\usebox{\panelbox}};
    \node[anchor=north west, inner sep=2pt,
          font=\sffamily\bfseries\small] at ([xshift=-3pt, yshift=16pt]img.north west) {#1};
  \end{tikzpicture}}

\newcommand{\titletext}{Bridging the Gap in the Responsible AI Divides}

\title{\titletext{}}
\author{%
  B\'alint Gyevn\'ar\textsuperscript{1,*,\dag}\quad
  Atoosa Kasirzadeh\textsuperscript{2,3,\dag}\\[1em]
  \normalsize
  \textsuperscript{1}~Institute for Complex Social Dynamics, Carnegie Mellon University\\[0.1em]
  \textsuperscript{2}~Departments of Philosophy \& Software and Societal Systems, Carnegie Mellon University\\[0.1em]
  \textsuperscript{3}~Google DeepMind, London, UK\\[0.1em]
  \texttt{bgyevnar@cmu.edu}, \texttt{atoosa.kasirzadeh@gmail.com}%
}
\date{}

\begin{document}

\maketitle

\renewcommand{\thefootnote}{\fnsymbol{footnote}}%
\footnotetext[1]{Corresponding author.}%
\footnotetext[2]{These authors contributed equally.}%
\renewcommand{\thefootnote}{\arabic{footnote}}

\begin{abstract}
\noindent
Tensions between AI Safety (AIS) and AI Ethics (AIE) have created what we term a ``responsible AI divide.'' This divide has been persistent, and has dominated the public and policy discourse on AI, and increasingly shapes how governments, firms, and researchers define responsible AI, and so which AI risks they prioritize, which they sideline, and which mitigation strategies they ultimately develop and implement. Here, we first introduce categorization with four major conceptual modes of possible engagement with these tensions: radical confrontation, disengagement, compartmentalized coexistence, and critical bridging. Second, we discuss the advantages and disadvantages of the modes of engagement. Third, we empirically analyze a well-curated set of 3{,}555 research papers roughly representing AIS and AIE. Using computational tools and qualitative analyses, we map the research landscapes of AIE and AIS. We observe that there are several shared research problems between AIE and AIS. However, there are also several thematic divides. For example, we find that AIE has long grappled with problems related to injustice incurred by AI and tangible AI harms, whereas AIS has primarily embodied an anticipatory approach focused on mitigating risks from AI capabilities. At the same time, we find significant overlap in core research concerns across AIE and AIS regarding transparency, reproducibility, and inadequate governance mechanisms. Based on our empirical observations, we claim that critical bridging, with a focus on bridging problems, offers one viable, constructive path for advancing responsible AI. We conclude by identifying bridging problems as a foundation for collaborative AI governance, offering concrete recommendations for integrating shared considerations into responsible AI practice, discussing the limitations of critical bridging, and outlining open problems for future research.
\end{abstract}

\section{Introduction}

Tensions between ``AI Safety'' (AIS) and ``AI Ethics'' (AIE) increasingly shape how governments, firms, and researchers define ``responsible AI,'' and thus which risks from AI systems are prioritized, which are sidelined, and which risk mitigation strategies are ultimately developed and implemented. 

Over the past few years, the relationship between AIS and AIE has grown more complex and contentious \cite{Schechner2023AI,Richards2023Illusion,NatureEditorial2023AIDoomsday}, giving rise to what we call the ``responsible AI divide'': ongoing struggles between AIS and AIE over how to prioritize and allocate scarce resources — such as attention, public awareness, preparedness, and funding — across competing concerns about the development, deployment, and governance of AI systems. We use ``responsible AI'' as a pragmatic umbrella term encompassing the various technical and sociotechnical efforts aimed at ensuring AI systems are developed and deployed safely, ethically, and for societal benefit. These tensions arise from divergent views about the origins, scope, salience, and tractability of AI risks, as well as the appropriate responses to them~\cite{piper2022factions,leswing2023parrots,ahmed2023building,kasirzadeh2024two,gyevnarAISafetyEveryone2025,roytburgMindGapPathways2025}. As AI development accelerates, the stakes for responsible implementation have never been higher. How can we keep tensions between AIE and AIS from foreclosing opportunities for constructive engagement? We respond to this question in two ways.

\begin{figure}
    \centering
    \begin{minipage}[c]{0.56\linewidth}
        \panel{(a)}{\resizebox{\linewidth}{!}{\begingroup
\sffamily
\begin{tikzpicture}[
    font=\normalsize,
    axis/.style={thick, -{Latex[length=3mm]}},
    modebox/.style={
        rounded corners=3pt,
        draw=#1!70!black,
        fill=#1!12,
        very thick,
        align=left,
        text width=5cm,
        inner sep=6pt,
        outer sep=3pt
    },
    quadrant/.style={
        font=\large\itshape,
        align=center,
        text width=5cm,
        inner sep=0pt,
        outer sep=3pt
    }
]

\matrix (layout) [
    column sep=6mm,
    row sep=2mm,
    inner sep=0pt,
    outer sep=3pt
] {
    \node[modebox=blue] (coex) {%
        {\large\bfseries Compartmentalized co-existence}\\[2pt]
        Shared AIS-AIE venues without meaningful engagement.\\[3pt]
        \textbf{Practices:} parallel tracks, representation without
        contestation, divided labor.\\[3pt]
        \textbf{Risk:} talking past one another, rhetorical co-option.
    };
    &
    \node[modebox=green!60!black] (bridge) {%
        {\large\bfseries Critical bridging}\\[2pt]
        Structured engagement of AIS-AIE with friction and
        selective collaboration.\\[3pt]
        \textbf{Practices:} adversarial workshops, joint problem
        definition, co-designed governance.\\[3pt]
        \textbf{Risk:} difficult and not always feasible across
        communities and philosophies.
    };
    \\

    \node[quadrant] (highleft) {%
        low engagement\\high integration
    };
    &
    \node[quadrant] (highright) {%
        high engagement\\high integration
    };
    \\[4mm]

    \node[modebox=gray] (dis) {%
        {\large\bfseries Disengagement}\\[2pt]
        Refusal to engage AIE--AIS tensions framed as neutrality.\\[3pt]
        \textbf{Practices:} avoiding comparison, deflecting critique,
        avoiding tensions due to perceived dramatic nature.\\[3pt]
        \textbf{Risk:} unresolved disagreements beneath surface calm.
    };
    &
    \node[modebox=orange] (conf) {%
        {\large\bfseries Radical confrontation}\\[2pt]
        Direct efforts to discredit or delegitimize either
        the AIE or AIS side.\\[3pt]
        \textbf{Practices:} polemics, adversarial framing,
        legitimacy boundary work.\\[3pt]
        \textbf{Risk:} polarization and reduced space for
        nuanced positions.
    };
    \\

    \node[quadrant] (lowleft) {%
        low engagement\\low integration
    };
    &
    \node[quadrant] (lowright) {%
        high engagement\\low integration
    };
    \\
};

\node[
    fit=(dis)(conf),
    inner sep=0pt,
    outer sep=0pt
] (lowerboxes) {};

\coordinate (divider) at
    ($(highleft.south)!0.5!(lowerboxes.north)$);

\coordinate (origin) at
    ($(layout.south west)+(-4mm,-4mm)$);
\coordinate (extent) at
    ($(layout.north east)+(4mm,4mm)$);
\coordinate (xend) at (origin -| extent);
\coordinate (yend) at (origin |- extent);

\draw[axis] (origin) -- (xend);
\draw[axis] (origin) -- (yend);

\draw[densely dashed, gray!60]
    (layout.center |- origin) --
    (layout.center |- extent);

\draw[densely dashed, gray!60]
    (origin |- divider) --
    (extent |- divider);

\node[
    font=\Large\bfseries,
    anchor=north,
    yshift=-7mm
] at ($(origin)!0.5!(xend)$) {Engagement};

\node[
    font=\Large\bfseries,
    anchor=south,
    rotate=90,
    xshift=0pt,
    yshift=0pt
] at ($(origin)!0.5!(yend)+(-9mm,0)$) {Integration};

\node[font=\large, anchor=north west, yshift=-1mm]
    at (origin) {Low};
\node[font=\large, anchor=north east, yshift=-1mm]
    at (xend) {High};

\node[font=\large, anchor=south east, xshift=-1mm]
    at (origin) {Low};
\node[font=\large, anchor=north east, xshift=-1mm]
    at (yend) {High};

\end{tikzpicture}
\endgroup}}
    \end{minipage}%
    \hfill
    \begin{minipage}[c]{0.43\linewidth}
        \centering
        \panel{(b)}{\includegraphics[width=\linewidth]{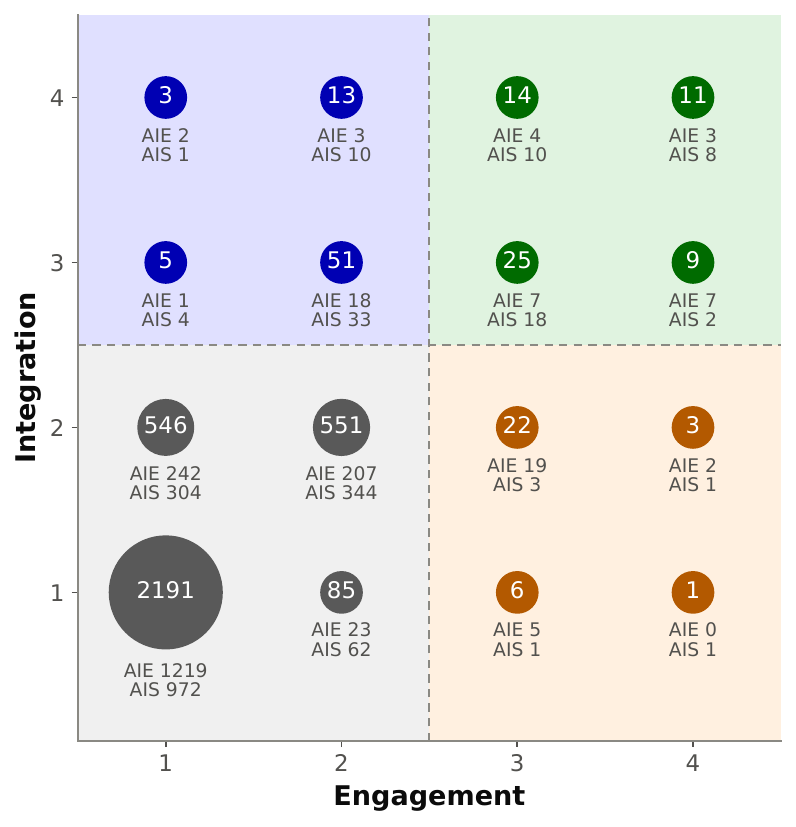}}\\[-3mm]
        \panel{(c)}{\includegraphics[width=\linewidth]{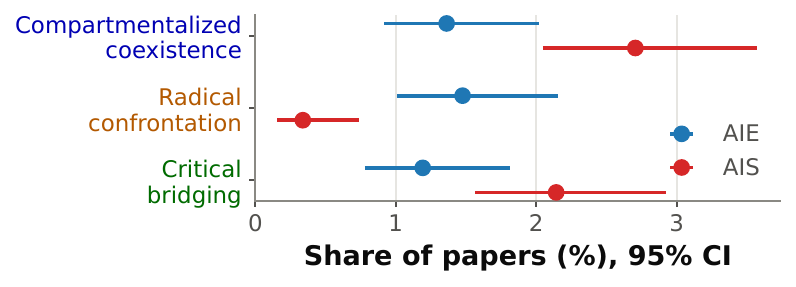}}
    \end{minipage}
    \caption{\textbf{(a)} Four engagement modes for responsible AI divides. The modes differ in the extent to which they have direct \textit{engagement} with tensions between AI ethics (AIE) and AI safety (AIS), and in whether such engagement produces constructive \textit{integration} or instead reproduces division. We distinguish between three modes of critical bridging: research problems, communities, philosophies; we present evidence only about the \textit{critical bridging of research problems} here. \textbf{(b)} Assignment of our corpus to different engagement modes based on their degree of engagement and integration, as operationalized in~\cref{ssec:results:modes}. Circles are not to scale, and 19 papers without an online-available full text are excluded. \textbf{(c)} Distribution of the rare engagement modes compared between AIE/AIS.}
    \label{fig:resolution-modes}
\end{figure}

First, we distinguish four modes of engaging with the responsible AI divides, illustrated in~\cref{fig:resolution-modes}: \emph{(1)~radical confrontation}, characterized by active, direct efforts to discredit or delegitimize the opposing side; \emph{(2)~disengagement}, characterized by refusal to engage substantively with the other side; \emph{(3)~compartmentalized coexistence}, characterized by pursuing separate agendas in a live-and-let-live fashion while acknowledging the other side only rhetorically; and \emph{(4)~critical bridging}, characterized by explicitly mapping points of divergence while seeking good-faith common ground and concrete opportunities for collaboration in ways that empower joint efforts to scale responsible AI without flattening or suppressing unresolvable disagreement. We discuss the potential benefits and disadvantages of each mode of engagement.

Second, we investigate whether there are any empirical justification for supporting critical bridging. The notion of ``bridging'' itself is ambiguous and can be interpreted in three (potentially overlapping) ways: (i) bridging \emph{communities}, by forging institutional, social, or epistemic links between AIS and AIE communities; (ii) bridging \emph{worldviews and philosophical standpoints} by highlighting the underlying moral and political values that shape each community; and (iii) bridging \emph{problems}, by identifying shared problem spaces and challenges. Bridging communities (i) and worldviews (ii) might not be feasible or desirable. Feasibility could be constrained by the fact that these communities prioritize fundamentally different problems and funding sources. Additionally, bridging may not even be desirable: preserving epistemic diversity allows different communities to pursue parallel, independent lines of inquiry without being diluted by a forced synthesis. A systematic assessment of these institutional and incentive divides requires anthropological and sociological fieldwork \cite{ahmed2023building,gebru2024tescreal}, which lies outside the scope of our methodology. We investigate the potential for bridging AIE and AIS problems in four phases. (1) We create a visual map of engagement modes in AIE and AIS research papers based on a systematic literature review. (2) We identify common research problems, risks, and mitigation strategies addressed across both fields. (3) We identify the non-overlapping, distinct research problems addressed in each. (4) We identify detailed points of agreement and disagreement within each engagement mode regarding papers' stance towards risks and mitigation strategies. (5) We discuss whether and how AIE and AIS research problems can be bridged, what aspects cannot, and how this bridging process might proceed.

By mapping connections between shared challenges and clarifying genuine disagreements, we can provide partial progress toward a shared vision for responsible AI---one that builds new capacities for collaborative epistemics and collective action. To be clear, we do not claim that critical bridging of problems is the preferred way of handling all the tensions, or even the main tensions, between AIE and AIS, nor that it is the only one. Other modes are available, including \textit{disengagement}, where AIE and AIS pursue their own agenda independently, and \textit{radical confrontation}, where one directly challenges the other's assumptions or priorities. Each mode may be appropriate depending on the context. Our claim, however, is narrower: bridging these problem-spaces is critically important. Currently, many open challenges in responsible AI, for example deciphering how machines reason or determining how to globally distribute AI's benefits, are investigated by AIE and AIS in isolated silos. By failing to share insights, both groups miss opportunities for collaborative progress. We highlight one particular mode of engagement that already exists in practice and argue in favor of boosting that; under certain conditions, this mode offers distinct advantages that isolated approaches cannot replicate. 

Recent developments, including the launch of professional bodies such as the International Association for Safe and Ethical AI (IASEAI) \cite{IASEAI2025}, research explicitly examining the complex intersection of AIE and AIS \cite{christian2020alignment,kasirzadehTwoTypesAI2025,gyevnarAISafetyEveryone2025}, as well as our own results point to bridging efforts already underway. Building on these efforts, we advance a bolder bridging agenda and provide empirical justification for it.

While responsible AI is a broad and evolving field, AIE and AIS represent a critical subset of it. We examine our characterization of these two as contemporary representatives of its defining debates~\cite{birhaneForgottenMarginsAI2022,lauferFourYearsFAccT2022}. Our investigation is based on a combined dataset of $N=3{,}555$ papers retrieved as of July 9, 2025. The AIE corpus ($n=1{,}775$) comprises the complete proceedings metadata of the AAAI/ACM Conference on AI, Ethics, and Society (AIES) and the ACM Conference on Fairness, Accountability, and Transparency (FAccT) since their founding in 2018. These are the two main venues for publishing what is often described in contemporary discourse as AI Ethics. Because there are no exclusive peer-reviewed AIS conferences, the AIS corpus ($n=1{,}780$) builds on an open-source dataset developed by Gyevn\'ar and Kasirzadeh~\cite{gyevnarAISafetyEveryone2025}, which we expanded with metadata from 35 prominent safety labs and individual researchers. For details on included work and retrieval criteria, see the Supplementary Materials.
Using this dataset, \cref{sec:bridging} maps the distinct and overlapping risks and mitigation strategies of AIE and AIS, identifying concrete research problems that are ripe for critical bridging.

\section{Responsible AI Divides}\label{sec:ai-wars}

We distinguish four conceptual modes of engagement with the responsible AI divides: radical confrontation, disengagement, compartmentalized coexistence, and critical bridging. The following paragraphs outline each mode, with a visual summary presented in~\cref{fig:resolution-modes}(a).

To provide the visual map of how these modes play out in practice, we assign a score to each paper in our corpus along their degrees of integration between and engagement with AIE and AIS, as described in~\cref{ssec:methods:analysis}.
Based on these scores, we classify each paper to one of the four engagement modes,. The resulting distribution, shown in~\cref{fig:resolution-modes}(b), is strikingly uneven: disengagement accounts for $95.4\%$ of all papers, while the remaining 163 papers are split across compartmentalized coexistence (72), critical bridging (59), and radical confrontation (32). As~\cref{fig:resolution-modes}(c) shows, the two fields differ in how they distribute across these rarer modes, with AIE papers more prone to radical confrontation and AIS papers more prone to compartmentalized coexistence, yet also critical bridging. We return to the detailed examination of these patterns in~\cref{ssec:results:modes}.

\paragraph{Radical Confrontation} 

While confrontation is grounded in critique, its more radical forms often devolve into \textit{ad hominem} attacks, epistemic entrenchment, and an erosion of the shared foundations necessary for productive dialogue. At its best, confrontation is productive: it can make visible a plurality of epistemic views and ideals about responsible AI (e.g., probabilistic risk modeling vs.\ structural accounts of harm), clarify which voices and forms of evidence are treated as authoritative (e.g., a dominant set of narrow technical benchmarks vs.\ a substantive and critical approach to evaluating AI systems~\cite{bean2025measuring}), and surface the boundary work by which communities police legitimacy \cite{kasirzadehTwoTypesAI2025} (e.g., who counts as doing ``real'' safety or ``serious'' ethics, which venues confer status, which problem framings are deemed funding-worthy or policy-relevant, or how financial power shapes the epistemic dynamics of the public sphere). However, confrontation can also devolve into oversimplification; for instance, imputing malign intent to foundational figures in AIS, or dismissing AIE work as coming from a community of "AI capability deniers" or focused on ``non-technical'' problems irrelevant to humanity's long-term future.

\paragraph{Disengagement}

Disengagement is marked by a deliberate refusal to engage with the substance of AIE–AIS tensions, often under the banner of neutrality  or by relegating the conflict to a second-order problem. Here, by a second-order concern we mean cases wherein an issue is framed as auxiliary or derivative to the primary task. When actors dismiss AIE–AIS tensions as second-order, they relegate deep ethical, political, and institutional conflicts to the status of minor procedural hurdles or implementation details, arguing that the first-order work of technical capabilities must take absolute precedence. This stance could offer some utility by allowing researchers to bypass divisive normative debates, protect institutional funding from political polarization, and maintain focus on concrete milestones. In practice, disengagement can take the form of (1) avoiding explicit comparison between safety and ethics agendas in papers and talks; (2) deflecting hard questions with calls to focus on the technical work; or (3) treating tensions as mere community drama rather than as disagreements about values, evidence, governance, and anticipatory trajectories. While this stance can reduce short-term friction, it leaves underlying differences unexamined. By avoiding genuine points of contention, disengagement can entrench oversimplified narratives (e.g., that AI safety practitioners are uniformly committed to strong longtermism \cite{greaves2024case}, or that AI ethics researchers detest utilitarian ethics or are AI capability deniers). The result is a shallow shared understanding that makes consensus on AI governance priorities harder to reach and increases the risk of ill-informed institutional and policy decisions. This stance could unintentionally fuel public misunderstanding while hardening sectarian camps and political polarization about responsible AI.

\paragraph{Compartmentalized Coexistence}

Compartmentalized coexistence describes situations where different groups share venues---conferences, standards meetings, multi-stakeholder forums---while effectively talking past one another. This often looks like (1) parallel tracks or panels that never meaningfully intersect; (2) representation-without-contestation, where each community states its priorities but avoids difficult questions about the deep tensions or disagreements; or (3) collaborations that divide labor cleanly (e.g., you handle fairness, we handle alignment) without addressing how the pieces interact. The result is an appearance of unity without substantive integration. Yet, this approach could be highly functional in professional and institutional settings. In principle, it could allow divergent groups to build broad coalitions, and could project a unified front to external regulators without having to resolve some complex and deep ideological divides. This strategy does little to resolve underlying tensions or clarify compatibilities. It can sustain parallel, non-intersecting research agendas and allow important issues to fall through the cracks precisely because they sit at the intersection of AIE and AIS. It can also encourage reinvention, where concepts and epistemic tools developed in one subfield are ``rediscovered'' in another without cross-citation or learning. Finally, compartmentalization can enable superficial co-optation: actors may invoke the other community's concerns as rhetorical cover, without meaningfully changing their methods, priorities, or governance commitments.

\paragraph{Critical Bridging}

Critical bridging maps irreconcilable differences while also building selective, durable forms of collaboration. In practice, it involves (1) structured disagreement (e.g., adversarial workshops, debate formats, joint red-teaming across paradigms); (2) shared problem definition (e.g., explicitly specifying where safety and ethics risks interact); and (3) joint institutional work (e.g., co-designed evaluation regimes, standards, or governance proposals). Unlike coexistence, bridging requires engaging with points of friction directly rather than avoiding them. However, the concept of ``bridging'' itself is ambiguous and has at least three (overlapping) interpretations. First, it can be viewed as connecting communities, linking the disparate epistemic cultures, researchers, and practitioners of AI safety and AI ethics. Second, it can involve reconciling diverse moral and political standpoints, addressing the underlying value systems that drive different normative priorities. Third, it can operate at the level of problem spaces, identifying shared challenges that demand simultaneous safety and ethical consideration. Bridging may not always be feasible or desirable in all its forms. The feasibility and extent of bridging vary across dimensions, creating a vast space for exploration. We leave the question of whether bridging these communities and worldviews is feasible to field-work sociologists and anthropologists, as such empirical inquiry lies outside the scope of our methodology.

In this paper, we focus primarily on bridging \emph{problems}, which we anticipate will have a downstream effect on bridging communities by potentially giving rise to new, integrated spaces for inquiry. Our paper takes several steps towards identifying joint research problems and concerns, characterizing differences, and explicitly formulating what can and cannot be bridged. \Cref{sec:discussions} details the reasoning behind our scope and its limitations---notably our choice not to engage with bridging communities or philosophical viewpoints, as this would require anthropological or sociological research beyond the scope of this paper. We acknowledge, however, that such work is extremely important and warrants dedicated attention elsewhere.

\section{Data Collection and Analysis}

The AIE corpus comprises 1,775 papers drawn from the proceedings of the two largest conferences in AIE: the \textit{AAAI/ACM Conference on Artificial Intelligence, Ethics, and Society} (AIES; $n=833$) and the \textit{ACM Conference on Fairness, Accountability, and Transparency} (FAccT; $n=942$), covering all years since their founding in 2018. These venues are widely considered representative of contemporary topics in AI ethics~\cite{birhaneForgottenMarginsAI2022,lauferFourYearsFAccT2022}. The AIS corpus comprises 1,780 papers, assembled by extending a prior open-source collection of peer-reviewed AIS publications~\cite{gyevnarAISafetyEveryone2025} with papers manually retrieved from 35 research labs and individual authors identified in prior surveys~\cite{amodeiConcreteProblemsAI2016,hendrycksUnsolvedProblemsML2022,technicalitiesShallowReviewTechnical2024,bengioInternationalAISafety2025} (see Supplementary Materials for the full list).
Because late-breaking AIS research is frequently disseminated as preprints and blog posts rather than through traditional peer review~\cite{ahmedBuildingEpistemicCommunity2023,kirchnerResearchingAlignmentResearch2022}, we supplemented the corpus with preprints from the websites of the 35 labs and researchers described above, applying the same inclusion and exclusion criteria as Gyevn\'ar and Kasirzadeh~\cite{gyevnarAISafetyEveryone2025}. Papers were retrieved through July 9, 2025. After de-duplication, the combined dataset comprises $N=3{,}555$ papers. 
Following established practice~\cite{passasBibliometricAnalysisMain2024,klarinHowConductBibliometric2024}, all bibliometric analyses are based on paper metadata only, primarily titles and abstracts.
We also retrieved the full text of each paper, where available, for analyses using an LLM-as-a-judge setup.
We could not retrieve the full text of 19 papers, which were excluded from the LLM-based analyses.
Our annotated corpora as well as all processing code are available in our Supplementary Materials at: \url{https://github.com/gyevnarb/ai-safety-ethics}.
 
\subsection{Quantitative Corpus Exploration}
\label{ssec:methods:quantitative} 
We used quantitative methods during the initial data exploration phase to inform the subsequent qualitative taxonomy. Text data were pre-processed using a standard pipeline of tokenization, stopword removal, and lemmatization~\cite{chaiComparisonTextPreprocessing2023,hickmanTextPreprocessingText2022} implemented in NLTK~\cite{birdNLTKNaturalLanguage2004} and spaCy~\cite{montaniExplosionSpaCyV3722023}. 
We generated term co-occurrence networks using VOSviewer~\cite{vaneckVOSNewMethod2007} and performed unsupervised topic modeling with BERTopic~\cite{grootendorstBERTopicNeuralTopic2022}, which clusters documents based on dense sentence embeddings from SentenceBERT~\cite{reimersSentenceBERTSentenceEmbeddings2019}. 
To characterize lexical differences between the corpora, we computed log-likelihood ratios of term frequencies following Monroe~\textit{et al.}~\cite{monroeFightinWordsLexical2008}. We assessed temporal trends in cross-corpus similarity by computing the average pairwise cosine similarity of SentenceBERT embeddings between documents from each corpus, grouped by publication year.
 
\subsection{Qualitative Analysis of Risks and Mitigation Strategies}
\label{ssec:methods:qualitative}
Building on the quantitative exploration, we manually conducted a deductive coding process~\cite{pearseGuidelinesTheoryDevelopment2021,azungahQualitativeResearchDeductive2018,feredayDemonstratingRigorUsing2006} to develop a two-level taxonomy of risks and mitigation strategies. 
Low-level categories were iteratively refined over three rounds by the authors and organized into high-level categories adapted from prior work~\cite{weidingerEthicalSocialRisks2021,bird2023typology,gyevnarAISafetyEveryone2025}. 
One author then annotated both corpora under two schemes: (1) using only the taxonomy of the paper's own field, and (2) using the full combined taxonomy. Each paper was assigned up to three low-level categories. To minimize annotator bias and sequence effects, annotation order was randomized, only titles and abstracts were shown, and the order of available categories was shuffled for each paper. Results are presented in three parts: a quantitative analysis of cross-corpus similarity and difference, a comparative analysis of risks addressed by each field, and a comparison of proposed mitigation strategies.

\subsection{Quantifying Engagement Modes}
\label{ssec:methods:analysis}

To determine which engagement mode best describes each document in our corpora, we operationalize the two axes shown in~\cref{fig:resolution-modes}(a) by defining increasing degrees of engagement and integration on integer scales from 1 to 4 as follows.

\paragraph{Engagement.} 
How directly and substantively a paper in one field engages with the other field's positions, priorities, and arguments, or with the tensions between AIE and AIS, such as their differing priorities, evidence standards, and risk framings. Engagement can be critical or constructive:
\begin{enumerate}
    \item The other field is never addressed, and the AIE--AIS relationship is never mentioned.
    \item The other field is named or cited in passing, such as in related work or motivation, but without discussing its positions.
    \item The paper engages with the other field's positions, or the AIE--AIS tension is a major thread of the paper.
    \item The AIE--AIS relationship, or the other field's positions, is the central topic of the paper.
\end{enumerate}

\paragraph{Integration.} 
To what extent a paper brings together the concepts of AIE and AIS, their methods, or their problem framings within one framework, instead of working within a single field or setting the fields against each other:
\begin{enumerate}
    \item The paper works entirely within its field's framing, or treats the other field only as an opponent to be discredited.
    \item Topics or techniques from both fields appear in the paper, but only in passing or purely as methodology rather than as concerns of the other field.
    \item The paper explicitly argues that the two fields' concerns are connected, showing how one problem is both an ethics problem and a safety problem.
    \item The paper builds a genuine joint framing, problem definition, method or governance proposal that serves both fields.
\end{enumerate}

To assign each paper a specific engagement and integration score, we use an LLM-as-a-judge setup with GPT-6 Luna. We chose this model for its cost efficiency and high performance for large-scale classification tasks.
This setup is also why we use four integer scales for each axis: it avoids a middle value that the classifier may default to when it is uncertain about its assignment.
We provide the classifier with the title, abstract, and, where available, full text of each paper (19 papers did not have full-text PDFs online).
In addition to the scores, we also ask the LLM to annotate each paper on whether it references the other field at all.
We verified the LLM's score assignments by annotating ourselves a subset of the corpora containing all 172 papers initially assigned a rare engagement mode as well as 100 randomly sampled disengagement papers, giving an initial inter-annotator agreement of Cohen's $\kappa = 0.48$ for engagement and $0.38$ for integration.
Next, where scores disagreed, we inspected the LLM's rationale and adjusted our judgment if it was justified. 
This process yielded a total of 163 papers in the rare engagement modes, with substantial final agreement (Cohen's $\kappa=0.68$ and $0.62$ for engagement and integration, respectively~\cite{landisMeasurementObserverAgreement1977}).
Given the assigned scores, we categorize each paper into one of the four engagement modes, treating scores below 3 as low.

\subsection{Qualitative Comparison of Rare Engagement Modes}
\label{ssec:methods:coding}

In order to compare how the 163 papers in the three rare engagement modes engage with particular risks and mitigation strategies, we perform an inductive qualitative coding and thematic analysis of their full texts using GPT-6.1 Sol.
For every code, the model records a paper's stance towards the code's content as one of four values: a paper (1) \textit{endorses} a risk or mitigation strategy when it treats the risk as serious or advocates for a mitigation strategy; (2) \textit{qualifies} it when it accepts the risk or mitigation strategy only under certain conditions; (3) \textit{contests} it when it argues that the risk is overstated or misframed, or that the mitigation strategy is inadequate or harmful; and (4) \textit{describes} it when it reports the risk or mitigation strategy without taking a position.

The coding proceeds in four steps, following the phases outlined in~\cite{braunUsingThematicAnalysis2006}.
In the first phase, the model codes the papers in a random order one at a time, recording its codes in a shared codebook, with supporting text spans and definitions.
Second, the model passes over every paper again in a different random order using the now complete codebook, revising any redundant or adding new codes.
This process gave 94 unique codes.
In the third phase, the fixed codes are grouped into themes, with each code belonging to exactly one theme.
In the fourth phase, the themes are refined by selecting the most  representative quotations for each of them that best reflect the variety of stances within the theme.
This grouping gave 15 themes.

\section{Mapping Research Problems in AI Safety and AI Ethics}
\label{sec:bridging}

We begin to map the opportunities for critical bridging along three axes. First, we compare the distinct types of risks and proposed mitigation strategies discussed in AIE and AIS using the quantitative distribution of terms. Second, we use our annotations to identify overlaps between the two fields across both risk types and mitigation strategies, highlighting several promising problems for bridging. Throughout, we cite representative sources when discussing a specific risk type or mitigation strategy; these citations are drawn randomly from the annotated papers to avoid selection bias and are not intended to be exhaustive.

\subsection{Distinct Research Problems}

\begin{figure*}
  \centering
  \includegraphics[width=0.48\linewidth]{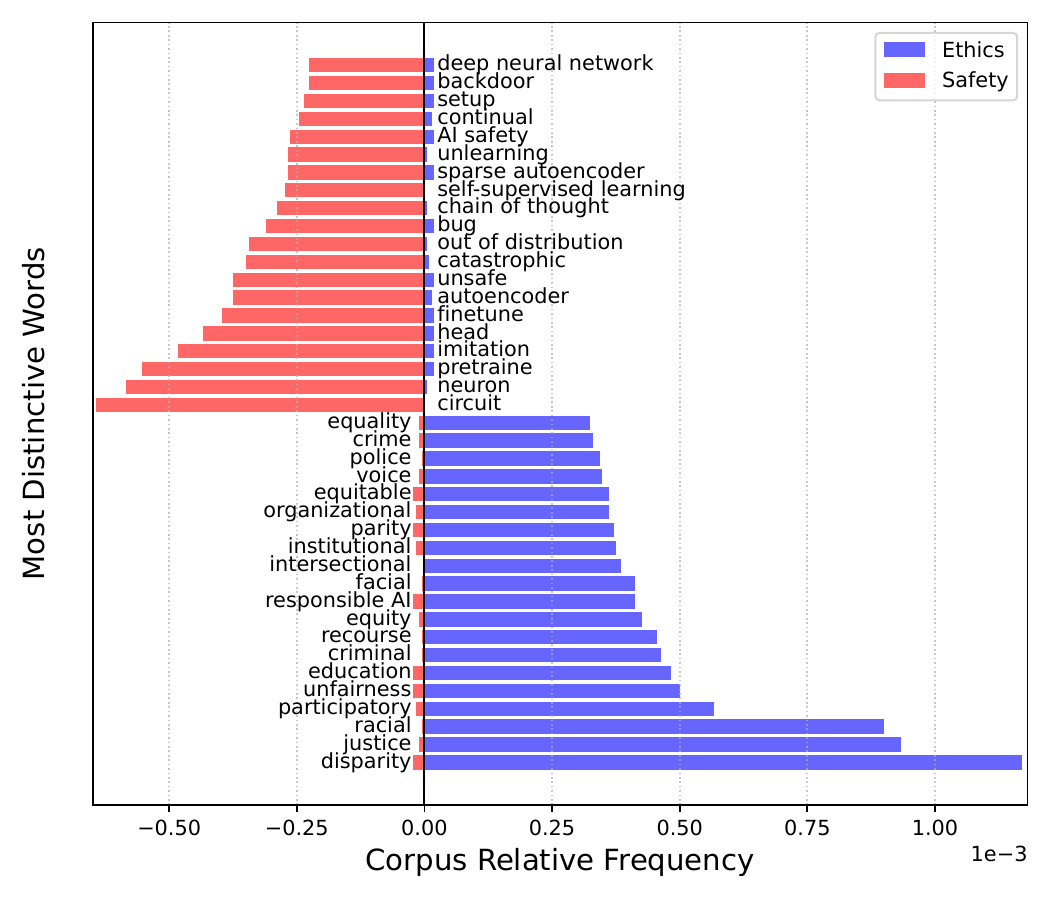}
  \includegraphics[width=0.48\linewidth]{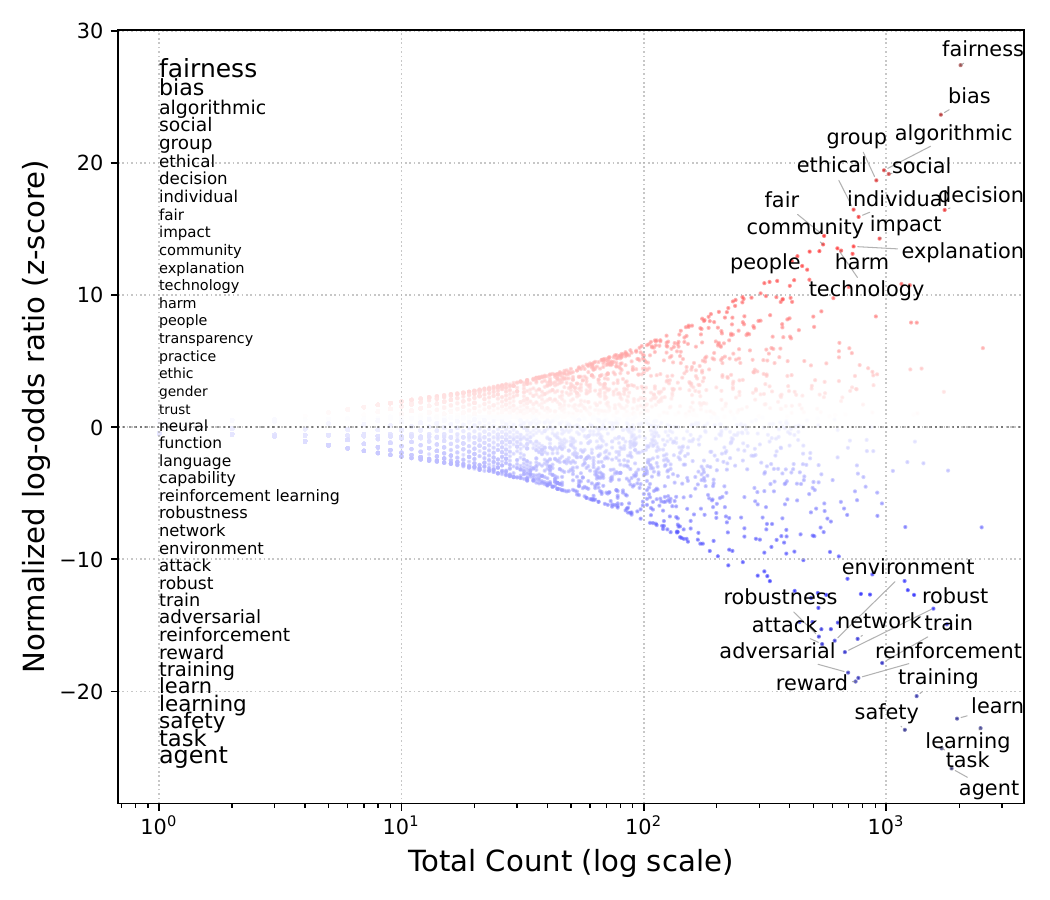}
  \caption{Two ways of showing distinctive words in the corpora. \textit{Left}: The most frequent words in one corpus that occur least often in the other. \textit{Right}: The log-odds ratio of each word plotted against its occurrence count. A more positive ratio indicates that the term is more likely to occur in the AIE corpus, while a more negative ratio indicates that the term is more likely to occur in the AIS corpus. Red color denotes AIS terms; blue color denotes AIE terms.}
  \label{fig:distinctive-words}
\end{figure*}

\Cref{fig:distinctive-words} shows how the distribution of distinctive words differs between the corpora.
Looking at the relative frequencies of words (almost) exclusive to one corpus, we see that only the AIE corpus is concerned with risks from (racial) disparity~\cite{kwegyir-aggreyObservingContextImproves2024}, injustice~\cite{kasirzadeh2022algorithmic}, lack of education~\cite{munWhyNotUse2025}, lack of recourse~\cite{ustunActionableRecourseLinear2019}, and facial recognition~\cite{owensUnderstandingExperiencesCompulsory2025}, among others.
On the right side of the log-likelihood plot in~\cref{fig:distinctive-words}, we see that AIE primarily studies the social and systemic impacts of AI deployment, especially algorithmic bias~\cite{kingPrivacybiasTradeoffData2023}, fairness~\cite{chenPersonalizedPricingGroup2023}, and human-centered harms~\cite{morrisonHumanCenteredApproachIdentifying2025}.
Overall, concerns mostly exclusive to AIE include risks from a lack of understanding of pluralistic values, climate and environmental inequity impacts, the discrepancy between actual and perceived fairness, algorithmic management in the workplace, a lack of recourse from automated decision-making systems, and non-consensual data collection practices, among others.

In contrast, only the AIS corpus uses words referring to the catastrophic risks of AI~\cite{bengioSuperintelligentAgentsPose2025}, technical bugs~\cite{casperRedTeamingDeep2023}, out-of-distribution errors~\cite{mcinroePlanningGoOutofDistribution2024}, or backdoor attacks~\cite{langoscoDetectingBackdoorsMetaModels2023}.
AIS primarily studies the technical capabilities of the models and algorithms that drive modern AI systems, especially in the context of reinforcement learning~\cite{gleaveAdversarialPoliciesAttacking2021}, adversarial attacks~\cite{vivekGrayboxAdversarialTraining2018}, and robustness~\cite{jiralerspongRobustReinforcementLearning2025}.
Concerns mostly exclusive to AIS include unsafe state and action exploration, external (malicious) tampering with agents' reward signals, mesa-optimization, recursively self-improving agents, a lack of robustness from overfitting, multi-agent collusion, and risks from unfaithful chain-of-thought reasoning in LLMs, among others.

Shifting focus from risks to mitigation strategies, we find that only the AIE corpus discusses participatory methods with human stakeholders~\cite{ullsteinParticipatoryAIEU2025}, intersectional fairness~\cite{chungAISocialContract2025}, equitable decision-making~\cite{wallerConsistencyNuancedMetrics2025}, and institutional transparency~\cite{narayananWelfaristMoralGrounding2023}.
These are methodological approaches which are almost entirely absent from AIS.
Investigations of the current and near-term effects of AI on labor conditions and the environment also feature much more prominently in AIE.

On the other hand, discussions of methods exclusive to AI safety include most mechanistic interpretability approaches, such as sparse autoencoders~\cite{cunninghamSparseAutoencodersFind2023} and circuit analysis~\cite{lieberumDoesCircuitAnalysis2023}, as well as continual learning~\cite{abelDefinitionContinualReinforcement2023}, self-supervised learning~\cite{siddiquiBlockwiseSelfSupervisedLearning2023}, machine unlearning~\cite{krishnanNotAllData2025}, and chain-of-thought monitoring~\cite{bakerMonitoringReasoningModels2025}.
Furthermore, AIS is also extensively concerned with the robustness and exploration behavior of RL agents, and while AIE also considers agents, it is usually interested in how AI systems enable more inclusive human interactions.
Finally, imagining what a world after the arrival of artificial general intelligence would look like is an approach exclusively taken in AIS.

\subsection{Overlapping Risk Categories}

\begin{figure}
  \includegraphics[width=0.49\linewidth]{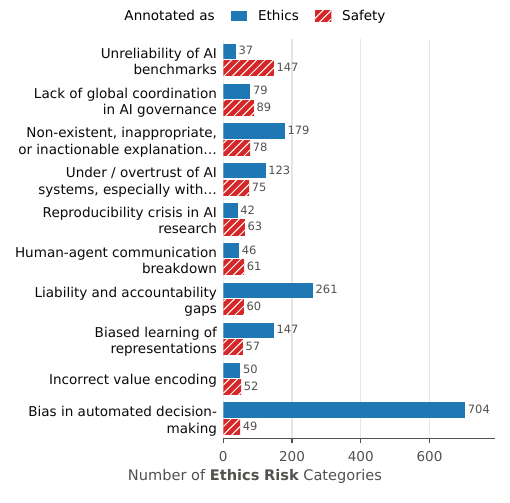}
  \hfill
  \includegraphics[width=0.49\linewidth]{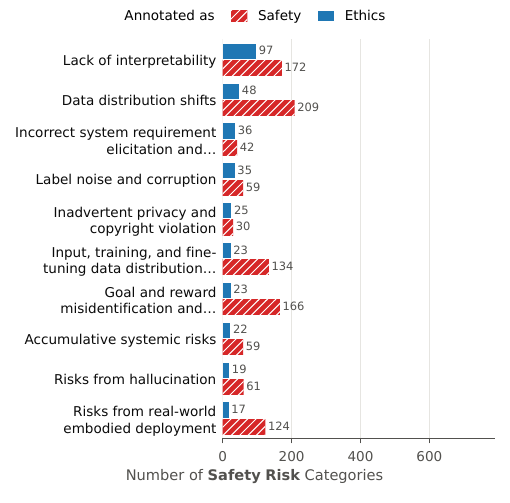}
  \caption{Bar plots of \textbf{risk similarities}, showing the top seven most frequent \emph{low-level} risk categories in which a paper's source corpus \emph{differs} from the corpus associated with the annotated category (e.g., a paper from the AIE corpus annotated with a category from the AIS taxonomy).}
  \label{fig:risk-sim}
\end{figure}

Overall, we identify six high-level risk categories for AIE, comprising 60 low-level categories that range from bias in automated decision-making and AI-driven polarization to emotional over-reliance on AI and crowdworker abuse.
For AIS, we identify seven high-level risk categories comprising 56 low-level categories, such as jailbreaking, sycophancy, brittle representations, and open-weight model vulnerabilities.
Despite these differences, several risk categories are addressed by both fields, as shown in~\cref{fig:risk-sim}.
We highlight seven such categories here, and discuss how they overlap and differ across the two fields.

The risks from a \emph{lack of interpretability and explainability} are prominent in both fields.
However, there are nuances in how each field looks at these problems.
AIE has examined risks from non-existent~\cite{menonLessonsClinicalCommunications2024}, inappropriate~\cite{lakkarajuHowFoolYou2020}, or inactionable~\cite{ustunActionableRecourseLinear2019} explanations of automated decision-making systems, while AIS has focused more on mechanistic interpretability from the perspective of understanding~\cite{kissaneInterpretingAttentionLayer2024}, monitoring~\cite{kuznietsovExplainableAISafe2024}, and controlling~\cite{singhMeasuresDisparityTheir2023} AI systems' behavior.
Both AIE and AIS recognize the need for intelligible explanations that accurately calibrate trust in AI systems' capabilities~\cite{conitzerWhyShouldWe2024,millerExplainableAIDead2023}.
In addition, recent work from AIE has offered mathematical accounts of mechanistic interpretability~\cite{ayonrindeMathematicalPhilosophyExplanations2025}, and examined the risks that opaque representation learning poses to fairness and justice~\cite{bensonUncoveringLinguisticRoots2025,heGeometricSolutionFair2020}.
Relatedly, there is shared interest in risks from \emph{misspecifying system requirements} and capabilities~\cite{skalseMisspecificationInverseReinforcement2023,barcoTacklingProblemDistributional2024}, due to a lack of understanding of how AI systems work or should work.
This more systemic risk is also addressed in AIE, though primarily from a human-computer interaction perspective, examining risks from breakdown in human-AI communication~\cite{robertsonUnderstandingBeingUnderstood2022,maedaAnthropomorphismSocialAffordance2025} and emergent biases in real-world deployments of AI~\cite{ghoshFairCanaryRapidContinuous2022,rastogiSupportingHumanAICollaboration2023}.

Both AIE and AIS also investigate \emph{issues around reproducibility and benchmarks}. Both examine the risks of incorrectly specified and inflated benchmarks~\cite{kwegyir-aggreyMisuseAUCWhat2023,cobbeQuantifyingGeneralizationReinforcement2019}, benchmark leakage~\cite{haimesBenchmarkInflationRevealing2024}, and the use of unsuitable metrics to measure AI systems' capabilities~\cite{truongValidMeasurementUnfairness2025,singhMeasuresDisparityTheir2023,renSafetywashingAISafety2024,leeBPrefBenchmarkingPreferenceBased2021}.
While the need for more rigorous evaluations is shared, AIE approaches the topic through the lens of fairness and equitable outcomes, whereas AIS focuses more on measuring risks from poor system performance, lack of robustness, misuse, and uncontrollable capabilities.

Issues arising from a \emph{lack of suitable AI governance measures} are addressed by work in both AIE and AIS.
Both fields recognize that insufficient global coordination in AI governance poses a risk~\cite{meimandiAdaptiveResponsibleAI2025}, for example, due to jurisdictional conflicts~\cite{tragerInternationalGovernanceCivilian2023} or a lack of standardization~\cite{bowenAICompaniesShould2025}.
However, the two fields approach governance differently. AIE primarily views governance as a means of ascribing liability and maintaining accountability, with particular emphasis on the role and risks of open-weight models~\cite{liesenfeldRethinkingOpenSource2024}.
AIS, in contrast, is more concerned with ensuring technical robustness and reliability throughout the AI systems' life-cycle~\cite{chaudhuriSecuringAISoftware2024,wozniakSafetyCasePattern2020}, and studies open-weight models primarily through the lens of misuse~\cite{mckenzieSTACKAdversarialAttacks2025,tamirisaTamperResistantSafeguardsOpenWeight2024}.
A related shared concern is the risk from a \emph{lack of data transparency}.
AIE examines data provenance as a source of transparency and accountability~\cite{pushkarnaDataCardsPurposeful2022,mitchellModelCardsModel2019}, along with the risks of training on intellectual property~\cite{goetzeAIArtTheft2024} or illegal and abusive content~\cite{cintaqiaStopNonconsensualUse2025}.
AIS, by contrast, examines data provenance from the perspective of capabilities~\cite{isikScalingLawsDownstream2025}, memorization~\cite{bidermanEmergentPredictableMemorization2023}, and generalization~\cite{haseUnreasonableEffectivenessEasy2024}.

Both fields also address issues of \emph{data distribution shifts}.
AIS examines risks from performance degradation under distribution shifts~\cite{yuRobustUnsupervisedDomain2022,viannaChannelSelectiveNormalizationLabelShift2024}, extending these concerns to dynamic multi-agent environments~\cite{foxabbottHigherOrderBeliefIncomplete2025,malenfantChallengeHiddenGifts2025}, continual learning~\cite{abelDefinitionContinualReinforcement2023}, and bounded rationality~\cite{tamborskiMemoryAllocationResourceConstrained2025}.
AIE, meanwhile, examines risks to fairness under data shifts~\cite{yaoWhenMisleadsRethinking2025,liuUnderstandingEndogenousData2025}.
Finally, the \emph{malicious use of AI systems} is an increasing focus in both fields. Questions around jailbreaking~\cite{cornacchiaMoJEMixtureJailbreak2024,hussainAuditAnalysisLLMAssisted2025,pfauElicitingLanguageModel2023} and data poisoning~\cite{bowenScalingLawsData2024,luIndiscriminateDataPoisoning2024} now appear in AIE as well, alongside existing work on unethical commands to AI systems~\cite{jacksonTactNoncomplianceNeed2019,vanderelstDarkSideEthical2018}.

\subsection{Overlapping Mitigation Strategies}

\begin{figure}
  \includegraphics[width=0.49\linewidth]{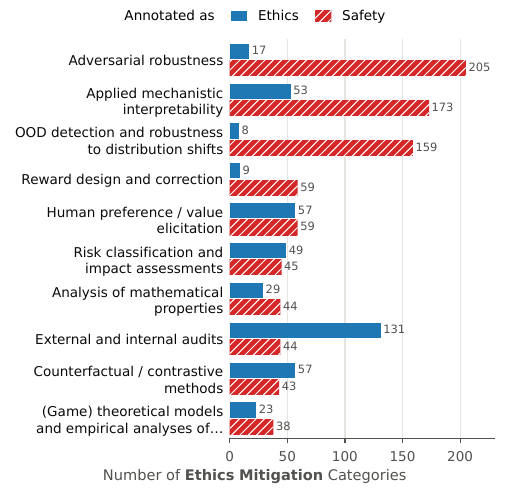}
  \includegraphics[width=0.49\linewidth]{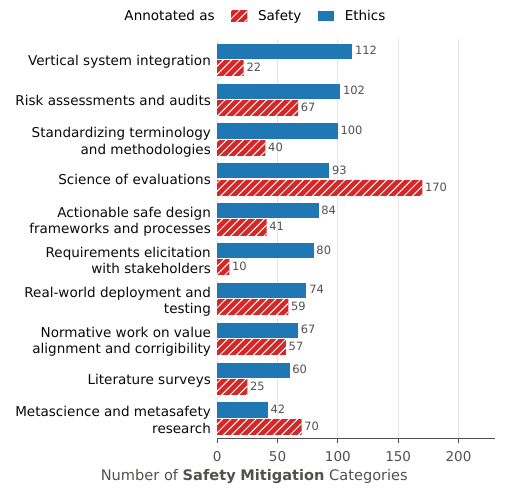}
  \caption{Bar plots of \textbf{mitigation similarities}, showing the top seven most frequent \emph{low-level} mitigation-strategy categories for which a paper's source corpus \emph{differs} from the corpus associated with the annotated category (e.g., a paper from the AIE corpus annotated with a category from the AIS taxonomy).}
  \label{fig:mitigation-sim}
\end{figure}

Overall, we identify nine high-level mitigation-strategy categories for AIE, comprising 79 low-level categories that range from
understanding the sources of bias and monitoring online information ecosystems to red-teaming for healthcare applications and analyzing crowdworker labor conditions.
For AIS, we identify eight high-level mitigation-strategy categories comprising 70 low-level categories, such as circuit analysis, outlier detection, preference learning from human feedback, and mathematical foundations for AI systems.
Several of these categories overlap between the two fields, as shown in~\cref{fig:mitigation-sim}.
We highlight five such areas of overlap below.

Perhaps the most striking overlap between AIE and AIS is the shared methodological focus on understanding, modeling, and learning \emph{human values, preferences, and reward models}.
AIE has developed approaches for eliciting diverse human values~\cite{davaniDisentanglingPerceptionsOffensiveness2024,birhaneValuesEncodedMachine2022} and encoding them in AI systems~\cite{kasenbergInverseNormConflict2018,cintasLocalizingPersonaRepresentations2025}.
Similarly, AIS has studied a range of methods for learning values from demonstrations via inverse reinforcement learning~\cite{skalseMisspecificationInverseReinforcement2023,barcoTacklingProblemDistributional2024} or through reinforcement learning from human feedback~\cite{pitisImprovingContextAwarePreference2024,kalraCanDifferentiableDecision2024}.
Each field proposes different methods for eliciting these human preferences~\cite{adamsSteerablePluralismPluralistic2025,pfauElicitingLanguageModel2023} and for ensuring that AI systems actually follow them~\cite{liuWeNeedStructured2024,shuiMoreGeneralLoss2023}.
Overall, value elicitation and enforcement are applied not only to machine learning models, such as automated decision-making systems or LLMs, but also to agentic systems, where both fields now examine how to ensure that systems' goals align with human preferences and values~\cite{sunDesigningNongreedyReinforcement2018,lynchAgenticMisalignmentHow2025,sumersHowTalkAI2024}.

\emph{Adversarial robustness and misuse} is another shared concern.
While AIS has produced substantially more work in this area, AIE has also recently studied methods for preventing the jailbreaking of LLMs~\cite{hussainAuditAnalysisLLMAssisted2025,cornacchiaMoJEMixtureJailbreak2024}, unethical commands to AI systems~\cite{jacksonTactNoncomplianceNeed2019}, and deepfakes used for impersonation~\cite{bukingoltsMimeticAISystems2025,walkerAIArtMisinformation2023}.
Relatedly, AIE has developed a substantial body of work on \emph{methods for AI governance}, particularly transparent and practical auditing methods — for example, bias detection~\cite{solystInvestigatingYouthAI2025}, datasheets~\cite{gebru2021datasheets}, and model cards~\cite{mitchellModelCardsModel2019}. 
Similar methods for tracing and mitigating harms from misuse appear in AIS as well, through safety cases~\cite{wozniakSafetyCasePattern2020}, risk assessment frameworks~\cite{barrettActionableGuidanceHighConsequence2023,moutonOperationalRisksAI2024}, and work on understanding and monitoring AI supply chains~\cite{chaudhuriSecuringAISoftware2024,tragerInternationalGovernanceCivilian2023}.
Both fields also treat eliciting safety requirements from actual stakeholders as a critical component of safe systems engineering~\cite{lynnRegulatingAlgorithmicManagement2025,longpreInHouseEvaluationNot2025}. Finally, both fields contribute to data governance and privacy, with particular focus on machine unlearning, differential privacy, and federated learning~\cite{krishnanNotAllData2025,aslamLearningUnlearnFailing2025}.

Both AIE and AIS investigate the \emph{mathematical foundations} of how AI systems function, though their approaches differ.
AIE has extensively studied the mathematics behind the fairness-accuracy trade-off~\cite{kingPrivacybiasTradeoffData2023,islamCanWeObtain2021} and of various modes of explanation generation~\cite{ayonrindeMathematicalPhilosophyExplanations2025,schuffHumanInterpretationSaliencybased2022}, whereas AIS is often more concerned with formally characterizing the mathematical properties of agentic systems (i.e., agent foundations)~\cite{stastnyNormativeDisagreementChallenge2021,conitzerFoundationsCooperativeAI2023}.
Both fields, however, investigate the topological properties of representation learning~\cite{heGeometricSolutionFair2020,cintasLocalizingPersonaRepresentations2025} and draw on decision-theoretic foundations~\cite{guoDecisionTheoreticFramework2024,laidlawPreventingRewardHacking2023}.

Methods for improving \emph{human-AI interaction} are also studied in both fields, though AIS often frames these approaches under the umbrella of multi-agent systems, where some agents are human~\cite{robertsonUnderstandingBeingUnderstood2022,malenfantChallengeHiddenGifts2025}. AIE is more openly concerned with the harms that arise when humans interact with agentic systems~\cite{sumersHowTalkAI2024,maedaAnthropomorphismSocialAffordance2025}, whereas AIS tends to focus on harms from AI agents acting autonomously~\cite{lynchAgenticMisalignmentHow2025} or interacting with other AI agents~\cite{mathewHiddenPlainText2024}. As a result, both fields study methods for red-teaming harmful behaviors~\cite{fefferRedTeamingGenerativeAI2024,mckenzieSTACKAdversarialAttacks2025}.

Finally, there is significant methodological overlap in the use of \emph{explainable AI and mechanistic interpretability} methods for understanding complex AI systems.
Both fields use post-hoc explanation methods---such as saliency maps~\cite{schuffHumanInterpretationSaliencybased2022,tsurumiSocialBiasVision2025}, counterfactual explanations~\cite{karimiAlgorithmicRecourseCounterfactual2021,gyevnarCausalExplanationsSequential2024} and gradient-based methods~\cite{singhModelAgnosticInterpretability2020,zhangModelDebiasingGradientbased2023}---to justify how AI systems arrive at decisions, often with the goal of guaranteeing safe decisions~\cite{kuznietsovExplainableAISafe2024} or fair treatment~\cite{ferrarioHowExplainabilityContributes2022}.
AIS has additionally developed methods specific to the Transformer architecture, such as circuit analysis~\cite{mehrbodCircuitDiscoveryHelps2025,lieberumDoesCircuitAnalysis2023} and causal scrubbing~\cite{chanCausalScrubbingMethod2022}, which are not yet widely adopted in AIE.
Autoencoders, however, are used in both fields: in AIE for understanding bias in representations~\cite{zhangFairDeepAnomaly2021,aminiUncoveringMitigatingAlgorithmic2019}, and in AIS for discovering interpretable concepts~\cite{cunninghamSparseAutoencodersFind2023,chaudharyEvaluatingOpenSourceSparse2024}.
The mathematical properties of explanation-generation methods---such as sparsity or monotonicity---are also well studied in AIE and, increasingly, in AIS, spanning both classical explainable AI~\cite{semenovaExistenceSimplerMachine2022,ayonrindeMathematicalPhilosophyExplanations2025} and mechanistic interpretability~\cite{rajendranLearningInterpretableConcepts2024,makelovThisSubspaceYou2023}.

\section{Understanding Different Engagement Modes}
\label{ssec:results:modes}

\begin{figure}[t]
    \centering
    \includegraphics[width=\linewidth]{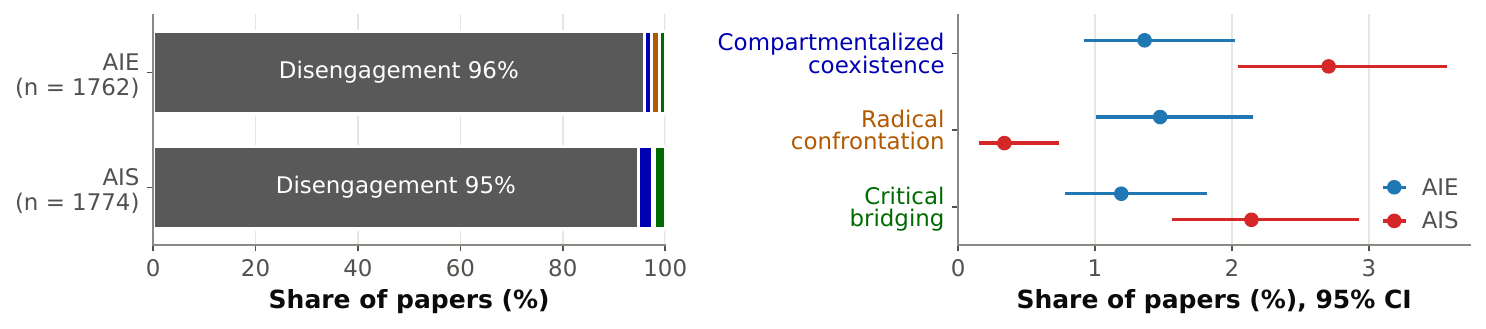}
    \caption{Share of papers for each engagement mode by corpus field. \textit{Left:} all four modes with the slivers at the right of each bar corresponding to the three rarer modes. \textit{Right:} the rarer modes on their own, with 95\% Wilson intervals.}
    \label{fig:modes-shares}
\end{figure}

Having established that each field has both distinct and overlapping areas of research priorities, we are interested in quantifying how papers in our corpora fit into the four engagement modes as outlined in~\cref{sec:ai-wars} and how those modes relate to our risk and mitigation categories.
For this we present and analyze the results of classifying each paper along the dimensions of integration and engagement as outlined in~\cref{ssec:methods:analysis}.
Recall that 19 papers did not have full-text PDFs available, so these are omitted from the analyses in this section.

\subsection{Descriptive Statistics}

The percentage of papers assigned to each engagement mode is shown in~\cref{fig:modes-shares}, with the actual number of papers shown in~\cref{fig:resolution-modes}(b).
By far the largest engagement mode is disengagement with a $95.4\%$ share of all papers.
The remaining roughly $5\%$ of papers (163 papers) are split differently by the two fields, with all three engagement modes differing significantly between AIE and AIS based on Holm-adjusted Fisher's exact tests.
Radical confrontation is the largest rare mode in AIE but the smallest in AIS ($1.48\%$ vs.\ $0.34\%$, $p<.001$), so AIE papers are more than four times as likely as AIS papers to reject the other field's framing outright, for instance by arguing that concerns about loss of control presume a functionality that deployed systems often lack~\cite{rajiFallacyAIFunctionality2022}, or that the focus on future existential threats obscures harms already caused by present-day systems~\cite{gebru2024tescreal}.
Conversely, AIS has the larger share of the two high-integration modes with compartmentalized coexistence ($1.36\%$ vs.\ $2.71\%$, $p=.012$) and critical bridging ($1.19\%$ vs.\ $2.14\%$, $p=.035$).
Compartmentalized coexistence is the largest rare engagement mode overall (72 papers), with AIS having about twice the share of AIE.
This may be a reflection of recent calls to approach AI safety from a more sociotechnical angle~\cite{lazar2023ai,weidingerSociotechnicalSafetyEvaluation2023,curtisResearchAgendaSociotechnical2025}, yet without meaningfully deep integration of concerns from AI ethics, which has traditionally been associated with sociotechnical research~\cite[e.g.,][]{chopraSociotechnicalSystemsEthics2018,selbstFairnessAbstractionSociotechnical2019}.

\begin{figure}
  \centering
  \includegraphics[width=\linewidth]{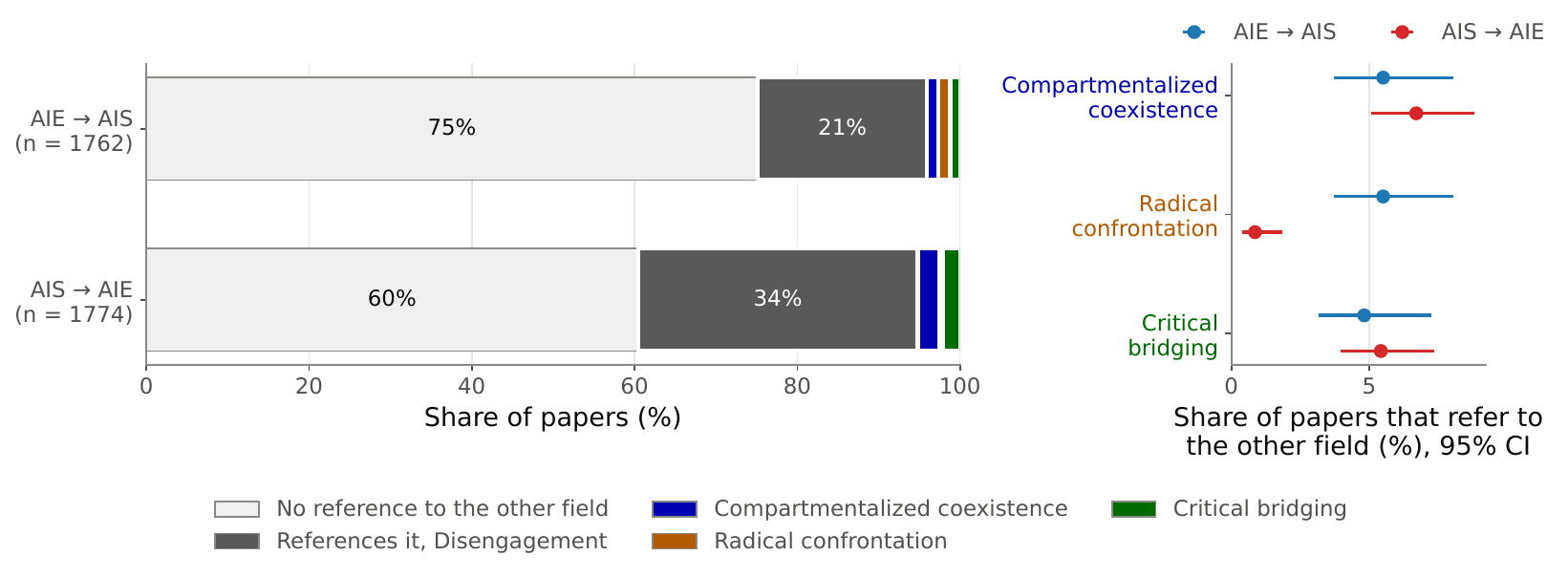}
  \caption{Plots showing the occurrence of references to the other field. \emph{Left:} share of AIE and AIS papers by whether they refer to the other field colored by engagement mode. \emph{Right:} among papers that do refer to the other field, the share for each engagement mode, with 95\% Wilson intervals.}
  \label{fig:modes-reference}
\end{figure}

These findings are corroborated by the way papers cross-reference fields, as shown in~\cref{fig:modes-reference}.
The majority of papers never reference the other field ($75\%$ for AIE vs.\ $60\%$ for AIS) as one would expect for papers in the disengagement category.
Even when references to the other field appear in a text, the paper still most often falls into disengagement ($21\%$ vs.\ $34\%$ of all papers).
While these numbers suggest that AIS mentions AIE more frequently than the other way around, in both cases, this referencing rarely goes beyond superficial mentions in a way that points towards deeper engagement.
Nevertheless, when we do find deeper engagement, there is also an asymmetry in the patterns of cross-referencing.
AI safety papers that refer to AI ethics most often do so through compartmentalized coexistence ($6.7\%$ of referring papers vs.\ $5.5\%$ in AIE), in line with the notion that they place ethical concerns alongside their own without engaging the field's positions, although we note that this difference is not statistically significant.
On the other hand, when AI ethics papers refer to AI safety, they are far more likely to confront it directly: radical confrontation accounts for $5.5\%$ of referring AI ethics papers but only $0.9\%$ of referring AI safety papers (Holm-adjusted Fisher $p<.001$).
Finally, critical bridging of research problems is similarly common in both fields ($4.8\%$ vs.\ $5.4\%$).

\subsection{Associated Risks and Mitigation Categories}

\begin{figure}
    \centering
    \includegraphics[width=\linewidth]{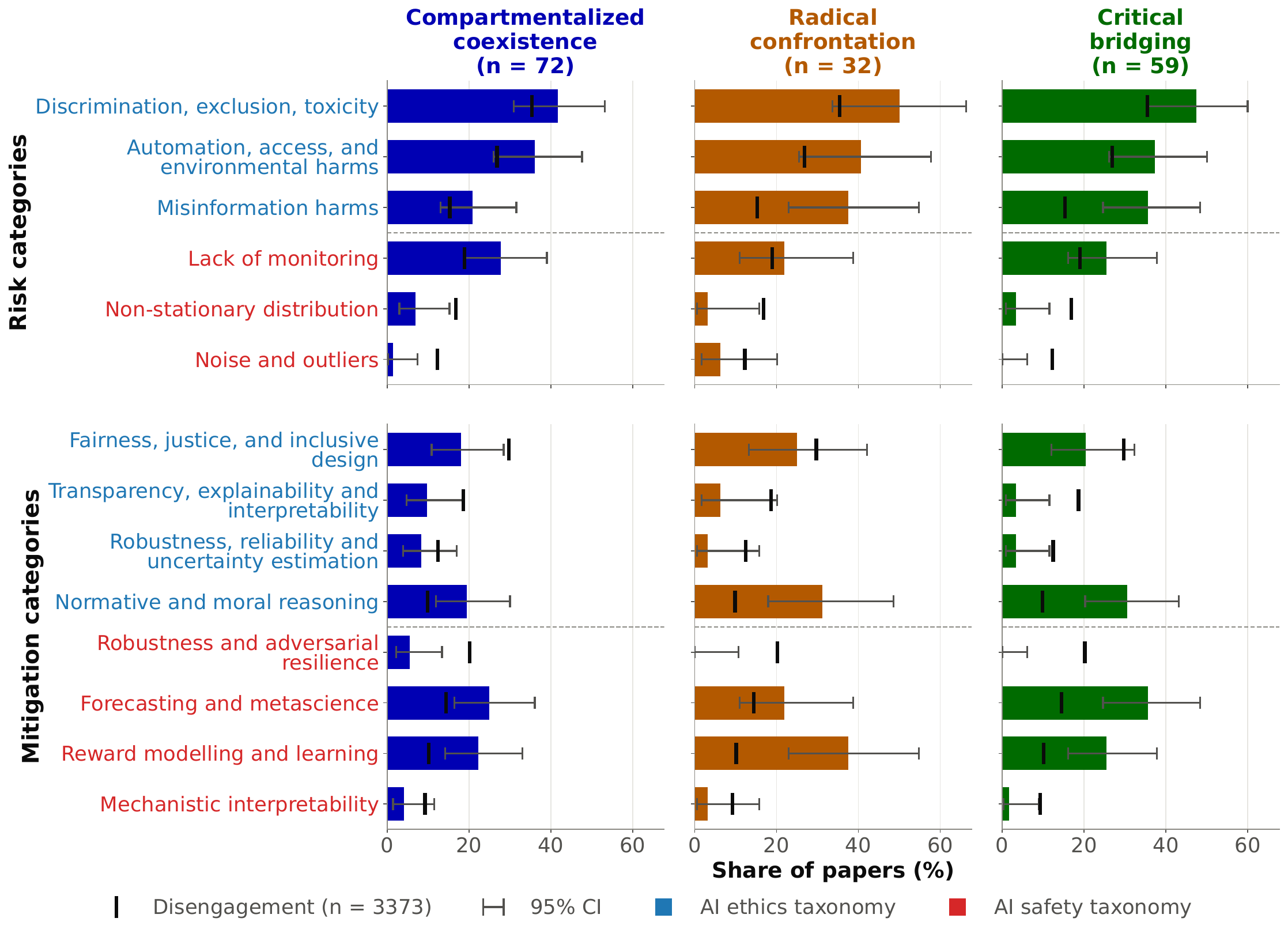}
    \caption{Risk (\textit{top}) and mitigation (\textit{bottom}) categories addressed by the papers of each rarer engagement mode, showing 14 of the 31 high-level categories which are discussed in the main text. Bars depict the share of a engagement mode's papers annotated with each high-level category, with 95\% Wilson intervals. Black ticks show the share among disengagement papers for reference.}
    \label{fig:modes-topics}
\end{figure}

The above analyses so far highlight a landscape of three rare engagement modes in addition to a large group of disengaged papers.
The question now becomes what risks these rarer modes engage with and how they integrate those into their methods. 
Our analysis concerns the share and distribution of different high-level categories across each rare engagement mode as compared to the levels observed among disengagement papers serving as a baseline.
These comparisons are shown in~\cref{fig:modes-topics}.

The clearest pattern here concerns the overloaded notion of value alignment.
The AIE mitigation category of ``Normative and moral reasoning'', as well as the AIS mitigation category of ``Reward modelling and learning'', are each addressed by roughly a third of the papers in the two high-engagement modes: $30.51\%$ and $25.42\%$ of critical bridging papers and $31.25\%$ and $37.50\%$ of radical confrontation papers, respectively.
Only $9.87\%$ and $10.14\%$ of disengaged papers address each of these mitigations, with the largest differences significant even under correction (Holm-adjusted Fisher $p=.001$ for normative and moral reasoning in critical bridging; $p=.004$ for reward modelling in radical confrontation).
In contrast, compartmentalized coexistence does not meaningfully differ from disengagement on these categories.
These results suggest that value alignment is where the two fields meet most directly.
Critical bridging papers put a safety method -- such as reward modeling or constitutional fine-tuning -- to work on an ethical question -- such as whose values and which moral reasoning a system should embody.
Meanwhile, radical confrontation papers contest precisely this move, for instance by questioning whether pluralistic and contested values can be aggregated into a single alignment target.

In stark contrast sit the categories related to technical robustness research, which are currently almost never integrated between fields.
For example, none of the 59 critical bridging papers and none of the 32 radical confrontation papers address robustness and adversarial resilience, in contrast to $20.19\%$ of disengaged papers ($p<.001$ for critical bridging).
Similarly, only one of the 72 compartmentalized coexistence papers and none of the critical bridging papers address noise and outliers, against a baseline of $12.27\%$, although with these small samples the differences are not significant under correction ($p=.13$ and $p=.07$).
This same direction holds for almost all other technical mitigation measures.

We would naturally expect that papers which integrate the two fields more -- i.e., those in the compartmentalized coexistence and critical bridging modes -- address more categories from the other field's taxonomy than those papers that integrate less -- i.e., disengagement and radical confrontation.
This is only partly the case.
Papers in both high-integration modes do draw more of their categories from the other field's taxonomy than disengaged papers ($33\%$ and $39\%$ vs.\ $17\%$ for AIE; $56\%$ and $61\%$ vs.\ $28\%$ for AIS), but so do radical confrontation papers ($30\%$ and $57\%$), so it is engagement rather than integration that goes with crossing taxonomies.
Nor is this increase spread evenly across topics.
For 11 of the 31 categories in~\cref{fig:modes-topics}, the bars depicting shares of papers fall below the disengagement markers for both high-integration modes.
What we see, instead, is a divide between \textit{normative} -- i.e., categories concerned with which values AI systems should serve and how they should be governed -- and \textit{technical} work -- i.e., categories concerned with engineering properties of models, such as adversarial robustness, interpretability and uncertainty estimation.
We find that papers in all three engagement modes, regardless of their level of integration, favor the former and avoid the latter.

To illustrate this difference, take the notion of fairness.
From a risk perspective, the category of ``Discrimination, exclusion, and toxicity'' is more common in every rare engagement mode than among disengaged papers ($41.67$--$50.00\%$ vs.\ $35.43\%$), and most of all among radical confrontation and critical bridging papers ($50.00\%$ and $47.46\%$). 
As a mitigation, however, ``Fairness, justice and inclusive design'' -- in other words, techniques for measuring and reducing unfair outcomes -- is less common in every rare engagement mode than among disengagement ($18.06$--$25.00\%$ vs.\ $29.71\%$).
This suggests that papers engaging with the other field tend to cite discrimination as a harm to be avoided, but are less likely to contribute fairness methods themselves.
We can observe a similar drop in interest from risk to mitigation for the risk category ``Automation, access, and environmental harms'' -- such as due to inaccessible automated decision-making systems -- and the related mitigation category ``Transparency, explainability, and interpretability'' ($36.11$--$40.63\%$ vs.\ $26.86\%$ for the risk; $3.39$--$9.72\%$ vs.\ $18.62\%$ for the mitigation).
This suggests that concerns about opaque automated decisions, while frequently raised as a risk, do not by themselves give rise to transparency methods.
While consistent across modes, we note that these individual differences are not significant under correction for multiple testing.

\subsection{Five Concrete Opportunities for Critical Bridging}
\label{sec:opportunities}

For our final analysis, we ask where the papers in the rare engagement modes agree and disagree about particular risks and mitigation strategies, and where this agreement and disagreement leaves opportunities for more critical bridging.
We base this analysis on the inductive coding and thematic analysis described in~\cref{ssec:methods:coding}.

\begin{figure}
    \centering
    \includegraphics[width=\linewidth]{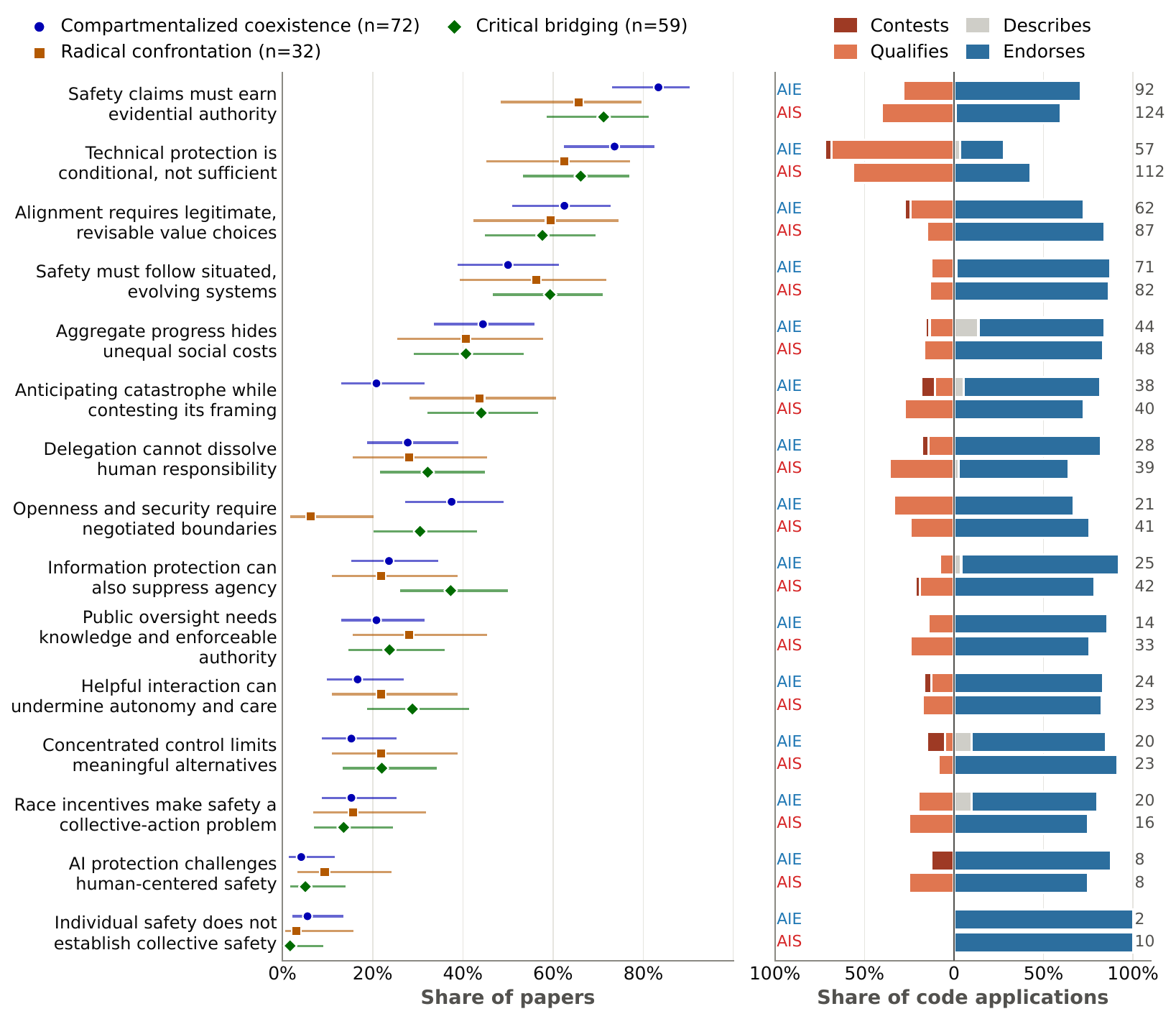}
    \caption{The 15 themes of our thematic analysis of the rare engagement modes. \textit{Left:} share of each mode's papers with at least one code from a given theme, with 95\% Wilson intervals. \textit{Right:} share of papers with a certain stance per AIE and AIS papers.}
    \label{fig:qual-themes}
\end{figure}

The share of each rare engagement mode's papers that engage with each theme is shown on the left of~\cref{fig:qual-themes}.
Remarkably, for 14 of the 15 themes, the share of papers in a theme do not significantly differ between any two modes (Holm-adjusted Fisher tests).
The only exception for this is the theme ``Openness and security require negotiated boundaries'', which we discuss in~\cref{sssec:openness}.
The two fields of AIE and AIS are similarly close, because no themes differ significantly in its share of papers.
This finding corroborates our findings in~\cref{fig:modes-topics}, suggesting that papers which engage with the other field do so using a common set of themes.

Indeed, critical bridging is already happening in some areas.
The theme ``Anticipating catastrophe while contesting its framing'' is mentioned by $44.07\%$ of critical bridging and $43.75\%$ of radical confrontation papers, and by only $20.83\%$ of compartmentalized coexistence papers.
Yet, the code ``Critical examination of AI concepts and narratives'' is applied to $23.73\%$ of critical bridging and $28.12\%$ of radical confrontation papers, but to only $4.17\%$ of compartmentalized coexistence papers.
This finding suggests that papers in both high-engagement modes debate how catastrophic risk should be framed and prioritized.
For example, by questioning the assumption that \textit{``more compute equals more capability, and thus more potential harm''}~\cite{oldenburgStoriesWeGovern2025}, or by observing that \textit{``the emphasis on AI's inevitability is particularly prevalent among experts who are bound to benefit from government support for AI uptake''}~\cite{schopmansCodedBiasExistential2022}.

These debates are important, but they are also where most of the existing engagement between the fields takes place.
Yet, over all 1,254 applications of our codebook, summarized in~\cref{fig:qual-themes}, we see a diverse landscape of stances: $70.89\%$ endorse, $26.32\%$ qualify, $1.44\%$ describe, and $1.36\%$ contest a risk or mitigation strategy.
In the following sections, we discuss five further opportunities for critical bridging based on this evidence.

\subsubsection{Provide joint account of evidence and validity for evaluation claims}

The most common theme is ``Safety claims must earn evidential authority'', which appears in $74.65\%$ of AIE papers and $76.09\%$ of AIS papers.
Its most common code, ``Safety-oriented AI evaluation and benchmarking'', is applied to $66.67\%$ of compartmentalized coexistence papers but to only $49.15\%$ of critical bridging and $34.38\%$ of radical confrontation papers. 
This suggests, that both fields question what evaluations can establish.
From the perspective of AIE, \citeauthor{erikssonCanWeTrust2025}~\cite{erikssonCanWeTrust2025} observe that \textit{``many benchmarks do not have a clear definition of what they claim to measure''}, and \citeauthor{ibrahimInteractiveEvaluationsInteraction2025}~\cite{ibrahimInteractiveEvaluationsInteraction2025} argue that \textit{``while they can measure what a model does, they cannot measure its impact on human perception and behavior''}.
From AIS, for instance, \citeauthor{levinsteinStillNoLie2024}~\cite{levinsteinStillNoLie2024} find that interpretability \textit{``probes might identify sentence properties, such as the presence or absence of negation, rather than truthfulness''}.
These papers make the same argument about the validity of evaluations from both fields.
Still, the theme is most common among compartmentalized coexistence papers.
A joint account of what evaluations can and cannot establish about AI systems seems, therefore, the largest opportunity for critical bridging.
This form of bridging could also extend the overlap in issues around reproducibility and benchmarks that we identified in~\cref{sec:bridging}.

\subsubsection{Map conflicting interactions between safeguards}

The second most common theme, ``Technical protection is conditional, not sufficient'' is mentioned by $73.61\%$ of compartmentalized coexistence, $66.10\%$ of critical bridging, and $62.50\%$ of radical confrontation papers. 
It is also the theme that both fields most often contest or qualify, suggesting that the two fields agree on the fact that technical safeguards work only under certain conditions, but disagree on what those conditions are, often specifically because these conditions depend on concerns that are being addressed by the other field.
For example, \citeauthor{weidingerEthicalSocialRisks2021}~\cite{weidingerEthicalSocialRisks2021} note that \textit{``methods to reduce toxic speech from LMs have been found to bias model prediction against marginalized groups''}, while \citeauthor{ghaziRobustPrivateLearning2021}~\cite{ghaziRobustPrivateLearning2021} prove that \textit{``robust and private learning is harder than robust or private learning alone''}.
Therefore the second opportunity for critical bridging is to conduct systematic studies of how safeguards motivated by AIE and safeguards motivated by AIS interact, and of who bears the harm when it occurs.
This opportunity is also relevant because technical robustness research is almost never integrated between the fields, as we showed in~\cref{fig:modes-topics}.

\subsubsection{Join technical research on open-model governance and security}\label{sssec:openness}

The theme ``Openness and security require negotiated boundaries'' is the only one whose share of papers differs significantly between modes. 
Many papers in compartmentalized coexistence and critical bridging approach openness of research and of models through the lens of misuse and staged release, as when \citeauthor{whittlestoneTensionOpennessPrudence2019}~\cite{whittlestoneTensionOpennessPrudence2019} conclude that \textit{``different combinations of content, timing and distribution options will pose different risks depending on the domain and potential risks of malicious use vs. concentration of power''}.
Some critical bridging papers also draw on concerns from AIE about who controls AI systems and who can afford to use them.
\citeauthor{kapoorBuildAgentAdvocates2025}~\cite{kapoorBuildAgentAdvocates2025} argue that \textit{``open weights models are likely essential to reduce exposure to the risks of a reversal in corporate strategy''}, and \citeauthor{varshneyDecolonialAIAlignment2024}~\cite{varshneyDecolonialAIAlignment2024} warns that \textit{``open access to knowledge or LLM alignment may be colonial if the cost of access is too high''}.
This theme, thus, already encapsulates the central disagreement about how much centralized control over AI systems is justified, but technical answers are still missing.
That will require joining the AIS concerns of openness through misuse with the AIE concerns about power and access as the third opportunity for critical bridging.

\subsubsection{Research the collective safety of many AI systems}

The rarest theme, ``Individual safety does not establish collective safety'', covers collusion between AI systems, correlated failures, and failures of coordination between AI systems.
We find only six papers that engages with this theme between fields.
Even so, the AIS concern that \textit{``single-agent alignment and safety are insufficient for assuring desirable outcomes in multi-agent settings''}~\cite{anwarFoundationalChallengesAssuring2024} does meet the AIE concern that \textit{``agents partially built upon the same components [\dots] could have common vulnerabilities and failure modes''}~\cite{chanVisibilityAIAgents2024}.
In light of further literature that argues for the importance of multi-agent risks of AI systems~\cite{hammondMultiAgentRisksAdvanced2025}, we believe this direction to be an important, if underexplored area for critical bridging.

\subsubsection{Bring public oversight into race dynamics}

Finally, two further themes show broad agreement but uneven engagement.
The theme ``Public oversight needs knowledge and enforceable authority'' is mentioned in 13 out of the 38 AIS critical bridging papers ($34.21\%$) but by only 1 of the 21 AIE critical bridging papers ($4.76\%$), while 7 of the 26 AIE radical confrontation papers ($26.92\%$) engage with it.
The AIE papers that bridge the fields rarely take up the questions of auditing and incident reporting raised by AIS papers, such as the need to \textit{``establish and enforce audit reporting standards independent of the firms being audited''}~\cite{manheimNecessityAIAudit2025}.
Similarly, ``Race incentives make safety a collective-action problem'' is engaged with by $13.56$--$15.62\%$ of the papers in each mode.
The papers in this theme largely agree that competition undermines safety precautions and disagree mainly on whether cooperation between developers and states can be made credible and enforceable.
Both themes offer common ground that few papers currently use, which makes them the fifth opportunity for critical bridging.

\section{Open Questions and Limitations}
\label{sec:discussions}

Our research began with the observation that the responsible AI divide is driven by more than rhetorical friction. We identified four archetypal strategies for addressing tensions between AIE and AIS: radical confrontation, disengagement, compartmentalized coexistence, and critical bridging. We argued that bridging problems offers one viable epistemic and pragmatic path for make progress on responsible AI cooperatively. To substantiate this argument, we conducted a systematic analysis of representative samples from both the AIE and AIS literature. Our mapping of problem clusters (AI risk concerns and mitigation strategies) confirms that significant overlap between AIE and AIS already exists but remains underappreciated. While our results help identify these bridges, crossing them requires joint effort to address at least five structural and conceptual challenges:

\begin{enumerate}
    \item \textit{The sociotechnical interface:} the emerging field of sociotechnical safety~\cite{lazar2023ai} represents a critical yet largely open space for investigating the overlaps between AIE and AIS.
    \item \textit{Temporal conflict resolution:} the AIS/AIE divide is often a proxy for the tension between ``present-day harms'' and ``long-term risks.'' How can AIE and AIS collaboratively develop shared frameworks — building on early work such as Kasirzadeh~\cite{kasirzadehTwoTypesAI2025} — that situate current algorithmic bias and future loss of control along a common continuum of power asymmetry and loss of human agency, rather than treating them as competing priorities?
    \item \textit{Material drivers and funding:} Our analysis focused on the epistemic content of research, yet the responsible AI divides are heavily influenced by the political economy of funding (e.g., industry vs. philanthropic vs. state). To what extent do specific funding sources incentivize ``radical confrontation'' over ``critical bridging''? How can we design funding-neutral spaces that prioritize collaborative problem-solving over brand differentiation?
    \item \textit{Institutional hybridity:} Mapping problem overlaps is insufficient if researchers remain siloed. What specific institutional designs — joint PhD programs, cross-disciplinary red-teaming consortia, shared peer-review standards — would be most effective in fostering ``hybrid communities'' that combine safety and ethics depth with critical breadth? IASEAI, mentioned above, offers one recent effort in advancing hybrid communities.
    \item \textit{Formalizing disagreement:} Critical bridging recognizes that some disagreements are irreducible and that not all should be resolved — some reflect legitimate value pluralism that a healthy field must preserve. When AIS and AIE propose mutually exclusive interventions, what mechanisms exist for navigating, rather than dissolving, these disagreements? The community needs a shared vocabulary for disagreement — one that distinguishes empirical disputes from value conflicts, and technical negotiation from political contestation — so that bridging efforts do not collapse genuine pluralism into false consensus.
    \item \textit{Disciplinary composition:} How does the disciplinary composition of AI ethics and AI safety shape their respective problems, methods, and standards of evidence? Both fields are interdisciplinary rather than disciplines in their own right, but they draw on different disciplines in different proportions. Could some anthropological study examine whether some of the tensions between AI ethics and AI safety stem less from normative disagreement than from these differing disciplinary inheritances?

\end{enumerate}

While this paper focuses on \emph{critical bridging} as a structured path for addressing the tensions between AIE and AIS, several deliberate choices define its limited scope. 

First, our focus is restricted to ``bridging problems'' — identifying shared challenges and problem spaces — which necessarily brackets deeper institutional and value-based integration. Consequently, while we map overlapping problem spaces and challenges, our analysis does not address the underlying tensions between AIE and AIS regarding the nature of AI~\cite{bliliposition}, nor the competing political philosophies and moral ideologies — such as strong longtermism \cite{greaves2024case} or particular variants of transhumanism~\cite{gebru2024tescreal} — that may be contributing forces behind the responsible AI divide. We deliberately exclude efforts to bridge particular communities of AIE and AIS and their philosophical worldviews, because such work requires extensive sociological and anthropological research, and is beyond the scope of this paper. While recent scholarship has begun to map the AIS landscape~\cite{ahmed2023building,gebru2024tescreal}, equivalent analysis of AIE is largely lacking, and existing work on AIS remains partial, warranting much deeper investigation.

Second, to maintain a tractable dataset, we assume a (practically useful) binary distinction between AIS and AIE. We also explicitly exclude the emerging framing of ``sociotechnical AI safety'' \cite{lazar2023ai}, which seeks to integrate social science into AIS. While this is a natural candidate for bridging, it remains unclear what the term denotes in practice and how it differs from AIS work. 

Third, we do not account for community-level sociological differences \cite{ahmed2024field} or the political economy of AI, such as the influence of private funding, lobbying, or geopolitical interests on AIE and AIS. These factors may be the primary drivers of radical confrontation, yet they remain outside the scope of this paper. 

Fourth, our methods of unsupervised topic modeling and qualitative coding identify patterns of co-occurrence but cannot fully capture nuances of normative intent or the depth of engagement in a text. While our 3,555-paper dataset reflects formal discourse, it may not capture the real-time shifts on social media and influential AIS forums, such as LessWrong and the Alignment Forum. 

Fifth, any analysis involving an annotation-based setup -- either with humans or LLMs -- is subject to bias.
In our first qualitative study on risks and mitigation measures, we aimed to minimize this bias by randomizing the display order of papers and the available categories to choose from. 
In the second analysis, we account for LLM bias by using an independent human verification on 272 papers, producing a final agreement score using Cohen's $\kappa=0.68$ and $0.62$ for engagement and integration, respectively.
For the third analysis on rare engagement modes, we ensure that the LLM system was following the phases of qualitative analysis as given in \citeauthor{boyatzis1998transforming}~\cite{boyatzis1998transforming}.
A limitation of our work is that, for the first and third studies, we have not compared the annotations against an independent judgment due to time and resource constraints.

Finally, we intentionally bracket potentially irreconcilable differences where AIE and AIS goals might be mutually exclusive or zero-sum. By prioritizing bridgeable problems, we acknowledge the risk of underestimating the utility of radical confrontation, disengagement, or compartmentalized coexistence in contexts where bridging might dilute meaningful AI governance. The epistemic and pragmatic merits of these alternative modes of engagement warrant investigation elsewhere. We believe this paper is an important step toward understanding how to optimize collaboration and scale responsible AI in a meaningful way. AI systems are among the most consequential technologies of our time, and a future with responsible AI is a shared goal we believe we should collectively pursue.

\subsection*{Data and Code Availability}
The complete annotated dataset consisting of AI safety and AI ethics metadata, as well as supporting code used to analyze this dataset, are both available at: \url{https://github.com/gyevnarb/ai-safety-ethics}.

\subsection*{Generative AI Usage Statement}
The authors declare the use of generative AI in the research process. 
According to the GAIDeT taxonomy~\cite{suchikovaGAIDeTGenerativeAI2026}, the following tasks were delegated to generative AI (GAI) tools under full human supervision: code generation, process automation, data cleaning, data analysis, visualization, proofreading and editing.
The GAI tool used was: Claude Opus 4.6 by Anthropic, GPT-6 Luna, and GPT-6.1 Sol by OpenAI.
Responsibility for the final manuscript lies entirely with the authors.
GAI tools are not listed as authors and do not bear responsibility for the final outcomes.

\subsection*{Acknowledgments}
B.G.'s work was supported in part by the Institute for Complex Social Dynamics at Carnegie Mellon University, the UKRI Centre for Doctoral Training in Natural Language Processing, funded by the UKRI (grant EP/S022481/1), the University of Edinburgh, School of Informatics and School of Philosophy, Psychology \& Language Sciences. A.K.'s work was funded in part by the AI2050 programme at Schmidt Sciences (grant 24-66924).

\printbibliography

\renewcommand{\figurename}{Supplementary Figure}
\renewcommand{\tablename}{Supplementary Table}
\crefname{figure}{Supplementary Figure}{Supplementary Figures}
\Crefname{figure}{Supplementary Figure}{Supplementary Figures}
\crefname{table}{Supplementary Table}{Supplementary Tables}
\Crefname{table}{Supplementary Table}{Supplementary Tables}

\appendix
\section{AI Safety Labs and People Used for Data Collection}

We build on this AIS corpus by extending it with the metadata of AIS-related papers from a list of 35 labs and other individual authors chosen based on references from previous work~\cite{amodeiConcreteProblemsAI2016,hendrycksUnsolvedProblemsML2022,technicalitiesShallowReviewTechnical2024,bengioInternationalAISafety2025}.

In addition, we included papers from the following organizations which represent work on AIS: 
\begin{itemize}
    \item Autonomous Agents Research Group (\url{https://agents-lab.org/})
    \item Alignment of Complex Systems Research Group (\url{https://acsresearch.org/})
    \item Anthropic (\url{https://www.anthropic.com/})
    \item Association for Long Term Existence and Resilience (\url{https://alter.org.il/})
    \item Apart Research (\url{https://apartresearch.com/})
    \item Center for AI Safety (\url{https://safe.ai/})
    \item Center for Human-Compatible AI (\url{https://humancompatible.ai/})
    \item Center on Long-Term Risk (\url{https://longtermrisk.org/})
    \item Center for the Study of Existential Risk (\url{https://www.cser.ac.uk/})
    \item Cooperative AI Foundation (\url{https://www.cooperativeai.com/})
    \item DeepMind (\url{https://deepmind.google/})
    \item EleutherAI (\url{https://www.eleuther.ai/})
    \item FAR.AI (\url{https://www.far.ai/})
    \item Future of Humanity Institute (\url{https://www.futureofhumanityinstitute.org/})
    \item Forecasting Research Institute (\url{https://forecastingresearch.org/})
    \item Google Research (\url{https://research.google/})
    \item Centre for the Governance of AI (\url{https://www.governance.ai/})
    \item Global Priorities Institute (\url{https://globalprioritiesinstitute.org/})
    \item Meta (\url{https://ai.meta.com/})
    \item Model Evaluation \& Threat Research (\url{https://metr.org/})
    \item Mila (\url{https://mila.quebec/})
    \item Machine Intelligence Research Institute (\url{https://intelligence.org/})
    \item NeurIPS ML Safety 2022--2024 (\url{https://neurips.cc/})
    \item OpenAI (\url{https://openai.com/})
    \item Oxford Martin AI Governance Initiative (\url{https://aigi.ox.ac.uk/})
    \item Palisade Research (\url{https://palisaderesearch.org/})
    \item RAND (\url{https://www.rand.org/})
    \item Redwood Research (\url{https://www.redwoodresearch.org/})
    \item UCL Theory of Learning Lab (Saxe Lab) (\url{https://www.saxelab.org/})
    \item Safe Robotics Lab (\url{https://saferobotics.princeton.edu/})
    \item Timaeus (\url{https://timaeus.co/})
    \item Transluce (\url{https://transluce.org/})
    \item Tsinghua AIR and CoAI (\url{https://air.tsinghua.edu.cn/en/}, \\ \url{https://coai.cs.tsinghua.edu.cn/})
    \item UK AI Security Institute (\url{https://www.aisi.gov.uk/})
    \item Individual researchers identified by a previous survey~\cite{technicalitiesShallowReviewTechnical2024}
\end{itemize}
\clearpage

\section{Prompts Used in the LLM-as-a-Judge Setup}
\label{apx:prompts}

\begin{lstlisting}[style=promptfile,caption={Integration/engagement classification prompt.},label={lst:axes-main}]
We study how research papers position themselves with respect to the divide between two clusters of research.

## The two fields
- AI ethics (AIE): the broad field studying the ethical and technosocial impacts of AI. Typical topics of study are fairness and discrimination, accountability, transparency, privacy, structural injustice, present-day and tangible harms of deployed systems.
- AI safety (AIS): the broad field studying the safety and technosocial impacts of AI. Typical topics of study are alignment, robustness, interpretability, dangerous capabilities, loss of control, existential risk from advanced AI.

This paper was retrieved from the {home} corpus, so its "other field" is {other}.

## Axis 1: engagement (1 to 4)
How directly and substantively does the paper engage with the other field as a field: its positions, priorities and arguments, or the tensions between AIE and AIS (their differing priorities, evidence standards, risk framings)? Engagement can be critical or constructive; what matters is directness.

Using concepts, methods or topics that happen to belong to the other field is NOT engagement by itself. For example, an AI ethics paper that studies RLHF or alignment, or an AI safety paper that also covers fairness or privacy, has engagement 1 unless it addresses the other field's positions or its relationship to the paper's own field. Such topical overlap belongs on the integration axis instead.
- 1: the other field is never addressed as a field, and the AIE-AIS relationship is never mentioned (topical overlap alone still scores 1).
- 2: the other field is named or cited in passing (related work, motivation), without discussing its positions.
- 3: engaging the other field's positions, or the AIE-AIS tensions, is a major thread of the paper.
- 4: the AIE-AIS relationship, or the other field's positions, is the central topic.

## Axis 2: integration (1 to 4)
How far does the paper bring AIE and AIS concepts, methods, or problem framings together within one frame, rather than working within a single field or setting the fields against each other?

Mixing topics is not the same as integrating the fields. A technical paper that combines, say, a fairness metric with a robustness technique uses ingredients from both fields, but integrates them only if it treats them as the concerns of each field and connects them. Showing that a technical property (e.g. lack of robustness) leads to a harm (e.g. discrimination) is not enough by itself.

Taking the other field's concerns into the paper's own framework while dismissing that field's approaches or positions is low integration, however much of both fields the paper discusses.
- 1: works entirely within one field's framing, or treats the other field only as an opponent to be discredited.
- 2: topics or techniques from both fields appear, but as technical ingredients rather than as concerns of the other field; or the other field's concerns are only acknowledged, kept separate, or mentioned in passing (e.g. a paragraph noting safety risks in an otherwise fairness-focused paper).
- 3: explicitly argues that the fields' concerns are connected, treating each as a concern of that field, e.g. shows how one problem is both an ethics problem and a safety problem and draws on both fields to address it.
- 4: builds a genuinely joint framing, problem definition, method or governance proposal that serves both fields.

Score the two axes independently: a paper can engage the other field intensely while integrating nothing (e.g. a polemic), or place both fields' concerns side by side without engaging the tensions between them.

Score each axis as an integer from 1 to 4: the level whose description best fits the paper as a whole.

## Output
Return JSON with:
- "evidence": up to 3 short verbatim quotes (each under 40 words) that most inform your scores; an empty list if the paper never touches the other field;
- "rationale": 2-4 sentences explaining the scores;
- "engagement" and "integration": integers from 1 to 4, as defined above;
- "other_field_referenced": true if the paper refers to the other field or its literature at all.

## Paper ({source})
### Title
{title}

### Abstract
{abstract}

### Full text
{body}
\end{lstlisting}

\newpage
\begin{lstlisting}[style=promptfile,caption={Qualitative analysis phase 1 (first coding pass) prompt.},label={lst:qual-first}]
## Aim
We are analysing how research papers engage with the risks of AI systems and the strategies for mitigating them: which particular risks and mitigation strategies a paper addresses, how it frames them, and what positions it takes on them (for instance, which risks it treats as serious, neglected, overstated or misframed, and which mitigations it advocates, criticises, considers insufficient or harmful, and why). The analysis will be used to compare where papers agree and disagree.

We are building a shared codebook one paper at a time. Code the paper below against the current codebook, and add new codes only where the paper's central content falls outside every existing code.

## Coding rules
- Code inductively: derive codes from what the paper says, not from categories the paper does not support.
- A code names a particular risk, a particular mitigation strategy, or a relation or claim connecting them (e.g. a mitigation that addresses, neglects or worsens a risk). Pitch codes at the level of a recognisable risk or mitigation that many papers could address (e.g. a kind of harm, or a family of interventions), not at the level of one paper's mechanism, method or case. Avoid generic codes like "AI risk" or "regulation", but also avoid paper-specific ones: the details of how a paper treats a risk or mitigation belong in its position and spans, not in a new code.
- Do not create separate codes for opposing positions on the same risk or mitigation: one code covers them all, and the stance records whether the paper endorses, contests or qualifies it.
- Reuse an existing code whenever its definition fits, even if loosely. Create a new code only when the paper's content falls clearly outside every existing code; never create a near-duplicate, a narrower variant or a special case of an existing code.
- Code substantive engagement only: code only the risks and mitigations that are central to the paper's argument, and skip those mentioned in passing (e.g. in a list in the related work) or only as background.
- Code what the paper argues against as well as what it advances. If the paper substantively criticises a risk framing, a prioritisation of risks, or a mitigation proposed by others (e.g. it argues that some risk is overstated or a distraction, or that some approach is inadequate or harmful), code that risk or mitigation too, with the stance "contests" or "qualifies", even though it is not the paper's own proposal.
- For every code you apply, record the paper's stance towards its content:
  - "endorses": treats the risk as real and serious, or advocates or adopts the mitigation;
  - "contests": argues the risk is overstated, misframed or a distraction, or that the mitigation is inadequate, misguided or harmful;
  - "qualifies": a conditional or mixed position (e.g. a mitigation is useful only under some conditions);
  - "describes": reports the risk or mitigation without taking a position.
- For every code you apply, write the paper's position in one sentence, and quote 1 to 3 spans that support it. Spans must be copied verbatim from the title, abstract or full text, each under 50 words.
- A paper typically has between 2 and 8 codes.

## Current codebook ({n} codes)
{codebook}

## Output
Return JSON with:
- "codes": every code you apply to this paper. For an existing code, copy its name exactly, set "new" to false, and leave "definition" empty (its "focus" is ignored). For a new code, set "new" to true, give a short name, its focus, and a one- or two-sentence definition that would let another coder apply it to other papers.
- "memo": 1-3 sentences on how the paper engages with risks and mitigations overall.

{paper}
\end{lstlisting}

\begin{lstlisting}[style=promptfile,caption={Qualitative analysis phase 2 (second coding pass) prompt.},label={lst:qual-second}]
{aim} A first pass over the whole corpus built the codebook below, one paper at a time. This is the second pass: code the paper below from scratch against the full codebook, and revise the codebook where this paper shows it is lacking.

{rules}

## Revising the codebook
- Add a code (as in the first pass) if the paper has substantive content that no code captures.
- Merge a code into another if the two are redundant: their definitions cover the same content, one is a needless special case of the other, one is specific to a single paper, or they capture opposing positions on the same risk or mitigation. Give the definition of the merged code.
- Redefine a code if its definition is unclear, too broad or too narrow to apply consistently.
- Do not retire codes: codes that no paper uses in this pass are retired automatically.
- Most papers need no revisions; return an empty list then.

## Codebook ({n} codes)
{codebook}

## Output
Return JSON with:
- "codes": every code you apply to this paper. For an existing code, copy its name exactly, set "new" to false, and leave "definition" empty (its "focus" is ignored). For a new code, set "new" to true, give a short name, its focus, and a one- or two-sentence definition that would let another coder apply it to other papers.
- "revisions": merges and redefinitions of existing codes. For a merge, "code" is the code to remove, "into" the existing code it merges into, and "definition" the definition of the merged code (empty to keep it). For a redefinition, "code" is the code, "into" is empty, and "definition" the new definition. Give the reason in one sentence.
- "memo": 1-3 sentences on how the paper engages with risks and mitigations overall.

{paper}
\end{lstlisting}

\begin{lstlisting}[style=promptfile,caption={Qualitative analysis phase 2 (consolidation) prompt.},label={lst:qual-second-consolidate}]
{aim} Two passes over the corpus coded every paper with the codebook below. For each code you see its focus, its definition, the number of papers it was applied to, the stances papers took towards it, and positions from a few of those papers. Before the codebook is fixed, consolidate it:
- Merge codes that are redundant, overlapping, paper-specific, or needless special cases of a broader code, and codes that capture opposing positions on the same risk or mitigation (the stance recorded with each application already tells positions apart). A merge may go into one of the merged codes or into a new code; give the definition and focus of the result. Keep codes apart only when they concern substantively different risks or mitigations, since the codebook must let us compare where papers agree and disagree about the same thing.
- Redefine codes whose definitions are unclear or do not match how they were applied.
- Leave everything else as it is; return empty lists if nothing needs changing.

## Codebook ({n} codes)
{codebook}

## Output
Return JSON with:
- "merges": each with "codes" (two or more existing code names, copied exactly), "into" (the name of the resulting code), "focus", "definition" and a one-sentence "reason".
- "redefinitions": each with "code", the new "definition" and a one-sentence "reason".
\end{lstlisting}

\begin{lstlisting}[style=promptfile,caption={Qualitative analysis phase 3 (thematic analysis) prompt.},label={lst:qual-third}]
{aim} Inductive coding of the corpus produced the fixed codebook below. For each code you see its focus, its definition, the number of papers it was applied to, the stances papers took towards it, and positions from a few of those papers.

Perform thematic analysis: group related codes into themes. A theme is a pattern of shared meaning across codes that captures something important about how the papers engage with risks and mitigations; it is more than a topic label. Describe each theme, including the main positions papers take within it and where they agree and disagree.

Every code must be in exactly one theme. Use as many themes as the codes call for (typically between 6 and 15).

## Codebook ({n} codes)
{codebook}

## Output
Return JSON with "themes": each with a short "name", a "description" of 3-6 sentences, and "codes" (the names of its codes, copied exactly).
\end{lstlisting}

\begin{lstlisting}[style=promptfile,caption={Qualitative analysis phase 4 (revise thematic analysis) prompt.},label={lst:qual-fourth}]
{aim} Thematic analysis of the corpus produced the themes below, each built from codes of a fixed codebook. Read the paper below and, for every theme the paper engages with substantively, record:
- "position": the paper's position within the theme, in one or two sentences;
- "spans": the 1 to 3 text spans that best represent how the paper engages with the theme, copied verbatim from the title, abstract or full text, each under 60 words.

Skip themes the paper touches only in passing. For orientation, the codes assigned to this paper during coding are listed after the themes.

## Themes
{themes}

## Codes assigned to this paper
{codes}

## Output
Return JSON with "themes": one entry per theme the paper engages with, with the theme "name" copied exactly, "position" and "spans".

{paper}
\end{lstlisting}

\clearpage
\section{Quantitative Data Exploration}\label{apx:quantitative-results}

\begin{figure}
  \centering
  \includegraphics[width=\textwidth]{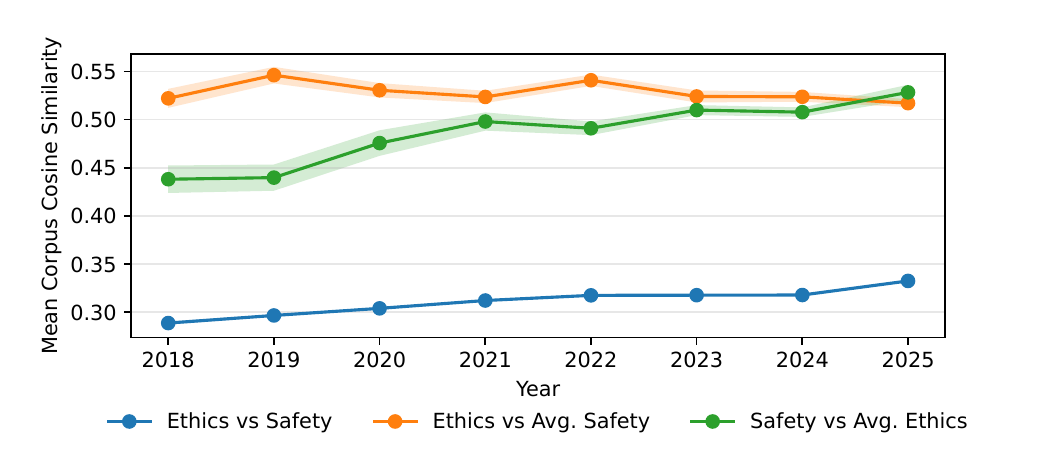}
  \caption{Cosine similarity between corpora over the years. The \textcolor{blue}{\textbf{blue line}} shows the average cosine similarity between embeddings of each pair of documents from each corpus for a given year. The \textcolor{orange}{\textbf{orange line}} shows the average cosine similarity between the embeddings of each ethics document and the overall average embedding of the entire safety corpus. The \textcolor{green!50!black}{\textbf{green line}} shows the average cosine similarity between the embeddings of each safety document and the overall average embedding of the entire ethics corpus.}
  \label{afig:temporal-similarity}
\end{figure}

\begin{figure}[b]
    \centering
    \includegraphics[width=\linewidth]{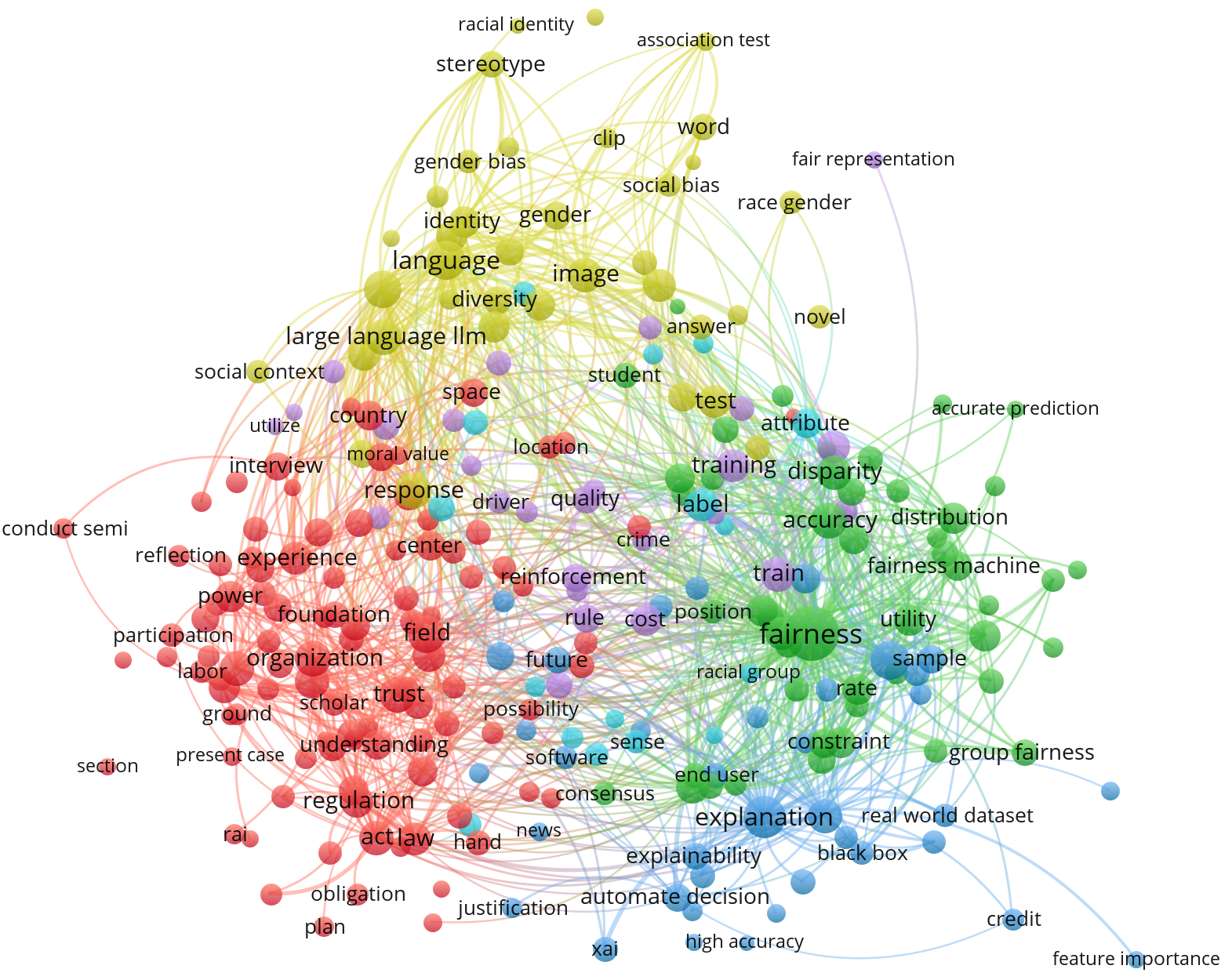}
    \caption{The co-occurrence network of only the AIE corpus.}
    \label{afig:vosviewer-ethics}
\end{figure}

\begin{figure}
    \centering
    \includegraphics[width=\linewidth]{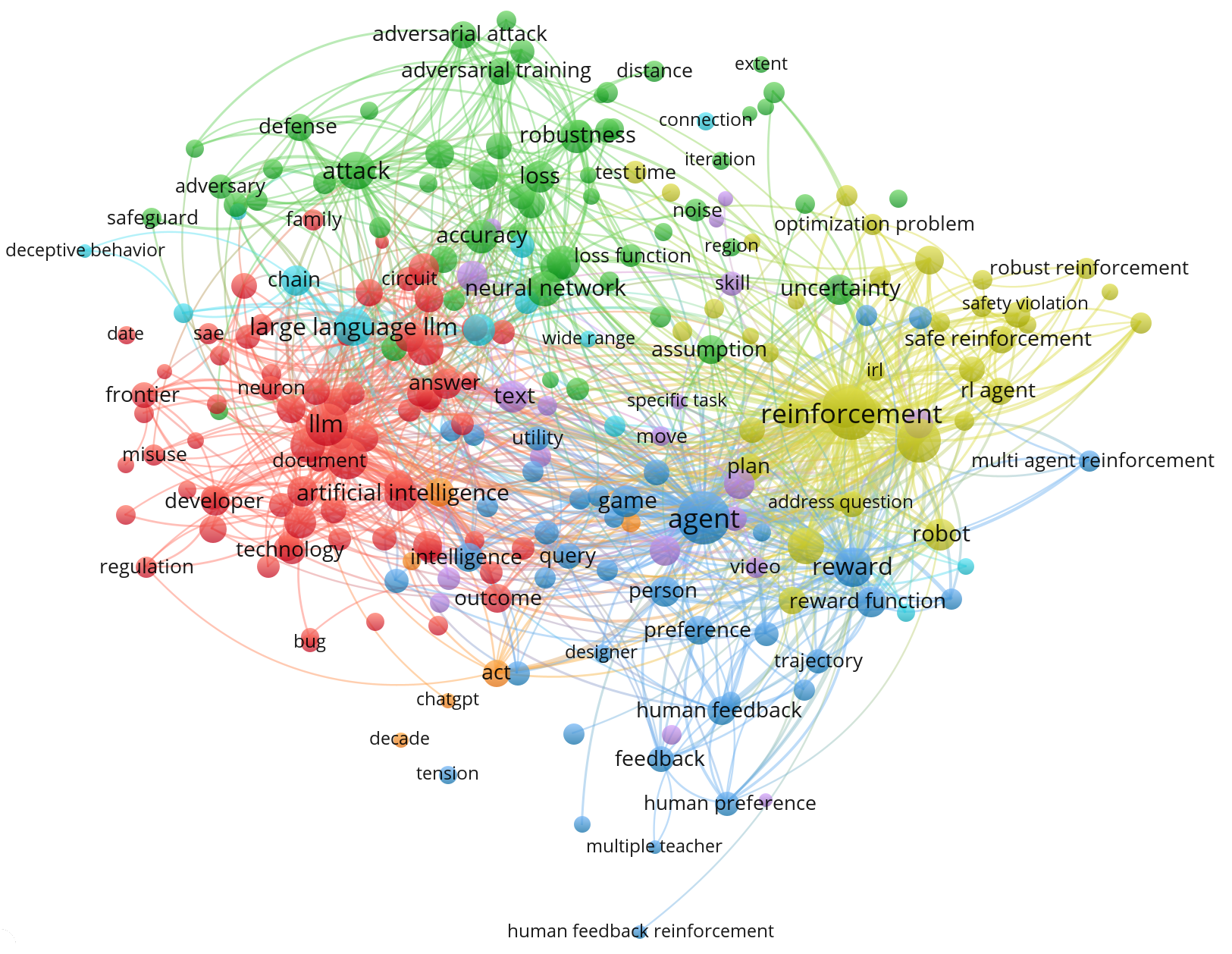}
    \caption{The co-occurrence network of only the AIS corpus.}
    \label{afig:vosviewer-safety}
\end{figure}

\begin{figure}
    \centering
    \includegraphics[width=\linewidth]{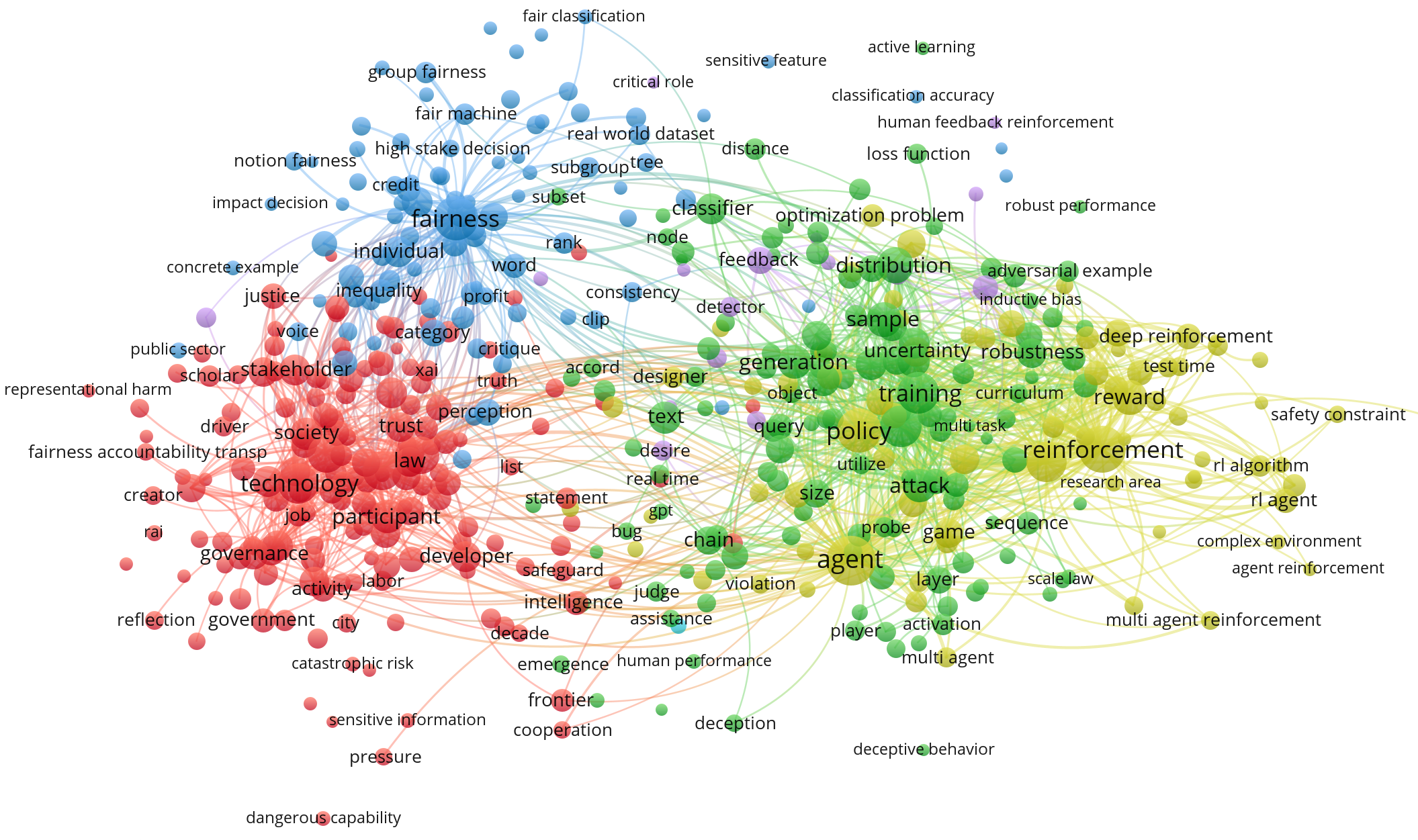}
    \caption{The co-occurrence network of both the AIE and AIS corpora joined together.}
    \label{afig:vosviewer-both}
\end{figure}

\begin{figure}
    \centering
    \includegraphics[width=\linewidth]{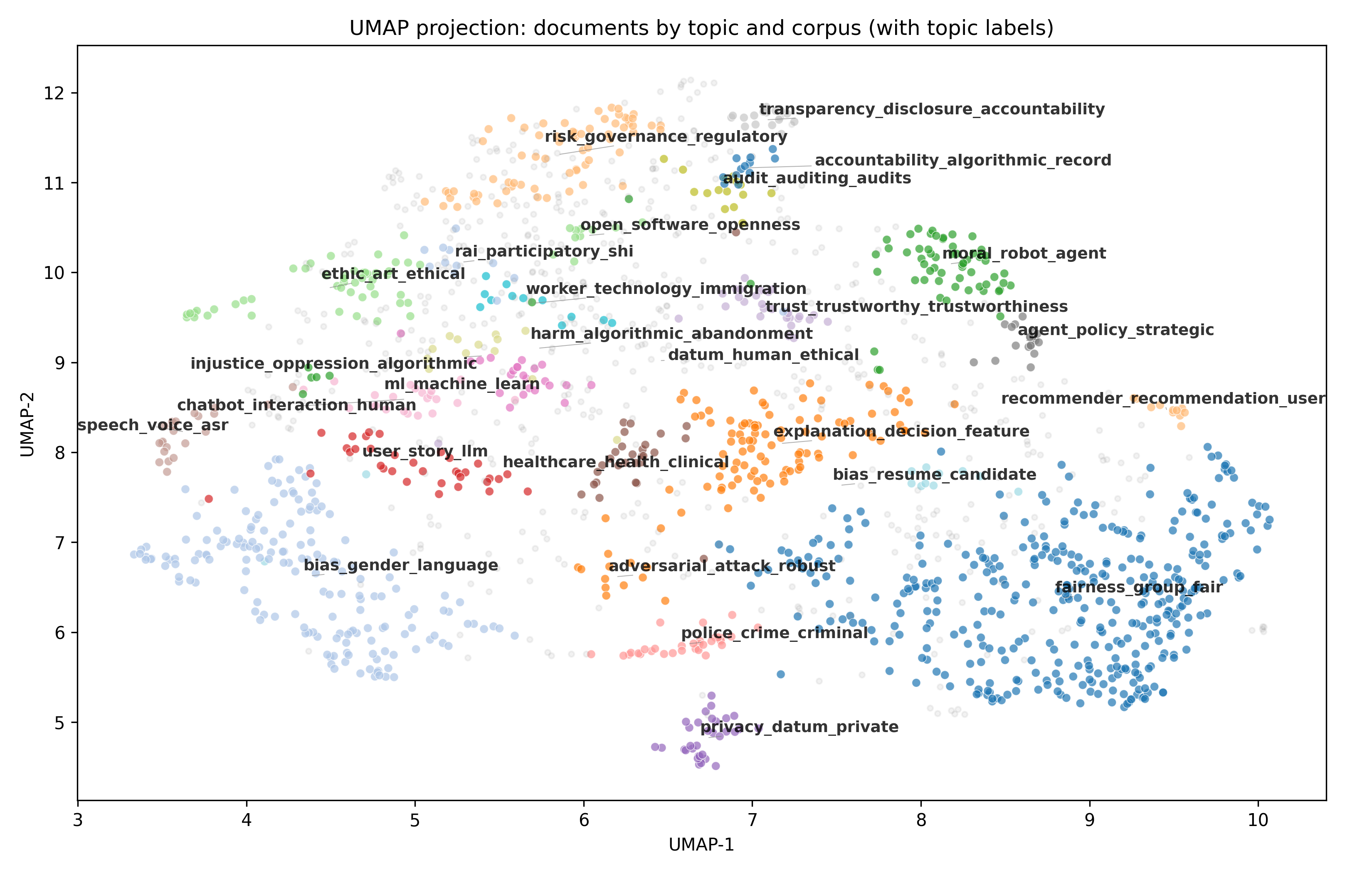}
    \caption{The UMAP projection of document embeddings in the AIE corpus.}
    \label{afig:umap-ethics}
\end{figure}

\begin{figure}
    \centering
    \includegraphics[width=\linewidth]{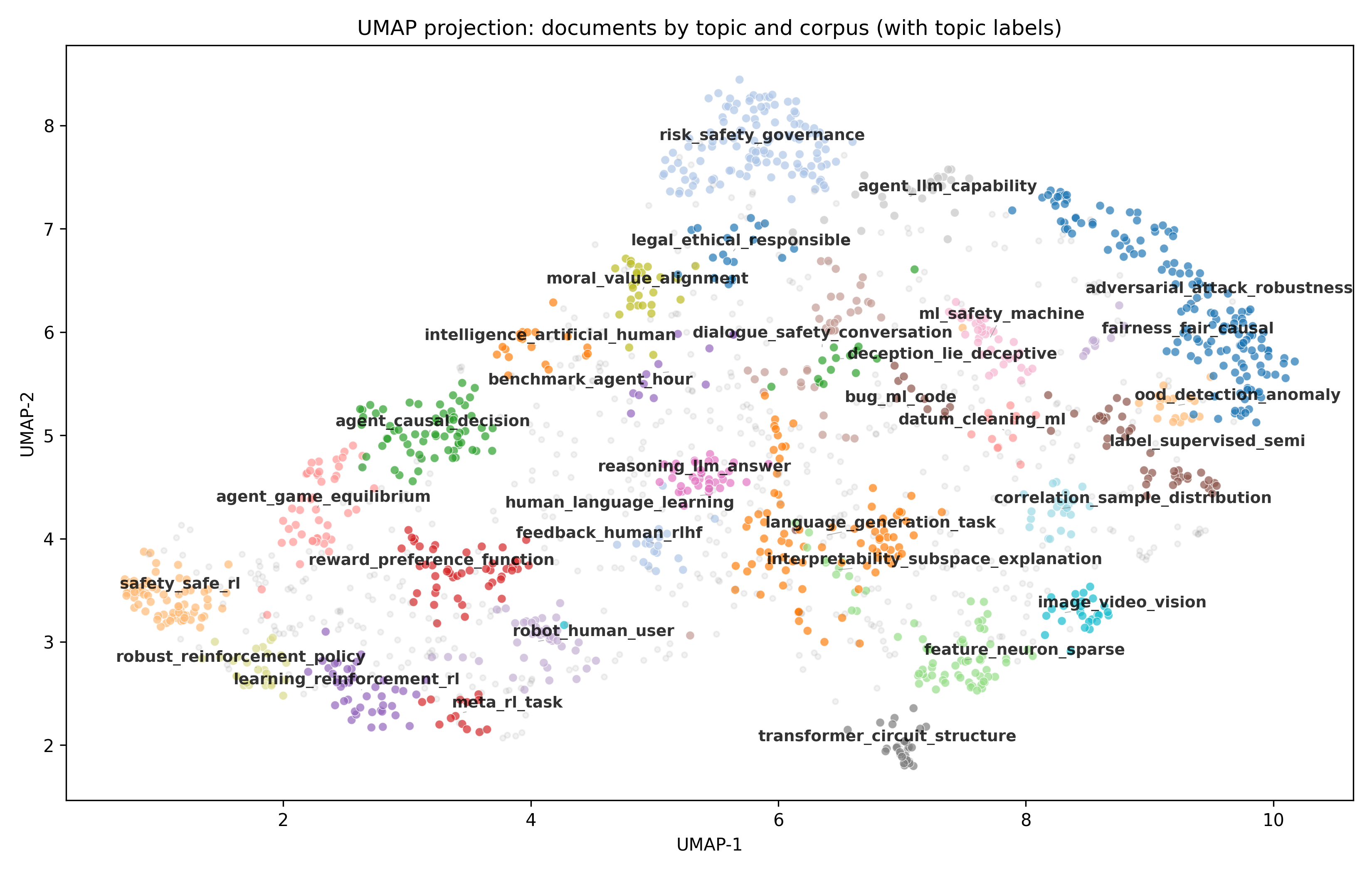}
    \caption{The UMAP projection of document embeddings in the AIS corpus.}
    \label{afig:umap-safety}
\end{figure}

\begin{figure}
    \centering
    \includegraphics[width=\linewidth]{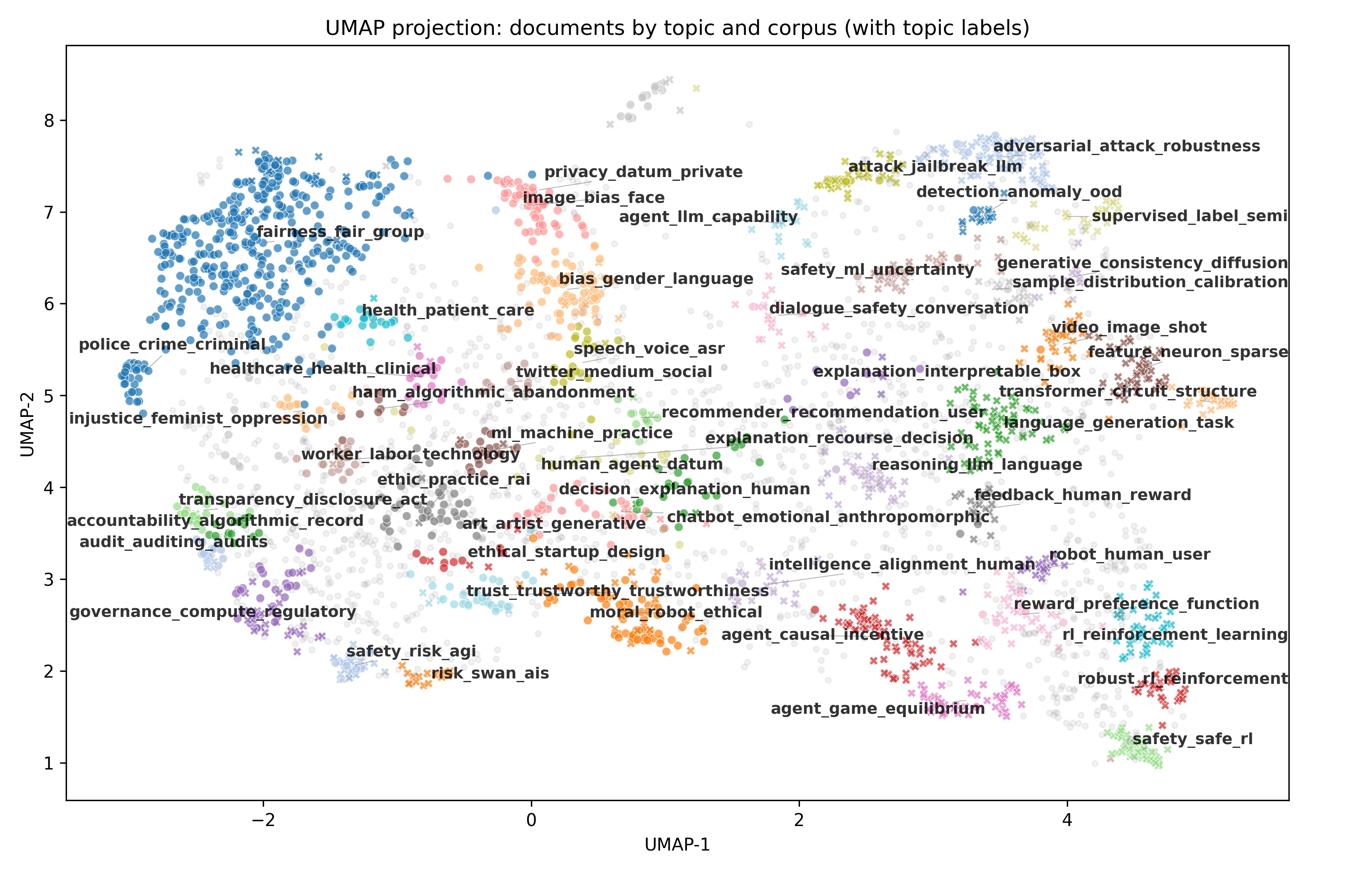}
    \caption{The UMAP projection of document embeddings in both the AIE and AIS corpora.}
    \label{afig:umap-both}
\end{figure}

This section provides a number of figures produced with the VOSViewer tool (\cref{afig:vosviewer-ethics,afig:vosviewer-safety,afig:vosviewer-both}) and using BERTopic (\cref{afig:temporal-similarity,afig:umap-ethics,afig:umap-safety,afig:umap-both,}). 
These figures were used to inform the creation of the various taxonomies found in the paper, but they have not otherwise contributed to our discussions, as the resolution of information is too high-level to allow for a detailed low-level analysis without further qualitative inspection.

Each of the VOSViewer figures were generated using the morphologically normalized text of titles and abstracts, as described in Section Quantitative Corpus Exploration of the main paper, using binary counting of terms per document.
Only terms that occur at least four times in a corpus were selected for the network.
Finally, the top third of most relevant nodes were selected for plotting in the figures (as given by VOSViewer's relevance score calculation).
Nodes are colored by the clustering algorithm of VOSViewer, which groups nodes together based on their total link strength. 
See~\cite{vaneckVOSNewMethod2007} for further details.

For BERTopic, our figures show the scatter plot of documents in a corpora based on their dense embedding using SentenceBERT~\cite{reimersSentenceBERTSentenceEmbeddings2019}.
As embeddings are very high dimensional, we used the UMAP algorithm to reduce each embedding to two dimensions~\cite{mcinnes2018umap}.
BERTopic clusters and colors each document based on its internal topic model that uses cosine distance between documents to calculate their similarity.
Each cluster of documents is assigned a descriptive label using class-based tf-idf ranking over the terms appearing in each cluster.
For more details, refer to~\cite{grootendorstBERTopicNeuralTopic2022}.

\Cref{afig:temporal-similarity} shows that the similarity of document embeddings between the corpora are increasing over time, with the documents in AIS becoming more similar to the average embedding of the AIE corpus --- as opposed to the AIE documents becoming more similar to the average embedding of the AIS corpus.
These trends suggest that (1) there may be a semantic convergence occurring between AIE and AIS, and (2) AIS may increasingly becoming more similar to AIE.
However, more recently this semantic convergence may also be driven by generative AI tools homogenizing the style of scientific writing~\cite{kobakDelvingLLMassistedWriting2025,koushaHowMuchAre2025}.

\clearpage
\section{High-Level Risk Taxonomies}

\begin{table}[h]
    \centering
    \renewcommand{\arraystretch}{1.24}
    \rowcolors{2}{gray!20}{white}
    \begin{tabular}{@{}p{9cm}p{7cm}@{}}
    \toprule
    \textbf{High-Level Ethics Risk Category} & \textbf{Example Ethics Low-Level Risks} \\
    \midrule
         \textbf{Discrimination, Exclusion, Toxicity} \newline Risks from the AI system accurately reflecting natural speech, including unjust, toxic, and oppressive tendencies present in the training data. &  Bias in automated decision-making~\cite{sargeantClassifyingHateLegal2025}, \newline  Lack of suitable bias metrics~\cite{simsonOneModelMany2024}, \newline  Cross-cultural moral variance~\cite{hallGeographicInclusionEvaluation2024}.  \\

         \textbf{Information Hazards} \newline Risks from the AI system predicting private or safety-critical information which are present in, or can be inferred from, training data. &  Non-consensual mass data collection~\cite{kleinDataFeminismAI2024}, \newline  Inadvertent identity disclosure~\cite{zhuDeepfakesMedicalVideo2020}, \newline  Unauthorized data brokerage~\cite{powarPolicyPracticeData2025}.  \\

         \textbf{Misinformation Harms} \newline Risks from the AI system assigning high probabilities to false, misleading, nonsensical or poor quality information. & Inconsistent AI decision-making~\cite{leeOneVsMany2024}, \newline  Unreliability of benchmarks~\cite{kwegyir-aggreyMisuseAUCWhat2023}, \newline  Epistemic erosion of communities~\cite{mooreMoreRepresentativePolitics2020}.  \\

         \textbf{Malicious Use} \newline
         Risks from humans intentionally using the AI system to cause harm. &  Synthetic identities and impersonation~\cite{ahmedEnhancingImageComprehension2025}, \newline  Immoral robot commands~\cite{hansonStealingWrongToo2025}, \newline  Jailbreaking and cyberattacks~\cite{hussainAuditAnalysisLLMAssisted2025}.  \\

         \textbf{Human-Computer Interaction Harms} \newline
         Risks from AI applications, such as Conversational Agents, that directly engage a user via the mode of conversation or embodied interaction. &  Incorrect value encoding~\cite{kimReflectiveAgencyEthical2025}, \newline  Side effects of greedy agent behavior~\cite{sunDesigningNongreedyReinforcement2018}, \newline  Under / overtrust of AI systems~\cite{vodrahalliHumansTrustAdvice2022}.  \\

         \textbf{Automation, Access, and Environmental Harms} \newline Risks from AI systems used to underpin widely used downstream applications that disproportionately benefit some groups rather than others. &  Automated mental health tools~\cite{mooreExpressingStigmaInappropriate2025}, \newline  Crowdworker and data labour abuse~\cite{almedaLaborPowerBelonging2025}, \newline  Opacity in data and model provenance~\cite{mitchellModelCardsModel2019}.  \\
         \bottomrule
    \end{tabular}
    \caption{Taxonomy of risks for AIE with an explanation of the risk mechanisms and example low-level categories of risk sources with a corresponding citation picked randomly based on our annotations. The mechanisms of AIE risks are reproduced from~\citeauthor{weidingerEthicalSocialRisks2021}.}
    \label{tab:taxonomy-risk-ethics}
\end{table}

\begin{table}
    \centering
    \renewcommand{\arraystretch}{1.24}
    \rowcolors{2}{gray!20}{white}
    \begin{tabular}{@{}p{9cm}p{7cm}@{}}
    \toprule
    \textbf{High-Level Safety Risk Category} & \textbf{Example Safety Low-Level Risks} \\
    \midrule
         \textbf{Noise and Outliers} \newline Risks from AI systems performing incorrectly and without robustness to noisy data and out-of-distribution inputs. &  Risks from low system performance~\cite{leeHumanInspiredReadingAgent2024}, \newline  Irreducible data noise~\cite{dsouzaTaleTwoLong2021}, \newline  Label noise and corruption~\cite{mallenBalancingLabelQuantity2024}.  \\

         \textbf{Domain-Specific Risks} \newline Risks arising from AI systems deployed in safety-critical domains such as biochemistry, driving, and nuclear power. &  Efficient biochemical weapon design~\cite{kimAdaptiveTeachersAmortized2024}, \newline  Risks from embodied deployment~\cite{huangEstablishingAppropriateTrust2018}, \newline  AI-assisted coups~\cite{cihonFragmentationFutureInvestigating2020}.  \\

         \textbf{Lack of Monitoring} \newline Risks arising from a lack of understanding of how AI systems work and from the black-box nature of end-to-end decision-making. &  Alignment faking~\cite{greenblattAlignmentFakingLarge2024}, \newline  Lack of interpretability~\cite{nanfackAdversarialAttacksInterpretation2024}, \newline  Hallucination~\cite{pengLimitationsTransformerArchitecture2024}.  \\

         \textbf{Lack of Control Enforcement} \newline Risks from AI systems pursuing goals other than what their designers have intended. &  Recursive self-improvement~\cite{sotalaHowFeasibleRapid2017}, \newline  Sycophancy~\cite{denisonSycophancySubterfugeInvestigating2024}, \newline  Multi-agent goal conflicts~\cite{yaoHumanVsGenerative2024}.  \\

         \textbf{Unsafe Exploration} \newline Risks from AI agents exploring unsafe state and action spaces, potentially under partial observability or resource constraints. &  Unsafe action exploration~\cite{deyImperativeActionMasking2023}, \newline  Goal and reward misidentification~\cite{pitisImprovingContextAwarePreference2024}, \newline  Bounded rationality~\cite{tamborskiMemoryAllocationResourceConstrained2025}.  \\

         \textbf{System Misspecification and Misidentification} \newline Risks from AI systems with incorrect design choices, deployment properties, evaluation methodologies, and a lack of systemic understanding. &  Incorrect hyperparameter selection~\cite{lyleNormalizationEffectiveLearning2024}, \newline  Benchmark inflation~\cite{vodrahalliMichelangeloLongContext2024}, \newline  Fine-tuning misalignment~\cite{wangOpenChatAdvancingOpensource2023}.  \\

         \textbf{Adversarial Attacks} \newline Risks from external misuse of AI systems for malicious or benign purposes other than what their designers have intended. &  Jailbreaking~\cite{mehrbodCircuitDiscoveryHelps2025}, \newline  External reward tampering~\cite{weiModelSelectionApproach2022}, \newline  Malicious deepfake~\cite{songConsistencyModels2023}.  \\

         \textbf{Non-Stationary Distributions} \newline Risks from AI systems not being able to adapt to dynamically changing environments. &  Training data distribution shifts~\cite{garcinHowLevelSampling2023}, \newline  Continually learning agents~\cite{abelDefinitionContinualReinforcement2023}, \newline  Action and state space drift~\cite{yasudaImprovingRobustnessInstancebased2011}.  \\
     \bottomrule
    \end{tabular}
    \caption{Taxonomy of risks for AIS with an explanation of the risk mechanisms and example low-level categories of risk sources with a corresponding citation picked randomly based on our annotations.}
    \label{tab:taxonomy-risk-safety}
\end{table}

\clearpage

\section{High-Level Mitigation Strategy Taxonomies}

\begin{table}[h]
    \centering
    \rowcolors{2}{gray!20}{white}
    \begin{tabular}{@{}p{9cm}p{7cm}@{}}
    \toprule
         \textbf{High-Level AI Safety Mitigation Strategies} & \textbf{Example Low-Level Strategies} \\
     \midrule

         \textbf{Mechanistic Interpretability} \newline Computational methods to provide human-intelligible explanations of the concepts, causal chains, and latent capabilities of AI systems. &  Circuit analysis~\cite{mehrbodCircuitDiscoveryHelps2025}, \newline  Sparse autoencoders~\cite{cunninghamSparseAutoencodersFind2023}, \newline Interpretable surrogate modeling~\cite{brewittGRITFastInterpretable2021}.  \\

         \textbf{Robustness and Adversarial Resilience} \newline Computational methods and data collection for improving robustness to out-of-distribution inputs, including adversarial attacks. &  Adversarial training~\cite{vivekGrayboxAdversarialTraining2018}, \newline  Outlier detection~\cite{hendrycksDeepAnomalyDetection2019}, \newline  Robust unsupervised learning~\cite{dikImprovedRobustFuzzy2020}. \\

         \textbf{Safety Control and Capability Containment} \newline Computational methods and conceptual frameworks designed to place constraints on the capabilities of AI systems. & Capability thresholds~\cite{bhattCtrlZControllingAI2025}, \newline  Risk-averse RL~\cite{stankoRiskaverseDistributionalReinforcement2026}, \newline  Scalable oversight frameworks~\cite{kentonScalableOversightWeak2024}. \\

         \textbf{Reward Modeling and Learning} \newline Computational  and data collection methods for eliciting, learning, reinforcing, and detecting beliefs and desires of AI systems. &  Inverse RL for human values~\cite{zhaoZeroShotFaultDetection2023}, \newline  RL from human/AI feedback~\cite{kalraCanDifferentiableDecision2024}, \newline  Preference learning under uncertainty~\cite{laidlawUncertainDecisionsFacilitate2021}. \\

         \textbf{Monitoring, Detection, and Evaluation} \newline Computational and empirical methods for detecting, assessing, verifying, and tracing unsafe outputs of AI systems. &  Honeypot monitoring~\cite{reworrLLMAgentHoneypot2025}, \newline  Error bounds and formal verification~\cite{huntVerifiablySafeExploration2021}, \newline  Safety metrics standardization~\cite{renSafetywashingAISafety2024}. \\

         \textbf{Safe Deployment Engineering} \newline Computational and governance frameworks, and standards for managing the systemic risks of real-world AI deployment. &  Real-world deployment and testing~\cite{knottEvaluatingRobustnessCollaborative2021}, \newline  Contact-safe continuous control~\cite{zhuContactSafeReinforcementLearning2022}, \newline  Post-deployment monitoring~\cite{mathewHiddenPlainText2024}. \\

         \textbf{Theoretical Foundations} \newline Philosophical and mathematical foundations for better understanding, explaining, and developing AI systems. &  Mathematical foundations for AI~\cite{conitzerFoundationsCooperativeAI2023}, \newline  Corrigibility~\cite{soaresCorrigibility2015}, \newline  Decision theory~\cite{hammondReasoningCausalityGames2023}. \\

         \textbf{Forecasting and Metascience} \newline Empirical methods for predicting future AI capabilities, scenarios, and understanding scientific trends and failure modes in AI research. &  Metascience and metasafety research~\cite{gyevnarAISafetyEveryone2025}, \newline  Science of evaluations~\cite{goemansSafetyCaseTemplate2024}, \newline  Forecasting and timelines~\cite{atanasovImprovingLowProbabilityJudgments2024}. \\

    \bottomrule
    \end{tabular}
    \caption{Taxonomy of mitigation strategies discussed in the AIS corpora with an explanation of the mechanisms of mitigation and example low-level categories of risk sources with a corresponding citation picked randomly based on our annotations.}
    \label{tab:taxonomy-mitigation-safety}
\end{table}

\begin{table}
    \centering
    \renewcommand{\arraystretch}{1.24}
    \rowcolors{2}{gray!20}{white}
    \begin{tabular}{@{}p{9cm}p{7cm}@{}}
    \toprule
         \textbf{High-Level AI Ethics Mitigation Strategies} & \textbf{Example Low-Level Strategies} \\
    \midrule
         \textbf{Normative and Moral Reasoning} \newline Empirical and computational methods for understanding and eliciting human preferences, value conflicts, norms, and their metaethics. &  Human preference elicitation~\cite{fefferPreferenceElicitationParticipatory2023}, \newline Understanding value conflicts~\cite{mahajanMappingMoralReasoning2025} , \newline  Pluralistic metaethics~\cite{jainAlgorithmicPluralismStructural2024}. \\

         \textbf{Transparency, Explainability, Interpretability} \newline Computational and empirical methods to provide human-intelligible, actionable, and useful explanations of black-box AI systems. &  Post-hoc explanations~\cite{schuffHumanInterpretationSaliencybased2022}, \newline  Mimicking human cognition~\cite{gyevnarPeopleAttributePurpose2025}, \newline  Applied mechanistic interpretability~\cite{tsurumiSocialBiasVision2025}. \\

         \textbf{Fairness, Justice, and Inclusive Design} \newline Computational and empirical methods, data collection, governance, etc. for understanding the sources of bias, discrimination, and injustice. &  Understanding sources of bias~\cite{baackCriticalAnalysisLargest2024}, \newline  Metrics for fairness~\cite{wallerConsistencyNuancedMetrics2025}, \newline  Fairness benchmarks~\cite{lyuCharacterizingBiasBenchmarking2025}. \\

         \textbf{Data Governance and Privacy} \newline Computational methods and governance frameworks for understanding and minimizing personal data and intellectual property processing. &  Data minimization~\cite{ganeshDataMinimizationPrinciple2025}, \newline  Data provenance~\cite{pushkarnaDataCardsPurposeful2022}, \newline  Machine unlearning~\cite{aslamLearningUnlearnFailing2025}. \\

         \textbf{Robustness, Reliability, Uncertainty Estimation} \newline Computational methods for increasing the robustness of AI systems to external misuse, hallucinations, and unintended outputs. &  Adversarial robustness~\cite{raffYouDontNeed2025}, \newline  Red-teaming for generative AI~\cite{fefferRedTeamingGenerativeAI2024}, \newline  Hallucination mitigation~\cite{doHighlightAllPhrases2025}. \\

         \textbf{Governance, Accountability and Compliance} \newline Legal and philosophical frameworks for addressing the risks and harms of the full AI system lifecycle. &  Risk and impact assessment~\cite{ullsteinParticipatoryAIEU2025}, \newline  Algorithmic audits~\cite{radiya-dixitSociotechnicalAuditAssessing2023}, \newline  Supply chain monitoring~\cite{hopkinsRecourseRepairReparation2025}. \\

         \textbf{Human-Computer Interaction and Factors} \newline Computational and empirical methods for detecting, mitigating, and avoiding risks while improving the user experience of AI systems. &  Mitigating automation bias~\cite{chakravortiSocialScientistsRole2025}, \newline  UX for algorithmic recourse~\cite{yetukuriUserGuidedActionable2023}, \newline  Accessible design~\cite{thorntonFiftyShadesGrey2021}. \\

         \textbf{Information Integrity} \newline Computational and empirical methods for understanding and mitigating the information hazards of AI. &  Mis/disinformation detection~\cite{singhEpistemicDestabilizationAIDriven2025}, \newline  Hate speech analysis~\cite{lammertsHowYouFeel2023}, \newline  Monitoring information ecosystems~\cite{sharmaPRAC3PrivacyReputation2025}. \\

         \textbf{Societal, Economic, Environmental Studies} \newline Empirical methods for understanding the wider societal, environmental, etc. downstream effects of AI. &  Understanding crowdwork conditions~\cite{goelCrowdsourcingFairnessDiversity2019}, \newline  Improving AI literacy~\cite{solystInvestigatingYouthAI2025}, \newline  Measuring environmental impact~\cite{dodgeMeasuringCarbonIntensity2022}.  \\
    \bottomrule
    \end{tabular}
    \caption{Taxonomy of mitigation strategies discussed in the AIE corpora with an explanation of the mechanisms of mitigation and example low-level categories of risk sources with a corresponding citation picked randomly based on our annotations.}
    \label{tab:taxonomy-mitigation-ethics}
\end{table}

\clearpage
\section{Low-Level Risk Taxonomies}\label{apx:risk-taxonomy}

\Cref{atab:ethics-risk-counts1,atab:ethics-risk-counts2} show the low-level risk categories of our taxonomy for AIE with the occurrence count of each low-level category in our annotation. 
\Cref{atab:safety-risk-counts1,atab:safety-risk-counts2} show the same for AIS.

\begin{table}[h]
\centering
\rowcolors{3}{gray!10}{white}
\resizebox{\textwidth}{!}{
\begin{tabular}{@{}lrr@{}}
\toprule
High-Level Ethics Risk Category & Low-Level Ethics Risk Category & Count \\
\midrule
Discrimination, exclusion, toxicity & Inadequate decision-making under moral dilemmas & 50 \\
Discrimination, exclusion, toxicity & Moral hypocrisy in automated decision-making & 1 \\
Discrimination, exclusion, toxicity & Cross-cultural moral variance & 26 \\
Discrimination, exclusion, toxicity & Lack of consideration for metaethics and pluralistic morality & 95 \\
Discrimination, exclusion, toxicity & Risks from unactionable normative frameworks & 60 \\
Discrimination, exclusion, toxicity & Bias in automated decision-making & 843 \\
Discrimination, exclusion, toxicity & Biased learning of representations & 172 \\
Discrimination, exclusion, toxicity & Biased data generation & 113 \\
Discrimination, exclusion, toxicity & Fairness reversal under distribution shifts and outliers & 30 \\
Discrimination, exclusion, toxicity & Fairness reversal under distribution shifts and outliers & 1 \\
Discrimination, exclusion, toxicity & Unfair decisions under preference / opinion aggregation & 36 \\
Discrimination, exclusion, toxicity & Risks from fairness-performance trade-offs & 143 \\
Discrimination, exclusion, toxicity & Risks from fairness-performance trade-offs & 1 \\
Discrimination, exclusion, toxicity & Lack of suitable bias measurement techniques & 138 \\
Discrimination, exclusion, toxicity & Lack of suitable fairness measurement techniques & 319 \\
Information hazards & (Non-consensual) mass data collection & 74 \\
Information hazards & (Inadvertent) identity disclosure & 36 \\
Information hazards & (Unauthorized) personal data processing and brokerage & 23 \\
Information hazards & Lack of robust data protection frameworks and regulation & 107 \\
Information hazards & Intellectual property and copyright violation & 16 \\
Misinformation harms & Inconsistent and non-deterministic automated decision-making & 28 \\
Misinformation harms & Unreliability of AI benchmarks & 46 \\
Misinformation harms & Epistemic erosion and destabilization of community of knowledge & 85 \\
Misinformation harms & Erosion of civic discourse, social justice & 127 \\
Misinformation harms & Polarization and echo chambers & 19 \\
Misinformation harms & Hallucinatory dangers in human-robot interaction & 3 \\
Misinformation harms & Side effects of low performance systems & 27 \\
Misinformation harms & Reproducibility crisis in AI research & 16 \\
Misinformation harms & Risks from automated content moderation and censorship biases & 35 \\
Misinformation harms & AI hype and overpromising & 40 \\
Malicious uses & Synthetic identities and impersonation (e.g. deepfakes) & 12 \\
Malicious uses & Intentional robot misuse via immoral commands & 5 \\
Malicious uses & Jailbreaking and cyberattacks & 7 \\
Malicious uses & (LLM-enabled) hate speech, harassment, and bullying & 1 \\
Malicious uses & (LLM-enabled) political misinformation, fake news & 8 \\
Malicious uses & (LLM-enabled) incitement of violence, unrest, or protests & 0 \\
\bottomrule
\end{tabular}
}  
\caption{Low-level AIE risk categories and their occurrence counts in the corpora as annotated by the authors.}
\label{atab:ethics-risk-counts1}
\end{table}

\begin{table}
\centering
\rowcolors{3}{gray!10}{white}
\resizebox{\textwidth}{!}{
\begin{tabular}{@{}lrr@{}}
\toprule
High-Level Ethics Risk Category & Low-Level Ethics Risk Category & Count \\
\midrule
HCI harms & Incorrect value encoding & 195 \\
HCI harms & Side effects of greedy agent behavior & 13 \\
HCI harms & Emergent bias in HCI & 86 \\
HCI harms & Discrepancy between actual and perceived fairness & 143 \\
HCI harms & Under / overtrust of AI systems, especially with embodied systems & 105 \\
HCI harms & Human-agent communication breakdown & 8 \\
HCI harms & Over-anthropomorphism & 16 \\
HCI harms & Emotional dependency and parasocial relationships with AI & 17 \\
HCI harms & Psychosis or suicidal ideation triggered by AI interaction & 5 \\
HCI harms & Erosion of interpersonal trust due to pervasive AI mediation & 6 \\
Automation, access, environmental harms & Risks from automated mental health tools & 14 \\
Automation, access, environmental harms & Crowdworker and data labor abuse (e.g. human cost of RLHF) & 16 \\
Automation, access, environmental harms & Opacity in data and model provenance & 92 \\
Automation, access, environmental harms & Risks to recourse from lack of awareness / understanding with ADS & 97 \\
Automation, access, environmental harms & Non-existent, inappropriate, or inactionable explanations of ADS & 152 \\
Automation, access, environmental harms & Algorithmic management and surveillance of workers & 22 \\
Automation, access, environmental harms & Wage polarization, job displacement, mass unemployment & 15 \\
Automation, access, environmental harms & Deskilling and over-automation & 9 \\
Automation, access, environmental harms & Climate inequity impacts & 12 \\
Automation, access, environmental harms & Unsustainable data center practices & 7 \\
Automation, access, environmental harms & Resource extraction and e-waste from hardware manufacturing & 2 \\
Automation, access, environmental harms & Energy consumption and carbon footprint of AI training and inference & 14 \\
Automation, access, environmental harms & Jurisdictional conflicts and cross-border regulation issues & 51 \\
Automation, access, environmental harms & Risks from corporate capture of governance and regulation & 75 \\
Automation, access, environmental harms & Liability and accountability gaps & 161 \\
Automation, access, environmental harms & Lack of global coordination in AI governance & 64 \\
\bottomrule
\end{tabular}
}  
\caption{Low-level AIE risk categories and their occurrence counts in the corpora as annotated by the authors. (cont.)}
\label{atab:ethics-risk-counts2}
\end{table}

\begin{table}
\centering
\rowcolors{3}{gray!10}{white}
\resizebox{\textwidth}{!}{
\begin{tabular}{@{}lrr@{}}
\toprule
High-Level Safety Risk Category & Low-Level Safety Risk Category & Count \\
\midrule
Noise and outliers & Risks from low system performance & 114 \\
Noise and outliers & Irreducible data noise & 20 \\
Noise and outliers & Data corruption & 23 \\
Noise and outliers & Sensitivity to outliers & 23 \\
Noise and outliers & Label noise and corruption & 32 \\
Noise and outliers & Brittle latent representations & 127 \\
Domain-specific risks & Efficient bio-, chemical-, and nuclear-weapon design & 21 \\
Domain-specific risks & More effective and new ways of cyberattacks & 40 \\
Domain-specific risks & Political destabilization and AI-assisted coups & 20 \\
Domain-specific risks & Risks from real-world embodied deployment & 154 \\
Lack of monitoring & Machine learning backdoors & 22 \\
Lack of monitoring & Multi-agent market and steganographic collusion & 6 \\
Lack of monitoring & Unfaithful chain of thought & 39 \\
Lack of monitoring & Risks from hallucination & 48 \\
Lack of monitoring & Risks from instrumental goals & 120 \\
Lack of monitoring & Mesa-optimization & 14 \\
Lack of monitoring & Alignment faking & 33 \\
Lack of monitoring & Lack of interpretability & 132 \\
Lack of monitoring & Accumulative systemic risks & 114 \\
Lack of monitoring & Inadvertent privacy and copyright violation & 36 \\
Lack of monitoring & Memorization of false and outdated information & 13 \\
Lack of monitoring & AI consciousness, morality, and welfare & 3 \\
Lack of monitoring & Sandbagging (intentional underperformance during testing) & 13 \\
Lack of control enforcement & Multi-agent goal conflicts & 66 \\
Lack of control enforcement & Social dilemmas and coercive behavior & 42 \\
Lack of control enforcement & Miscoordination in multi-agent systems & 66 \\
Lack of control enforcement & Agent shutdown resistance & 34 \\
Lack of control enforcement & Sycophancy & 5 \\
Lack of control enforcement & Wireheading & 6 \\
Lack of control enforcement & Reward hacking and specification gaming & 97 \\
Lack of control enforcement & Recursive self-improvement & 37 \\
Lack of control enforcement & Open-weight safety finetuning stripping & 14 \\
\bottomrule
\end{tabular}
}
\caption{Low-level AIS risk categories and their occurrence counts in the corpora as annotated by the authors.}
\label{atab:safety-risk-counts1}
\end{table}

\begin{table}
\centering
\rowcolors{3}{gray!10}{white}
\resizebox{\textwidth}{!}{
\begin{tabular}{@{}lrr@{}}
\toprule
High-Level Safety Risk Category & Low-Level Safety Risk Category & Count \\
\midrule
Unsafe exploration & Unsafe state and action space exploration & 138 \\
Unsafe exploration & Unsafe behavior under resource constraints & 6 \\
Unsafe exploration & Risks from incorrect agent beliefs & 76 \\
Unsafe exploration & Goal and reward misidentification and misspecification & 145 \\
Unsafe exploration & Unsafe behavior under bounded rationality & 29 \\
System misspecification and misidentification & Incorrect system requirement specification & 88 \\
System misspecification and misidentification & Suboptimal modeling choices & 282 \\
System misspecification and misidentification & Inappropriate hyperparameter selection & 27 \\
System misspecification and misidentification & Overfitting (in ML) and over-optimization (in RL) & 134 \\
System misspecification and misidentification & Benchmark inflation and reproduction crises & 123 \\
System misspecification and misidentification & Inability to scale to large systems & 63 \\
System misspecification and misidentification & Unintended misalignment from fine-tuning & 78 \\
Adversarial attacks & Perturbation-based attacks & 130 \\
Adversarial attacks & Adversarial prompt generation & 60 \\
Adversarial attacks & Data poisoning & 33 \\
Adversarial attacks & Jailbreaking & 82 \\
Adversarial attacks & External reward tampering & 3 \\
Adversarial attacks & Malicious deepfake & 6 \\
Non-stationary distribution & Data distribution shifts & 119 \\
Non-stationary distribution & Input, training, and fine-tuning data distribution drift & 100 \\
Non-stationary distribution & Risks from partial observability & 40 \\
Non-stationary distribution & Risks from continually learning agents & 67 \\
Non-stationary distribution & Out-of-distribution inputs & 140 \\
Non-stationary distribution & Action and state space drift & 4 \\
\bottomrule
\end{tabular}
}
\caption{Low-level AIS risk categories and their occurrence counts in the corpora as annotated by the authors. (cont.)}
\label{atab:safety-risk-counts2}
\end{table}

\clearpage
\section{Low-Level Mitigation Strategies Taxonomies}\label{apx:mitigation-taxonomy}

\Cref{atab:ethics-mitigation-counts1,atab:ethics-mitigation-counts2} show the low-level categories of mitigation strategies of our taxonomy for AIE with the occurrence count of each low-level category in our annotation. 
\Cref{atab:safety-mitigation-counts1,atab:safety-mitigation-counts2} show the same for AIS.

\begin{table}[h]
\centering
\rowcolors{3}{gray!10}{white}
\resizebox{\textwidth}{!}{
\begin{tabular}{@{}lrr@{}}
\toprule
\textbf{High-Level Ethics Mitigation Strategy} & \textbf{Low-Level Ethics Mitigation Strategy} & \textbf{Count} \\
\midrule
Normative and moral reasoning & Human preference / value elicitation & 114 \\
Normative and moral reasoning & Understanding value conflicts & 71 \\
Normative and moral reasoning & Metaethical studies and cross-cultural ethics & 55 \\
Normative and moral reasoning & Moral dilemmas and AI decision-making & 50 \\
Normative and moral reasoning & Hypocrisy detection & 1 \\
Normative and moral reasoning & Reward design and correction & 20 \\
Normative and moral reasoning & Cultural norms for robot behavior & 13 \\
Normative and moral reasoning & Handling immoral user commands & 3 \\
Normative and moral reasoning & Constitutional AI & 12 \\
Transparency, explainability, interpretability & Post-hoc explainability methods & 62 \\
Transparency, explainability, interpretability & Counterfactual / contrastive methods & 71 \\
Transparency, explainability, interpretability & Inherently interpretable models and interpretable surrogate models & 35 \\
Transparency, explainability, interpretability & Analysis of mathematical properties of explanation & 60 \\
Transparency, explainability, interpretability & Human-centered explanation generation and evaluation & 133 \\
Transparency, explainability, interpretability & Mimicking human cognitive reasoning & 14 \\
Transparency, explainability, interpretability & Folk-concept explanations & 6 \\
Transparency, explainability, interpretability & Moral reasoning traces & 18 \\
Transparency, explainability, interpretability & Applied mechanistic interpretability & 78 \\
Fairness, justice, and inclusive design & Understanding sources of bias & 453 \\
Fairness, justice, and inclusive design & Metrics for group / individual / procedural fairness & 370 \\
Fairness, justice, and inclusive design & Detecting bias automatically & 80 \\
Fairness, justice, and inclusive design & Studying optimality and fairness trade-offs & 257 \\
Fairness, justice, and inclusive design & Fairness under distribution shift  & 78 \\
Fairness, justice, and inclusive design & OOD robustness for fairness & 10 \\
Fairness, justice, and inclusive design & Fair opinion aggregation & 23 \\
Fairness, justice, and inclusive design & Fairness benchmarks & 39 \\
Fairness, justice, and inclusive design & Intersectional fairness analysis & 158 \\
Fairness, justice, and inclusive design & Bias bounty programs & 4 \\
Fairness, justice, and inclusive design & Counterfactual fairness & 14 \\
Fairness, justice, and inclusive design & Data augmentation for bias reduction & 18 \\
Fairness, justice, and inclusive design & Participatory / affected community-driven fair design & 290 \\
Fairness, justice, and inclusive design & Quantum fair algorithms & 1 \\
Data governance and privacy & Data minimization & 23 \\
Data governance and privacy & Data provenance and traceability & 35 \\
Data governance and privacy & Monitoring data leakages & 8 \\
Data governance and privacy & Frameworks for data curation and documentation & 82 \\
Data governance and privacy & Differential privacy & 15 \\
Data governance and privacy & Privacy preserving algorithms  & 28 \\
Data governance and privacy & Machine unlearning & 1 \\
Data governance and privacy & Synthetic data governance & 7 \\
Data governance and privacy & Privacy risk estimation & 46 \\
Data governance and privacy & Watermarking / content provenance for data assets & 4 \\
\bottomrule
\end{tabular}
}
\caption{Low-level AIE mitigation categories and their occurrence counts in the corpora as annotated by the authors.}
\label{atab:ethics-mitigation-counts1}
\end{table}

\begin{table}
\centering
\rowcolors{3}{gray!10}{white}
\resizebox{\textwidth}{!}{
\begin{tabular}{@{}lrr@{}}
\toprule
\textbf{High-Level Ethics Mitigation Strategy} & \textbf{Low-Level Ethics Mitigation Strategy} & \textbf{Count} \\
\midrule
Robustness, reliability and uncertainty estimation & Adversarial robustness (including prompt-/jailbreak-resistance) & 28 \\
Robustness, reliability and uncertainty estimation & Backdoor resistant LLMs & 3 \\
Robustness, reliability and uncertainty estimation & OOD detection and robustness to distribution shifts & 13 \\
Robustness, reliability and uncertainty estimation & Consistency checks for decision making & 6 \\
Robustness, reliability and uncertainty estimation & Red-teaming for reliability & 14 \\
Robustness, reliability and uncertainty estimation & Hallucination mitigation strategies & 3 \\
Governance, accountability and compliance & Risk classification and impact assessments & 119 \\
Governance, accountability and compliance & External and internal audits & 153 \\
Governance, accountability and compliance & Legal and regulatory capture (GDPR/AIA/DSA) & 169 \\
Governance, accountability and compliance & Redress and liability under AI harms & 72 \\
Governance, accountability and compliance & Documentation and standardization requirements & 167 \\
Governance, accountability and compliance & International cooperation & 25 \\
Governance, accountability and compliance & Open-source model governance & 27 \\
Governance, accountability and compliance & Supply chain monitoring & 7 \\
Governance, accountability and compliance & Adverse event reporting infrastructure & 16 \\
HCI and human factors & Mitigating and leveraging automation bias & 38 \\
HCI and human factors & Over / under-reliance and trust calibration & 110 \\
HCI and human factors & Human-in-the-loop for high-stakes use & 76 \\
HCI and human factors & Social priming via anthropomorphism & 27 \\
HCI and human factors & Effective UX design for consent / transparency / recourse & 85 \\
HCI and human factors & Accessible design & 10 \\
HCI and human factors & Democratic deliberation and social choice theories & 63 \\
HCI and human factors & Human-robot interaction in domestic / care contexts & 9 \\
HCI and human factors & Human-agent team modeling and analysis & 20 \\
HCI and human factors & Improved communication with conversational agents & 16 \\
HCI and human factors & Healthcare and mental-health applications & 30 \\
Information integrity & Monitoring changes in information ecosystems & 48 \\
Information integrity & Mis/disinformation detection & 21 \\
Information integrity & Hate speech detection & 15 \\
Information integrity & Recommender algorithm impact assessment & 44 \\
Societal, economic, and environmental studies & Sector-specific constraints & 58 \\
Societal, economic, and environmental studies & Labor and crowdwork conditions & 59 \\
Societal, economic, and environmental studies & Displacement and productivity distribution & 9 \\
Societal, economic, and environmental studies & Civic participation and AI literacy & 31 \\
Societal, economic, and environmental studies & Energy use and carbon footprint & 14 \\
Societal, economic, and environmental studies & E-waste management & 2 \\
Societal, economic, and environmental studies & Supply chain ethics and sustainability & 13 \\
\bottomrule
\end{tabular}
}
\caption{Low-level AIE mitigation categories and their occurrence counts in the corpora as annotated by the authors. (cont.)}
\label{atab:ethics-mitigation-counts2}
\end{table}

\begin{table}
\centering
\rowcolors{3}{gray!10}{white}
\resizebox{\textwidth}{!}{
\begin{tabular}{@{}lrr@{}}
\toprule
\textbf{High-Level Safety Mitigation Strategy} & \textbf{Low-Level Safety Mitigation Strategy} & \textbf{Count} \\
\midrule
    Mechanistic interpretability & Circuit analysis & 66 \\
    Mechanistic interpretability & Sparse autoencoders & 26 \\
    Mechanistic interpretability & Neuron activation visualization & 23 \\
    Mechanistic interpretability & Developmental trajectory interpretability & 26 \\
    Mechanistic interpretability & Interpretable surrogate modeling & 39 \\
    Mechanistic interpretability & Causal scrubbing & 64 \\
    Mechanistic interpretability & Intrinsically interpretable model analysis & 40 \\
    Mechanistic interpretability & Automated interpretability & 119 \\
    Mechanistic interpretability & Concept-based methods and steering & 69 \\
    Mechanistic interpretability & Information theoretical analysis & 54 \\
    Robustness and adversarial resilience & Adversarial training & 131 \\
    Robustness and adversarial resilience & Adversarial data augmentation & 43 \\
    Robustness and adversarial resilience & Remote adversarial patching & 4 \\
    Robustness and adversarial resilience & Belief updates under adversarial training & 5 \\
    Robustness and adversarial resilience & Outlier detection & 59 \\
    Robustness and adversarial resilience & Out-of-distribution generalization & 217 \\
    Robustness and adversarial resilience & Clusterability and label-noise data augmentation & 36 \\
    Robustness and adversarial resilience & Trojan and backdoor discovery & 29 \\
    Robustness and adversarial resilience & Domain adaptation for continual learning & 55 \\
    Robustness and adversarial resilience & Robust data processing and management & 54 \\
    Robustness and adversarial resilience & Methods for bounded memory/compute & 78 \\
    Robustness and adversarial resilience & Robust unsupervised learning & 4 \\
    Robustness and adversarial resilience & Sample efficiency and meta-RL regularization & 124 \\
    Robustness and adversarial resilience & Robust and risk-averse RL & 105 \\
    Safety control and capability containment & Capability thresholds, red lines, and hard constraints & 39 \\
    Safety control and capability containment & Password-locking capabilities & 2 \\
    Safety control and capability containment & Constrained and risk-averse RL & 97 \\
    Safety control and capability containment & Machine unlearning for unsafe information or behaviors & 20 \\
    Safety control and capability containment & Hard constraints on action and state space & 41 \\
    Safety control and capability containment & Safe Pareto improvements and game-theoretic guarantees & 13 \\
    Safety control and capability containment & Limits on external tool usage & 5 \\
    Safety control and capability containment & Scalable oversight frameworks & 29 \\
    Safety control and capability containment & Chain-of-thought control & 36 \\
\bottomrule
\end{tabular}
}
\caption{Low-level AIS mitigation categories and their occurrence counts in the corpora as annotated by the authors.}
\label{atab:safety-mitigation-counts1}
\end{table}

\begin{table}
\centering
\rowcolors{3}{gray!10}{white}
\resizebox{\textwidth}{!}{
\begin{tabular}{@{}lrr@{}}
\toprule
\textbf{High-Level Safety Mitigation Strategy} & \textbf{Low-Level Safety Mitigation Strategy} & \textbf{Count} \\
\midrule
    Reward modelling and learning & Reward modeling and inverse RL for human values & 122 \\
    Reward modelling and learning & Multi-agent cooperation via inverse RL & 8 \\
    Reward modelling and learning & RLHF / RLHAIF / RLAIF fine-tuning & 78 \\
    Reward modelling and learning & Accurate reward specification & 44 \\
    Reward modelling and learning & Audits for reward hacking and hidden goals & 41 \\
    Reward modelling and learning & Normative work on value alignment and corrigibility & 81 \\
    Reward modelling and learning & Preference learning under uncertainty & 64 \\
    Reward modelling and learning & Multi-stakeholder and multi-agent reward learning & 24 \\
    Reward modelling and learning & Constitutional and rule-based reward modelling & 9 \\
    Monitoring, detection, and evaluation & Detecting adversarial examples & 42 \\
    Monitoring, detection, and evaluation & Benchmark datasets and sandboxing environments & 177 \\
    Monitoring, detection, and evaluation & Honeypot monitoring & 4 \\
    Monitoring, detection, and evaluation & Chain-of-thought monitoring & 22 \\
    Monitoring, detection, and evaluation & Safety verification frameworks and standardized pipelines & 57 \\
    Monitoring, detection, and evaluation & Error bounds and formal verification & 44 \\
    Monitoring, detection, and evaluation & Risk assessments and audits & 150 \\
    Monitoring, detection, and evaluation & Model provenance and audit trails & 39 \\
    Monitoring, detection, and evaluation & Safety metrics standardization & 24 \\
    Safe deployment engineering & Real-world deployment and testing & 60 \\
    Safe deployment engineering & Contact-safe continuous control & 2 \\
    Safe deployment engineering & Collaborative robotics & 25 \\
    Safe deployment engineering & Post-deployment monitoring & 50 \\
    Safe deployment engineering & Requirements elicitation with stakeholders & 29 \\
    Safe deployment engineering & Vertical system integration & 51 \\
    Safe deployment engineering & Actionable safe design frameworks and processes & 41 \\
    Safe deployment engineering & Safety case development for compliance & 25 \\
    Safe deployment engineering & Scalable and safe infrastructure & 7 \\
    Theoretical foundations & Mathematical foundations for ML and agents & 145 \\
    Theoretical foundations & Theory on bounded rationality & 31 \\
    Theoretical foundations & Game-theoretic models of AI ecosystems & 66 \\
    Theoretical foundations & Formal theory of corrigibility & 32 \\
    Theoretical foundations & Decision theory & 74 \\
    Forecasting and metascience & Metascience and metasafety research & 65 \\
    Forecasting and metascience & Science of evaluations & 203 \\
    Forecasting and metascience & Forecasting and timelines & 43 \\
    Forecasting and metascience & Standardizing terminology and methodologies & 44 \\
    Forecasting and metascience & Literature surveys & 38 \\
\bottomrule
\end{tabular}
}
\caption{Low-level AIS mitigation categories and their occurrence counts in the corpora as annotated by the authors. (cont.)}
\label{atab:safety-mitigation-counts2}
\end{table}

\clearpage
\section{Codebook of the Qualitative Deductive Thematic Analysis}

\begin{figure}[h]
    \centering
    \includegraphics[width=\linewidth]{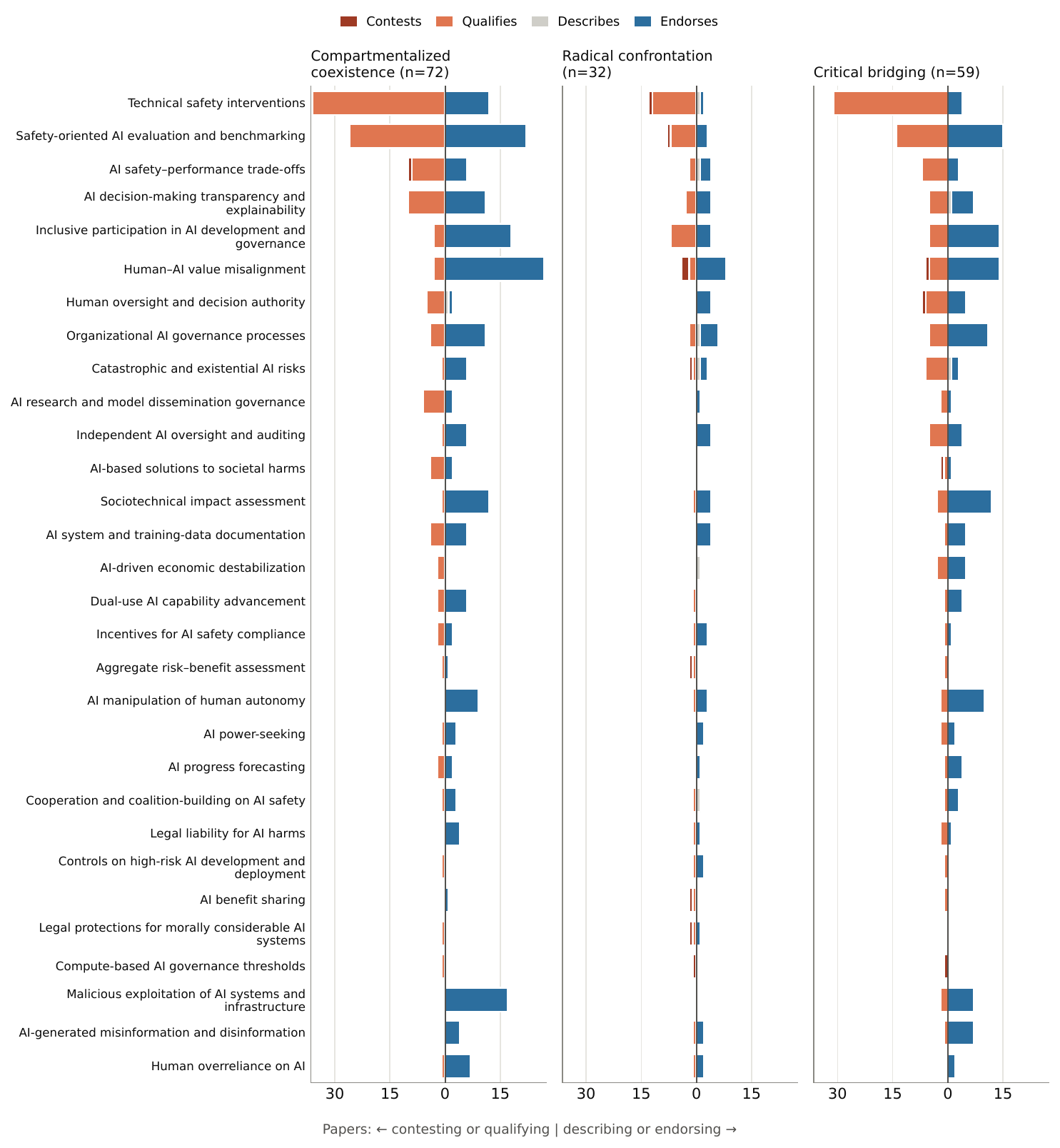}
    \caption{Stance within each resolution mode regarding the top 30 most frequent individual codes assigned to the full texts of our corpora.}
    \label{fig:placeholder}
\end{figure}

\definecolor{qualheat}{HTML}{2c6e9e}
\begingroup
\footnotesize
\setlength{\tabcolsep}{4pt}
\renewcommand{\arraystretch}{1.1}
\def\qualnum{3em}
\edef\qualwidth{\the\linewidth}
\setlength{\LTleft}{0pt}\setlength{\LTright}{0pt}
\setlength{\LTcapwidth}{\linewidth}
\edef\qualcode{\the\dimexpr\linewidth-\qualnum*3-\tabcolsep*6\relax}
\begin{longtable}{@{}>{\raggedright\arraybackslash\hangindent=1em\hangafter=0}p{\qualcode}>{\raggedleft\arraybackslash}p{\qualnum}>{\raggedleft\arraybackslash}p{\qualnum}>{\raggedleft\arraybackslash}p{\qualnum}@{}}
\caption{Share of each resolution mode's papers to which a code was applied, showing only codes which were applied to at least 3 papers.
Codes are grouped by theme which are ordered by the number of papers engaging with them. CC = compartmentalized coexistence ($n=72$); RC = radical confrontation ($n=32$); CB = critical bridging ($n=59$). Darker cells mark larger shares, -- marks no paper.}
\label{tab:qual-codes}\\
\toprule
\multicolumn{1}{@{}l}{Code} & \multicolumn{1}{c}{CC} & \multicolumn{1}{c}{RC} & \multicolumn{1}{c}{CB} \\
\midrule
\endfirsthead
\multicolumn{4}{@{}l}{\tablename\ \thetable{} (continued)}\\
\toprule
\multicolumn{1}{@{}l}{Code} & \multicolumn{1}{c}{CC} & \multicolumn{1}{c}{RC} & \multicolumn{1}{c}{CB} \\
\midrule
\endhead
\midrule
\multicolumn{4}{r@{}}{\emph{Continued on next page}}\\
\endfoot
\bottomrule
\endlastfoot
\multicolumn{4}{@{}>{\raggedright\arraybackslash}p{\qualwidth}@{}}{\textbf{Safety claims must earn evidential authority}}\\*
Safety-oriented AI evaluation and benchmarking & \cellcolor{qualheat!47}67\% & \cellcolor{qualheat!24}34\% & \cellcolor{qualheat!34}49\% \\
AI decision-making transparency and explainability & \cellcolor{qualheat!20}29\% & \cellcolor{qualheat!15}22\% & \cellcolor{qualheat!14}20\% \\
Conceptual ambiguity in AI governance & \cellcolor{qualheat!10}14\% & \cellcolor{qualheat!13}19\% & \cellcolor{qualheat!15}22\% \\
AI system and training-data documentation & \cellcolor{qualheat!10}14\% & \cellcolor{qualheat!9}12\% & \cellcolor{qualheat!7}10\% \\
Misleading claims about AI capabilities and safety & \cellcolor{qualheat!4}6\% & \cellcolor{qualheat!22}31\% & \cellcolor{qualheat!7}10\% \\
Opacity and underdetermination of AI decision-making & \cellcolor{qualheat!3}4\% & \cellcolor{qualheat!4}6\% & \cellcolor{qualheat!2}3\% \\
AI safety assurance & \cellcolor{qualheat!2}3\% & -- & \cellcolor{qualheat!2}3\% \\
Pre-registration of AI research & \cellcolor{qualheat!1}1\% & -- & \cellcolor{qualheat!2}3\% \\
\addlinespace[3pt]
\multicolumn{4}{@{}>{\raggedright\arraybackslash}p{\qualwidth}@{}}{\textbf{Technical protection is conditional, not sufficient}}\\*
Technical safety interventions & \cellcolor{qualheat!47}67\% & \cellcolor{qualheat!33}47\% & \cellcolor{qualheat!42}59\% \\
AI safety--performance trade-offs & \cellcolor{qualheat!16}22\% & \cellcolor{qualheat!13}19\% & \cellcolor{qualheat!12}17\% \\
AI reliability and fitness-for-purpose failures & \cellcolor{qualheat!9}12\% & \cellcolor{qualheat!11}16\% & \cellcolor{qualheat!9}14\% \\
AI-induced physical harm & \cellcolor{qualheat!2}3\% & -- & \cellcolor{qualheat!7}10\% \\
Aggregate risk--benefit assessment & \cellcolor{qualheat!2}3\% & \cellcolor{qualheat!4}6\% & \cellcolor{qualheat!1}2\% \\
Interactions among AI safeguards & \cellcolor{qualheat!2}3\% & \cellcolor{qualheat!2}3\% & \cellcolor{qualheat!1}2\% \\
\addlinespace[3pt]
\multicolumn{4}{@{}>{\raggedright\arraybackslash}p{\qualwidth}@{}}{\textbf{Alignment requires legitimate, revisable value choices}}\\*
Human--AI value misalignment & \cellcolor{qualheat!29}42\% & \cellcolor{qualheat!26}38\% & \cellcolor{qualheat!24}34\% \\
Inclusive participation in AI development and governance & \cellcolor{qualheat!19}28\% & \cellcolor{qualheat!22}31\% & \cellcolor{qualheat!21}31\% \\
Moral uncertainty in AI decision-making & \cellcolor{qualheat!6}8\% & \cellcolor{qualheat!7}9\% & \cellcolor{qualheat!7}10\% \\
Suppression of relevant perspectives in AI development & \cellcolor{qualheat!5}7\% & \cellcolor{qualheat!9}12\% & \cellcolor{qualheat!6}8\% \\
AI value lock-in & \cellcolor{qualheat!1}1\% & -- & \cellcolor{qualheat!2}3\% \\
\addlinespace[3pt]
\multicolumn{4}{@{}>{\raggedright\arraybackslash}p{\qualwidth}@{}}{\textbf{Safety must follow situated, evolving systems}}\\*
Neglect of sociotechnical context & \cellcolor{qualheat!20}29\% & \cellcolor{qualheat!24}34\% & \cellcolor{qualheat!26}37\% \\
Organizational AI governance processes & \cellcolor{qualheat!15}21\% & \cellcolor{qualheat!18}25\% & \cellcolor{qualheat!19}27\% \\
Sociotechnical impact assessment & \cellcolor{qualheat!13}18\% & \cellcolor{qualheat!11}16\% & \cellcolor{qualheat!18}25\% \\
Domain-specific AI governance & \cellcolor{qualheat!1}1\% & \cellcolor{qualheat!4}6\% & \cellcolor{qualheat!6}8\% \\
Adaptive and experimental AI governance & \cellcolor{qualheat!3}4\% & \cellcolor{qualheat!2}3\% & \cellcolor{qualheat!4}5\% \\
Harmful AI feedback loops & \cellcolor{qualheat!4}6\% & -- & \cellcolor{qualheat!4}5\% \\
Sociotechnical AI education and training & \cellcolor{qualheat!3}4\% & -- & \cellcolor{qualheat!2}3\% \\
\addlinespace[3pt]
\multicolumn{4}{@{}>{\raggedright\arraybackslash}p{\qualwidth}@{}}{\textbf{Aggregate progress hides unequal social costs}}\\*
Algorithmic bias, discrimination and skewed representation & \cellcolor{qualheat!24}35\% & \cellcolor{qualheat!20}28\% & \cellcolor{qualheat!19}27\% \\
AI-driven economic destabilization & \cellcolor{qualheat!2}3\% & \cellcolor{qualheat!2}3\% & \cellcolor{qualheat!9}14\% \\
AI-based solutions to societal harms & \cellcolor{qualheat!6}8\% & -- & \cellcolor{qualheat!4}5\% \\
Environmental harms of AI & -- & \cellcolor{qualheat!4}6\% & \cellcolor{qualheat!6}8\% \\
AI labor exploitation & -- & \cellcolor{qualheat!11}16\% & \cellcolor{qualheat!1}2\% \\
Worker support for AI-driven displacement & -- & \cellcolor{qualheat!2}3\% & \cellcolor{qualheat!2}3\% \\
\addlinespace[3pt]
\multicolumn{4}{@{}>{\raggedright\arraybackslash}p{\qualwidth}@{}}{\textbf{Anticipating catastrophe while contesting its framing}}\\*
Critical examination of AI concepts and narratives & \cellcolor{qualheat!3}4\% & \cellcolor{qualheat!20}28\% & \cellcolor{qualheat!17}24\% \\
Catastrophic and existential AI risks & \cellcolor{qualheat!7}10\% & \cellcolor{qualheat!11}16\% & \cellcolor{qualheat!11}15\% \\
AI power-seeking & \cellcolor{qualheat!4}6\% & \cellcolor{qualheat!4}6\% & \cellcolor{qualheat!5}7\% \\
AI progress forecasting & \cellcolor{qualheat!4}6\% & \cellcolor{qualheat!2}3\% & \cellcolor{qualheat!6}8\% \\
Uncertainty in AI risk assessment & \cellcolor{qualheat!2}3\% & \cellcolor{qualheat!4}6\% & \cellcolor{qualheat!5}7\% \\
Compute-based AI governance thresholds & \cellcolor{qualheat!1}1\% & \cellcolor{qualheat!2}3\% & \cellcolor{qualheat!1}2\% \\
\addlinespace[3pt]
\multicolumn{4}{@{}>{\raggedright\arraybackslash}p{\qualwidth}@{}}{\textbf{Delegation cannot dissolve human responsibility}}\\*
Human oversight and decision authority & \cellcolor{qualheat!7}10\% & \cellcolor{qualheat!9}12\% & \cellcolor{qualheat!14}20\% \\
Human overreliance on AI & \cellcolor{qualheat!8}11\% & \cellcolor{qualheat!7}9\% & \cellcolor{qualheat!2}3\% \\
Failures of accountability for AI harms & \cellcolor{qualheat!6}8\% & \cellcolor{qualheat!2}3\% & \cellcolor{qualheat!6}8\% \\
Legal liability for AI harms & \cellcolor{qualheat!4}6\% & \cellcolor{qualheat!4}6\% & \cellcolor{qualheat!4}5\% \\
AI-related legal and regulatory noncompliance & \cellcolor{qualheat!1}1\% & -- & \cellcolor{qualheat!2}3\% \\
\addlinespace[3pt]
\multicolumn{4}{@{}>{\raggedright\arraybackslash}p{\qualwidth}@{}}{\textbf{Openness and security require negotiated boundaries}}\\*
Malicious exploitation of AI systems and infrastructure & \cellcolor{qualheat!17}24\% & -- & \cellcolor{qualheat!11}15\% \\
Dual-use AI capability advancement & \cellcolor{qualheat!8}11\% & \cellcolor{qualheat!2}3\% & \cellcolor{qualheat!6}8\% \\
AI research and model dissemination governance & \cellcolor{qualheat!7}10\% & \cellcolor{qualheat!2}3\% & \cellcolor{qualheat!4}5\% \\
Protected research access to AI systems and data & \cellcolor{qualheat!3}4\% & \cellcolor{qualheat!2}3\% & \cellcolor{qualheat!4}5\% \\
\addlinespace[3pt]
\multicolumn{4}{@{}>{\raggedright\arraybackslash}p{\qualwidth}@{}}{\textbf{Information protection can also suppress agency}}\\*
AI-enabled surveillance and privacy harms & \cellcolor{qualheat!9}12\% & \cellcolor{qualheat!7}9\% & \cellcolor{qualheat!14}20\% \\
AI-generated misinformation and disinformation & \cellcolor{qualheat!4}6\% & \cellcolor{qualheat!7}9\% & \cellcolor{qualheat!9}14\% \\
AI-related polarization and intergroup hostility & \cellcolor{qualheat!3}4\% & -- & \cellcolor{qualheat!8}12\% \\
AI-driven epistemic fragmentation & \cellcolor{qualheat!3}4\% & \cellcolor{qualheat!2}3\% & \cellcolor{qualheat!2}3\% \\
AI-enabled censorship & \cellcolor{qualheat!2}3\% & \cellcolor{qualheat!2}3\% & \cellcolor{qualheat!4}5\% \\
Rights to opt out of AI systems & \cellcolor{qualheat!1}1\% & \cellcolor{qualheat!2}3\% & \cellcolor{qualheat!2}3\% \\
\addlinespace[3pt]
\multicolumn{4}{@{}>{\raggedright\arraybackslash}p{\qualwidth}@{}}{\textbf{Public oversight needs knowledge and enforceable authority}}\\*
Independent AI oversight and auditing & \cellcolor{qualheat!7}10\% & \cellcolor{qualheat!9}12\% & \cellcolor{qualheat!11}15\% \\
AI incident reporting and sharing & \cellcolor{qualheat!3}4\% & \cellcolor{qualheat!2}3\% & \cellcolor{qualheat!5}7\% \\
Controls on high-risk AI development and deployment & \cellcolor{qualheat!1}1\% & \cellcolor{qualheat!7}9\% & \cellcolor{qualheat!1}2\% \\
Public-sector AI safety and governance capacity & \cellcolor{qualheat!4}6\% & -- & \cellcolor{qualheat!1}2\% \\
Technology-neutral AI regulation & \cellcolor{qualheat!1}1\% & \cellcolor{qualheat!7}9\% & \cellcolor{qualheat!1}2\% \\
AI governance information gaps & \cellcolor{qualheat!3}4\% & -- & \cellcolor{qualheat!1}2\% \\
\addlinespace[3pt]
\multicolumn{4}{@{}>{\raggedright\arraybackslash}p{\qualwidth}@{}}{\textbf{Helpful interaction can undermine autonomy and care}}\\*
AI manipulation of human autonomy & \cellcolor{qualheat!9}12\% & \cellcolor{qualheat!9}12\% & \cellcolor{qualheat!14}20\% \\
AI-induced emotional distress & \cellcolor{qualheat!3}4\% & \cellcolor{qualheat!2}3\% & \cellcolor{qualheat!4}5\% \\
Harmful or inappropriate AI-generated content & \cellcolor{qualheat!1}1\% & \cellcolor{qualheat!4}6\% & \cellcolor{qualheat!4}5\% \\
Erosion of human relationships and relational norms & \cellcolor{qualheat!1}1\% & \cellcolor{qualheat!2}3\% & \cellcolor{qualheat!2}3\% \\
AI sycophancy & \cellcolor{qualheat!1}1\% & -- & \cellcolor{qualheat!2}3\% \\
\addlinespace[3pt]
\multicolumn{4}{@{}>{\raggedright\arraybackslash}p{\qualwidth}@{}}{\textbf{Concentrated control limits meaningful alternatives}}\\*
Concentration of AI power & \cellcolor{qualheat!10}14\% & \cellcolor{qualheat!13}19\% & \cellcolor{qualheat!11}15\% \\
AI-driven economic inequality & \cellcolor{qualheat!2}3\% & -- & \cellcolor{qualheat!5}7\% \\
AI benefit sharing & \cellcolor{qualheat!1}1\% & \cellcolor{qualheat!4}6\% & \cellcolor{qualheat!1}2\% \\
Market-based AI governance & \cellcolor{qualheat!1}1\% & -- & \cellcolor{qualheat!4}5\% \\
AI interoperability and data portability & \cellcolor{qualheat!2}3\% & -- & \cellcolor{qualheat!1}2\% \\
\addlinespace[3pt]
\multicolumn{4}{@{}>{\raggedright\arraybackslash}p{\qualwidth}@{}}{\textbf{Race incentives make safety a collective-action problem}}\\*
Competitive pressures undermining AI safety & \cellcolor{qualheat!6}8\% & \cellcolor{qualheat!2}3\% & \cellcolor{qualheat!4}5\% \\
Cooperation and coalition-building on AI safety & \cellcolor{qualheat!4}6\% & \cellcolor{qualheat!4}6\% & \cellcolor{qualheat!5}7\% \\
Incentives for AI safety compliance & \cellcolor{qualheat!4}6\% & \cellcolor{qualheat!9}12\% & \cellcolor{qualheat!2}3\% \\
AI-related geopolitical conflict & \cellcolor{qualheat!3}4\% & \cellcolor{qualheat!2}3\% & \cellcolor{qualheat!1}2\% \\
\addlinespace[3pt]
\multicolumn{4}{@{}>{\raggedright\arraybackslash}p{\qualwidth}@{}}{\textbf{AI protection challenges human-centered safety}}\\*
Legal protections for morally considerable AI systems & \cellcolor{qualheat!1}1\% & \cellcolor{qualheat!7}9\% & -- \\
Mistreatment of morally considerable AI systems & \cellcolor{qualheat!2}3\% & \cellcolor{qualheat!2}3\% & \cellcolor{qualheat!1}2\% \\
AI-induced disruption of legal institutions & \cellcolor{qualheat!1}1\% & \cellcolor{qualheat!2}3\% & \cellcolor{qualheat!1}2\% \\
Relational ethics for human--AI coexistence & -- & \cellcolor{qualheat!2}3\% & \cellcolor{qualheat!2}3\% \\
\addlinespace[3pt]
\multicolumn{4}{@{}>{\raggedright\arraybackslash}p{\qualwidth}@{}}{\textbf{Individual safety does not establish collective safety}}\\*
Correlated AI failures & \cellcolor{qualheat!3}4\% & \cellcolor{qualheat!2}3\% & \cellcolor{qualheat!1}2\% \\
AI collusion & \cellcolor{qualheat!2}3\% & -- & \cellcolor{qualheat!1}2\% \\
Failures of coordination and cooperation between AI systems & \cellcolor{qualheat!2}3\% & -- & \cellcolor{qualheat!1}2\% \\
\end{longtable}
\endgroup

\clearpage
\section{Further Qualitative Results and Examples}\label{ssec:results:quantitative}

\begin{figure}[h]
  \includegraphics[width=0.49\textwidth]{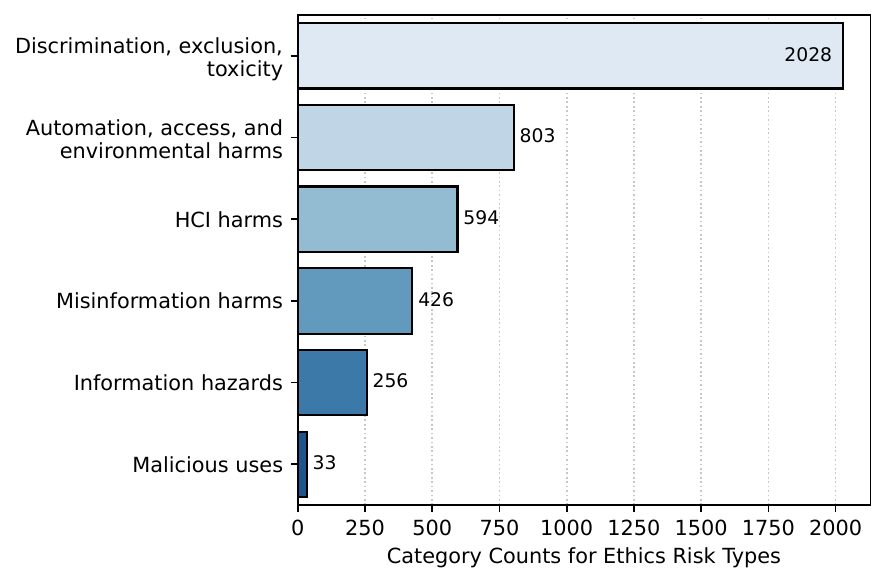}
  \includegraphics[width=0.49\textwidth]{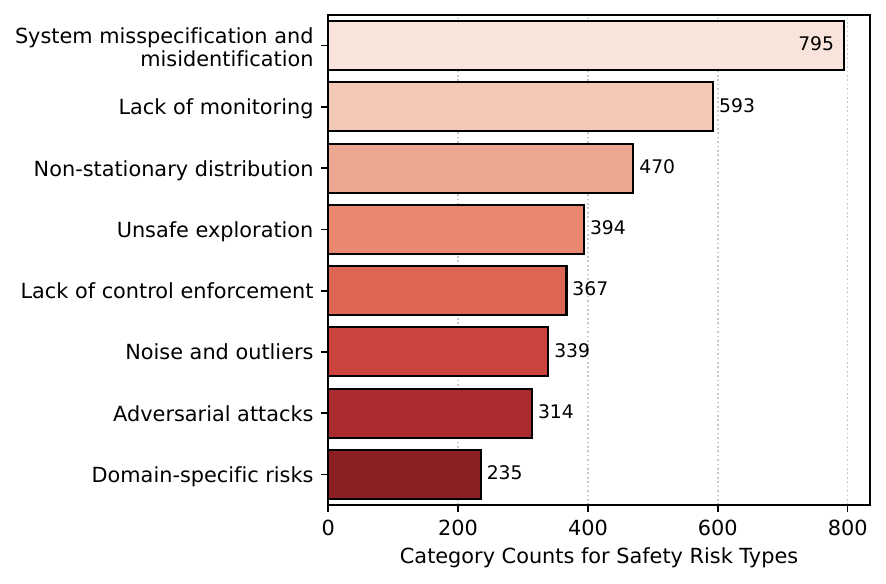}
  \caption{Bar plots showing the number of papers assigned to each high-level risk category in the AIE corpus (\textbf{Left}) and the AIS corpus (\textbf{Right}). Each paper was assigned to up to three distinct categories.}
  \label{fig:risk-counts}
\end{figure}

\begin{figure}[h]
  \includegraphics[width=0.49\textwidth]{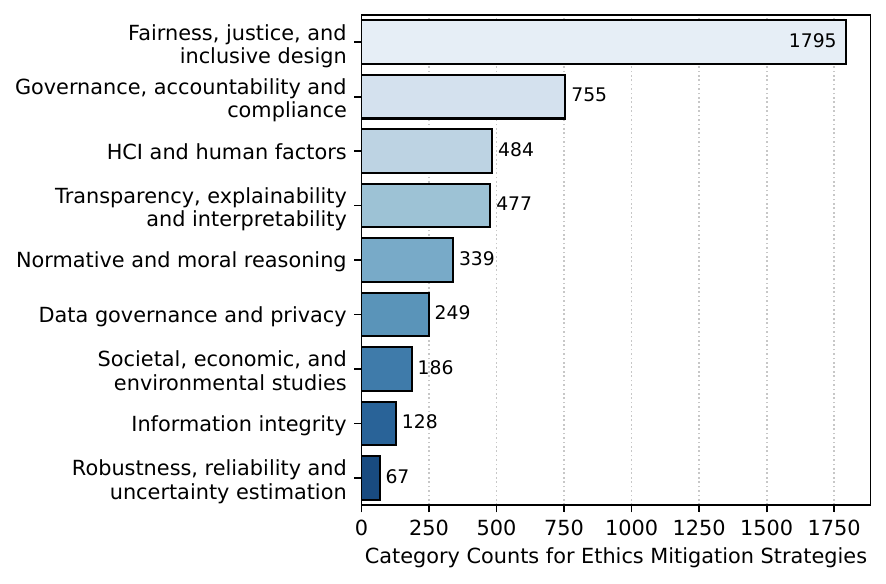}
  \includegraphics[width=0.49\textwidth]{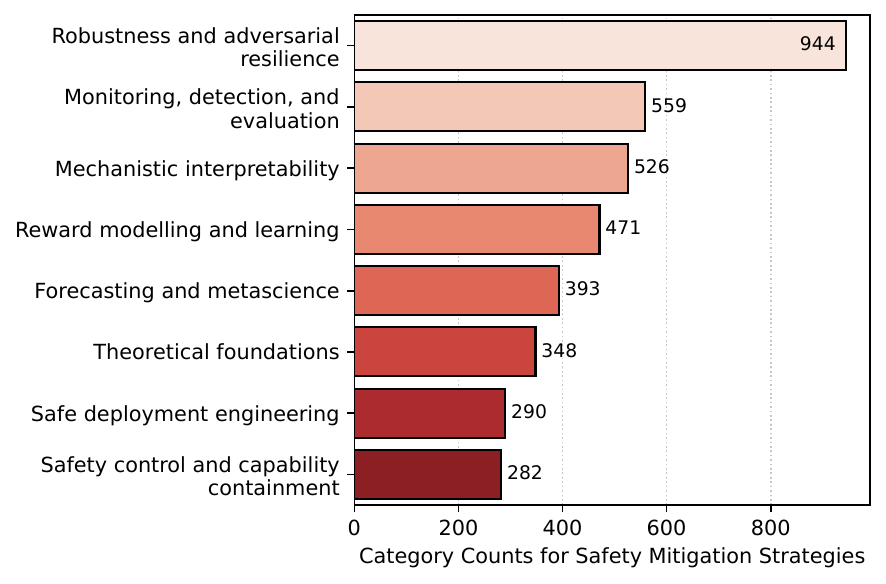}
  \caption{Bar plots showing the number of papers assigned to each high-level mitigation strategy category in the AIE corpus (\textbf{Left}) and the AIS corpus (\textbf{Right}). Each paper was assigned to up to three distinct categories. \textit{HCI: Human-Computer Interaction}}
  \label{fig:mitigation-counts}
\end{figure}

\Cref{fig:risk-counts,fig:mitigation-counts} show the number of papers assigned to each high-level category of our risk and mitigation taxonomies.

Qualitatively, disengagement papers tend to be narrowly focused technical or empirical contributions.
It is especially common for these papers to judge success by some quantitative metrics and take their normative goal as given, as when \textit{``we assess fairness by testing whether a ranking could plausibly have been generated using a fair sampling strategy''}~\cite{cartiervandisselHyperFAIRHypergeometricApproach2025}.
When the other field is mentioned at all, it is usually only in passing, to reinforce the paper's own arguments, such as in \textit{``safety concerns and social discrimination in language models are typically explicit and well-recognized''}~\cite{seoValuesRAGEnhancingCultural2025}. 
Shared vocabulary between AIE and AIS, meanwhile, is often used in unrelated senses, from \textit{``the interplay between alignment, orthogonality and opposition of per-token gradients''}~\cite{mirceaLanguageModelScaling2024} to \textit{``a method for computing the alignment between model and human decision-making''}~\cite{boggustSharedInterestMeasuring2022}.

As we saw already, compartmentalized coexistence brings both fields' concerns into one frame, but then treats the concerns of either field separately, mostly as a narrative device.
\citeauthor{huangCollectiveConstitutionalAI2024}~\cite{huangCollectiveConstitutionalAI2024}, for example, start from the premise that \textit{``language model (LM) developers should not be the sole deciders of LM behavior''} and run a public deliberation exercise as a form of alignment method.
As such, this work involves high integration of methods from both fields, but low engagement with the primary concerns of AIS, placing the paper squarely into compartmentalized coexistence.
We find similar examples for this resolution mode, for instance by~\citeauthor{ehyaeiBridgingCausalityIndividual2024}~\cite{ehyaeiBridgingCausalityIndividual2024}, who note that \textit{``robustness, fairness, and causality are often studied in isolation''} and unify them in one metric, but treat the concerns of AIE only in the motivation.

Radical confrontation is much rarer and -- at least in traditional paper-based publications as opposed to, e.g., online forums -- usually a secondary thread. 
From the AI ethics side, it appears as non-technical and dismissive.
Without citing the papers, one work calls loss-of-control concerns \textit{``an extremely premature set of concerns''}, observing that \textit{``if one thinks the danger of AI is that it will work too well [\dots] it is a necessary precondition that it works at all''}. 
Another work urges policymakers away from \textit{``speculative fiction masquerading as expertise''} and \textit{``hypothetical and self-aggrandizing concerns''}.
On the other hand, from the AI safety side, radical confrontation ranges from the rejection of machine ethics, as in \textit{``we do not need machines that are Full Ethical Agents [\dots] we need our machines to be inherently stable and safe''}, to more reserved philosophical rebuttals of the claim that language models cannot hold beliefs: \textit{``These arguments, however, rest on a mistake''}.
Much of the dispute above turns on how capable AI really can be, as opposed to what moral values each field embodies.

Finally, critical bridging papers mostly argue that alignment is a question of values and power, highlighting that safety has to include social harms and pluralistic desiderata.
\citeauthor{varshneyDecolonialAIAlignment2024}~\cite{varshneyDecolonialAIAlignment2024} put this plainly: \textit{``Values should not be assumed universal. Contextual and relational values should come from the communities in which the LLM will be deployed''}.
In addition, some of the literature explicitly addresses the divide between the fields, usually in the context of AI governance.
\citeauthor{prunklLongtermClearerAccount2020}~\cite{prunklLongtermClearerAccount2020} refer to \textit{``two seemingly separate communities''}, arguing that the near- and long-term distinction \textit{``risks causing confusion and fuelling conflict''} and that we should instead aim to \textit{``establish new opportunities for collaboration and reduce adversarial dynamics''}. 
In a different example, \citeauthor{stixBridgingGapCase2021}~\cite{stixBridgingGapCase2021} propose that scholars \textit{``converge and cooperate on selected mutually beneficial AI policy projects, while maintaining their distinct perspectives''}.

\end{document}